\documentclass[twoside]{article}

\usepackage{color}
\usepackage{amssymb,amsmath,amsthm} 
\usepackage{verbatim}
\usepackage{shadow}

\usepackage[hidelinks]{hyperref} 

\usepackage{boxedminipage}
\usepackage{fancybox}
\usepackage{graphicx,float}
\usepackage{newcent}
\usepackage{amssymb}%
\usepackage{pstricks}%
\usepackage{dsfont}
\usepackage{hyperref}
\usepackage{fancyhdr}
\usepackage{xcolor}
\usepackage{amsfonts}
\usepackage{algorithmic}
\usepackage{graphicx,epstopdf}
\usepackage[caption=false]{subfig}
\usepackage[official,right]{eurosym}
\usepackage{booktabs}
\usepackage{hyperref}

\usepackage{color}
\usepackage{amssymb,amsmath,amsthm} 
\usepackage{verbatim}
\usepackage{shadow}

\usepackage[hidelinks]{hyperref} 

\usepackage{boxedminipage}
\usepackage{fancybox}
\usepackage{graphicx,float}
\usepackage{newcent}
\usepackage{amssymb}%
\usepackage{pstricks}%
\usepackage{dsfont}
\usepackage{hyperref}
\usepackage{fancyhdr}
\usepackage{xcolor}
\usepackage{booktabs}
\usepackage{subcaption}

\numberwithin{equation}{section}

\newtheorem{theorem}{Theorem}[section]
\newtheorem{lem}{Lemma}[section]
\newtheorem{pro}{Proposition}[section]
\newtheorem{cor}{Corollary}[section]
\newtheorem{rem}{Remark}[section]
\newtheorem{rems}{Remarks}[section]
\newtheorem{ex}{Example}[section]
\newtheorem{defi}{Definition}[section]
\newtheorem{hyp}{Assumption}[section]

\newcommand{\bt}{\begin{theorem}}
\newcommand{\et}{\end{theorem}}
\newcommand{\bl}{\begin{lem}}
\newcommand{\el}{\end{lem}}
\newcommand{\bp}{\begin{pro}}
\newcommand{\ep}{\end{pro}}
\newcommand{\bcor}{\begin{cor}}
\newcommand{\ecor}{\end{cor}}

\newcommand{\bd}{\begin{defi} \rm }
\newcommand{\ed}{\end{defi}}
\newcommand{\brem }{\begin{rem} \rm }
\newcommand{\erem }{\end{rem}}
\newcommand{\brems }{\begin{rems} \rm }
\newcommand{\erems }{\end{rems}}
\newcommand{\bhyp }{\begin{hyp} \rm }
\newcommand{\ehyp }{\end{hyp}}
\newcommand{\bex}{\begin{ex} \rm }
\newcommand{\eex}{\end{ex}}

\newcommand{\PMM}{pricing martingale measure }

\newcommand{\bT}{\sigma_{\Td}}
\newcommand{\bS}{\sigma_{\Sd}}
\newcommand{\bU}{\sigma_{\Ud}}
\newcommand{\hbT}{\wh{\sigma}_{\Td}}
\newcommand{\hbS}{\wh{\sigma}_{\Sd}}
\newcommand{\hbU}{\wh{\sigma}_{\Ud}}

\newcommand{\buT}{\sigma_{\Td}(u)}
\newcommand{\sbuT}{\sigma^2_{\Td}(u)}
\newcommand{\buS}{\sigma_{\Sd}(u)}

\newcommand{\buU}{\sigma_{\Ud}(u)}

\newcommand{\sbvT}{\sigma^2_{\Td}(v)}

\newcommand{\btT}{\sigma_{\Td}(t)}

\newcommand{\btS}{\sigma_{\Sd}(t)}

\newcommand{\btU}{\sigma_{\Ud}(t)}

\newcommand{\hbuT}{\wh{\sigma}_{\Td}(u)}

\newcommand{\hbuS}{\wh{\sigma}_{\Sd}(u)}

\newcommand{\hsbvT}{\wh{\sigma}^2_{\Td}(v)}

\newcommand{\hbtT}{\wh{\sigma}_{\Td}(t)}

\newcommand{\hbtS}{\wh{\sigma}_{\Sd}(t)}

\newcommand{\hbtU}{\wh{\sigma}_{\Ud}(t)}

\newcommand{\btST}{\sigma_{\Sd,\Td}(t)}
\newcommand{\btUT}{\sigma_{\Ud,\Td}(t)}
\newcommand{\btSU}{\sigma_{\Sd,\Ud}(t)}

\newcommand{\hbtUT}{\wh{\sigma}_{\Ud,\Td}(t)}

\newcommand{\GSU}{\Gamma_{t,\Sd,\Ud}}
\newcommand{\GST}{\Gamma_{t,\Sd,\Td}}

\newcommand{\fbGUT}{\Gamma^{\beta,f,1}_{t,\Ud,\Td}}
\newcommand{\fbGT}{\Gamma^{\beta,f,1}_{t,\Td}}
\newcommand{\fbdGT}{\Gamma^{\beta,f,2}_{t,\Td}}
\newcommand{\fbtGSUT}{\Gamma^{\beta,f,3}_{t,\Sd,\Ud,\Td}}
\newcommand{\fbtGUT}{\Gamma^{\beta,f,3}_{t,\Ud,\Td}}
\newcommand{\fbtGT}{\Gamma^{\beta,f,3}_{t,\Td}}
\newcommand{\fbcGST}{\Gamma^{\beta,f,4}_{t,\Sd,\Td}}
\newcommand{\fbcGT}{\Gamma^{\beta,f,4}_{t,\Td}}

\newcommand{\Xbeta}{X^{\beta}}
\newcommand{\Xbetaf}{X^{f,\beta}}
\newcommand{\Xbetad}{X^{d,\beta}}
\newcommand{\Xbetap}{X^{p,\beta}}

\newcommand{\Xbetafx}{X^{f,\beta,1}}
\newcommand{\Xbetadx}{X^{d,\beta,1}}
\newcommand{\Xbetapx}{X^{p,\beta,1}}

\newcommand{\Xbetafy}{X^{f,\beta,2}}
\newcommand{\Xbetady}{X^{d,\beta,2}}
\newcommand{\Xbetapy}{X^{p,\beta,2}}

\newcommand{\Kdb}{K^{d,\beta}}
\newcommand{\Ifb}{I^{f,\beta}}
\newcommand{\Idb}{I^{d,\beta}}
\newcommand{\Ipb}{I^{p,\beta}}

\newcommand{\mart}{\simeq}

\newcommand{\gm}{\gamma}
\newcommand{\cq}{q}
\newcommand{\kpde}{\wt{\kappa}}
\newcommand{\diff}{\mathop{}\!d}
\newcommand{\wh}{\widehat}
\newcommand{\wt}{\widetilde}

\newcommand{\nnN}{N}
\newcommand{\hnnN}{\wh{N}}

\newcommand{\cqtT}{c^q_{t,T}}

\newcommand{\lqu}{\lambda^q_u}
\newcommand{\lqt}{\lambda^q_t}
\newcommand{\lqtT}{\lambda^q_{t,\Td}}
\newcommand{\LqtT}{\Lambda^{q}_{t,\Td}}
\newcommand{\LqtTo}{\Lambda^{q}_{t,\Td_0}}
\newcommand{\LqtTj}{\Lambda^{q}_{t,\Td_j}}
\newcommand{\LqtTn}{\Lambda^{q}_{t,\Td_n}}
\newcommand{\LqtS}{\Lambda^{q}_{t,\Sd}}

\newcommand{\Abst}{A^{\beta,d}_{s,t}}
\newcommand{\AbtT}{A^{\beta,d}_{t,\Td}}

\newcommand{\AbtTj}{A^{\beta,d}_{t,\Td_j}}
\newcommand{\AbtTn}{A^{\beta,d}_{t,\Td_n}}

\newcommand{\Abfst}{A^{\beta,f}_{s,t}}
\newcommand{\AbftT}{A^{\beta,f}_{t,\Td}}

\newcommand{\CF}{\textbf{CF}}
\newcommand{\CCBS}{\textbf{CCBS}}

\newcommand{\PSwn}{\textbf{PSwn}}
\newcommand{\RSwn}{\textbf{RSwn}}

\newcommand{\rhodf}{\rho_{12}}
\newcommand{\rhod}{\rho_{13}}
\newcommand{\rhof}{\rho_{23}}

\newcommand{\rd}{r^d}
\newcommand{\rf}{r^f}
\newcommand{\rc}{r^c}

\newcommand{\xd}{\bar{r}^d}
\newcommand{\xf}{\bar{r}^f}

\newcommand{\phiv}{\varphi}
\newcommand{\phid}{\varphi^d}
\newcommand{\phif}{\varphi^f}
\newcommand{\phiq}{\varphi^{\cq}}

\newcommand{\Phiq}{\Phi^{d,f,q}}

\newcommand{\zetad}{\zeta^d}
\newcommand{\zetaf}{\zeta^f}
\newcommand{\wtzetad}{\wt{\zeta}^d}
\newcommand{\wtzetaf}{\wt{\zeta}^f}
\newcommand{\whzetad}{\wh{\zeta}^d}
\newcommand{\whzetaf}{\wh{\zeta}^f}

\newcommand{\Xd}{X^d}
\newcommand{\Xf}{X^f}

\newcommand{\Bd}{B^d}
\newcommand{\Bdd}{B^d}
\newcommand{\Bf}{B^f}
\newcommand{\Rd}{R^d}
\newcommand{\Rf}{R^f}
\newcommand{\whB}{\wh{B}}

\newcommand{\FB}{B}
\newcommand{\Sd}{S}
\newcommand{\Td}{T}
\newcommand{\Ud}{U}

\newcommand{\alone}{\alpha_1}
\newcommand{\altwo}{\alpha_2}
\newcommand{\althree}{\alpha_3}

\newcommand{\sigmad}{\sigma_{d}}
\newcommand{\sigmaf}{\sigma_{f}}
\newcommand{\sigmaq}{\sigma_q}

\newcommand{\barsig}{\bar{\sigma}}

\newcommand{\Ww}{W}
\newcommand{\whW}{\wh{W}}

\newcommand{\Wone}{W^1}
\newcommand{\Wtwo}{W^2}
\newcommand{\Wthree}{W^3}

\newcommand{\Zone}{Z^1}
\newcommand{\Ztwo}{Z^2}
\newcommand{\Zthree}{Z^3}

\newcommand{\whZone}{\wh{Z}^1}
\newcommand{\whZtwo}{\wh{Z}^2}
\newcommand{\whZthree}{\wh{Z}^3}

\newcommand{\Fd}{F^d}

\newcommand{\Ff}{F^f}

\newcommand{\Fq}{F^q}

\newcommand{\Ffq}{F^{f,\cq}}

\newcommand{\nud}{\nu^d}

\newcommand{\nuf}{\nu^f}

\newcommand{\nufq}{\nu^{f,\cq}}

\newcommand{\ff}{{\mathbb F}}
\newcommand{\pp}{{\mathbb P}}
\newcommand{\rr}{{\mathbb R}}

\newcommand{\E}{{\mathbb E}}

\newcommand{\cE}{\mathcal{E}}
\newcommand{\cF}{\mathcal{F}}
\newcommand{\cT}{\mathcal{T}}

\newcommand{\Q}{\mathbb{Q}}

\newcommand{\EQ}{{\E}_{\Q}}

\newcommand{\whQ}{\wh{\Q}}
\newcommand{\EwhQ}{{\E}_{\whQ}}

\newcommand{\wtQ}{\wt{\Q}}
\newcommand{\EwtQ}{{\E}_{\wtQ}}

\DeclareRobustCommand{\qvar}[2]{\ensuremath{\langle{#1} \rangle}_{#2}}

\ifpdf
  \DeclareGraphicsExtensions{.eps,.pdf,.png,.jpg}
\else
  \DeclareGraphicsExtensions{.eps}
\fi

\makeatletter
\def\mathcolor#1#{\@mathcolor{#1}}
\def\@mathcolor#1#2#3{%
  \protect\leavevmode
  \begingroup
    \color#1{#2}#3%
  \endgroup
}
\makeatother

\title{{\Large \bf Multi-Curve Approach to Cross-Currency Basis Swaps \\
   Referencing Backward-Looking Term Rates \vskip 20 pt }}

\author{Yining Ding$\,^{a}$, Ruyi Liu$\,^{b}$ and Marek Rutkowski$\,^{a,c}$ \\ \\ \\ \\
\\
$^{a\,}$School of Mathematics and Statistics, University of Sydney \\ 
Sydney, NSW 2006, Australia \\ \\
$^{b\,}$School of Mathematics and Statistics, University of New South Wales \\ 
Sydney, NSW 2033, Australia \\ \\
$^{c\,}$Faculty of Mathematics and Information Science, Warsaw University of Technology \\ 
00-661 Warszawa, Poland
}

\date{\vskip 20 pt Final revision: October 30, 2025 \vskip 15 pt }

\hypersetup{final} 
\begin{document}

\maketitle

\begin{abstract}
The financial industry has undergone a significant transition from the London Interbank Offered Rates (LIBORs) to Risk Free Rates (RFRs) such as, e.g., the Secured Overnight Financing Rate (SOFR) in the U.S. and the Cash Rate (AONIA) in Australia, as primary benchmark rates for borrowing costs. The paper examines the pricing and hedging method for financial products in a cross-currency framework with the special emphasis on the Compound SOFR vs Average AONIA cross-currency basis swap (CCBS) where both reference rates are backward-looking and the swap is collateralized. While the SOFR and AONIA are used as particular instances of RFRs in a cross-currency basis swap, the proposed approach is able to handle backward-looking rates for any two currencies. We give explicit pricing and hedging results for a
constant notional cross-currency basis swap with either domestic or foreign collateralization using interest rate futures and currency futures as hedging instruments within an arbitrage-free cross-currency multi-curve setting.
\end{abstract}
\vskip 30 pt \noindent The research of R.~Liu, and M.~Rutkowski was supported by the Australian Research Council Discovery Project scheme
under grant DP200101550 {\it Fair pricing of superannuation guaranteed benefits with downturn risk}. This work has been accepted for publication in {\it SIAM Journal on Financial Mathematics} on October 30, 2025.

\newpage
{\small \tableofcontents}
%
%
%


\section{Introduction}

Interest rate benchmarks are central to the fixed income market, playing a crucial role in determining the cost of bank borrowing in wholesale money markets. Historically, the most widely used credit-based benchmarks for floating interest rates was the London Interbank Offered Rate (LIBOR) and similar forward-looking rates in other fixed income markets. The LIBOR reflected the expected future cost of interbank borrowing and lending and, traditionally, its quotations for maturities ranging from overnight to one year were based on submissions from a panel of banks.
However, the global financial crisis (GFC) of 2007-2009 exposed significant challenges within the financial industry, including a sharp decline in interbank financing transactions due to stricter regulations and increased credit risks. Moreover, LIBOR faced practical issues such as manipulation and sustainability concerns.

Consequently, in 2017, the Financial Conduct Authority (FCA) announced the phased discontinuation of LIBOR and, in response, several alternative benchmark rates were adopted in major economies. One prominent replacement are the backward-looking averages associated with the risk-free rates (RFRs). Notable examples of risk-free rates in major economies include the Secured Overnight Financing Rate (SOFR) in the United States, the Euro Short-Term Rate (\officialeuro{}STR) in the Eurozone, the Sterling Overnight Index Average (SONIA) in the United Kingdom,
the Tokyo Overnight Average Rate (TONAR) in Japan, and the Cash Rate (also informally known as the AONIA) in Australia.

The reform of interest rate benchmarks has garnered significant attention, particularly in light of the challenges posed by the transition from LIBOR to alternative rates like SOFR. For a general analysis of volatility adjustments for options on backward-looking term rates, based on short rate assumptions, we refer to \cite{Piterbarg2020}. One of the seminal contributions to this field is \cite{Lyashenko2019SOFR} where the LIBOR market model is extended to incorporate backward-looking rates such as SOFR. In the post-LIBOR landscape, the classical short-rate models have been revisited by several authors. Notably, the Hull and White model has been employed in \cite{Hofmann2020} and \cite{Turfus2020}. In the context of modeling SOFR futures, \cite{Skov2021rer} proposes a multifactor Gaussian Nelson-Siegel model, demonstrating its suitability for the SOFR futures market. Furthermore, in \cite{Backwell2020affine} and \cite{Fontana2023affine} the authors have developed models using affine term structure to represent the dynamics of RFRs, while \cite{Gellert2021} employs the Heath, Jarrow and Morton (HJM) model for instantaneous forward rates. In \cite{Macrina2020rfr}, the authors adopt a linear rational model for the savings account, enabling closed-form pricing formulae  for caplets, swaptions, and futures based on backward-looking benchmarks. Similarly, \cite{Bickersteth2025SOFR} also provide closed-form pricing results, while also developing explicit hedging strategies using market observables, such as SOFR futures. Given the complexities involved in hedging, we focus on Vasicek-type dynamics for the factor process, as outlined in \cite{Bickersteth2025SOFR}, due to their ability to produce analytic results while maintaining tractability for hedging applications. Additionally, the incorporation of stochastic discontinuities (spikes) in the dynamics of overnight rates has been explored in the literature. For a more
in-depth analysis of this topic, the interested reader is referred to \cite{Backwell2021jumps,Brace2024,Fontana2023spikes,Gellert2021}.

Since the GFC, numerous fundamental paradigms underpinning financial valuation have been rightly questioned. One notable change important for this work has been the widening spreads between certain interest rates, particularly between overnight rates and unsecured rates like LIBOR, as well as between these rates and those agreed upon in repurchase agreements (repo rates). Even prior to the GFC, repurchase agreements and collateralization were recognized as viable methods to finance cash flows, primarily aimed at managing counterparty risk. With the increased importance of collateralization agreements and the emergence of interest rate spreads, it has become clear that greater caution is needed in the context of valuation and hedging. In particular, when multiple sources of funding are utilized, the spreads between interest rates linked to different funding sources must be carefully scrutinized.

In \cite{Pallavicini2011funding}, the authors explore the effects of collateralization, factoring in the costs associated with different rates paid and received on the margin account. The important distinction in valuation between unsecured and collateralized swaps is analyzed in \cite{Piterbarg2010funding}. Furthermore, \cite{Crepey2015bilateralV1} and \cite{Crepey2015bilateralV2} propose a hedging framework that decomposes transaction value into different components, each linked to distinct funding accounts. For further insights, we refer to \cite{Brigo2014funding} and \cite{Castagna2013}.
The contributions mentioned above are limited to a single currency framework. However, when multiple currencies are considered, funding strategies and collateralization agreements become significantly more complex. Funding strategies in a multi-currency setting using FX swaps as basic collateralized instruments are studied in \cite{piterbarg2012cooking}. It is demonstrated in \cite{Fujii2011Choice} (see also \cite{Wolf2022}) that collateralization significantly impacts derivatives pricing, emphasizing that the choice of collateral currency can notably affect derivative prices, a finding that is verified in our model outlined in Remark \ref{Rem: choice}. In \cite{fujii2010note,fujii2010multi,Gnoatto2023cchjm,Gnoatto2021multi}, the authors also provide valuation formulae for contingent claims involving currency dislocations between contractual and collateral cash flows, although their approach uses the unsecured funding rate as the numeraire.

Our research is primarily inspired by \cite{Bickersteth2025SOFR}, who provided an explicit formula for pricing and hedging collateralized SOFR derivatives within a single currency framework. We extend their results to a cross-currency context, making our main contribution the provision of closed-form pricing formulae  and explicit hedging strategies for various fixed-income financial products based on existing futures contracts. The selection of instruments used to construct hedging strategies is also motivated by the studies presented in \cite{Jamshidian1994}, \cite{Lyashenko2019SOFR} and \cite{Mercurio2018}.

Consistent with \cite{Bickersteth2025SOFR}, we adopt a multi-curve framework but the possibility of default by either party is excluded. This approach enables us to capture discrepancies among funding rates, collateral rates, and repo rates across different economies. The central contribution of our work lies in the explicit analytic hedging strategy we propose. While similar pricing results have been presented in previous studies (e.g., \cite{Henrard2018} and \cite{Lyashenko2019SOFR}), those works did not address the crucial issue of hedging (see, however, \cite{Jamshidian1994} for hedging results for differential swaps and quanto options). Moreover, some authors have examined hedging strategies within the context of market frictions, providing numerical results (see, e.g., \cite{Buehler2019deephedging} and \cite{Rogers2010hedge}). Uniquely, our hedging strategies are articulated in terms of tradable assets, such as existing futures contracts on interest rate averages and currencies, rather than abstract model-specific processes. Although we focus
on explicit solutions for a {\it constant notional} CCBS, it is clear that the same approach and techniques can be applied with only minor modifications to other instances of a CCBS encountered in market practice, for instance, a {\it mark-to-market} CCBS or a {\it quanto} CCBS.

The paper is organized as follows. In Section \ref{sec2}, we present a general formulation of benchmark interest rates and we provide a formal definition of cross-currency financial products equipped with a payment scheme without imposing any specific model. In Section \ref{sec3}, we introduce trading strategies involving various types of futures contracts and we propose a suitable notion of a martingale measure for contracts with proportional collateralization. This allows us to obtain a general representation for arbitrage-free price of any cross-currency contract. In  Section \ref{sec4}, we develop a framework for modeling interest rate and exchange rate dynamics, employing two Vasicek's models (one for each economy) and the classical Garman and Kohlhagen model for the exchange rate, and explore the dynamics of different futures rates relevant for hedging.
In  Section \ref{sec5} and Section \ref{sec6}, we combine all the results derived previously to obtain closed-form results for the pricing and hedging of AONIA/SOFR cross-currency basis swaps with both domestic and foreign collateralization using the futures contracts.  Explicit pricing and hedging formulae are provided, with some proofs relegated to the appendix. We conclude the paper by presenting, in  Section \ref{sec7}, numerical results for cross-currency swaps and swaptions. We first verify by Monte Carlo method the correctness of our previously derived pricing and hedging results and, subsequently, we provide a detailed sensitivity analysis and an examination of risk exposure. For practical collateral guidelines and a detailed specification of deliverable assets, the reader is referred to the ISDA documents \cite{ISDA2003} and \cite{ISDA2005}.

\section{Overnight interest rates and related averages}  \label{sec2}

Cross-currency derivatives are financial instruments that enable investors to manage their exposure to foreign exchange (FX) risk. They belong to a broader category of derivatives, which are financial contracts whose value is derived from the value of an underlying asset or index. The use of cross-currency derivatives has become increasingly popular in recent years as businesses and investors seek to expand their operations globally and are exposed to a greater level of FX risk.

In this section, we  provide an overview of cross-currency derivatives with particular structure, their mechanics involving certain types of fixed income derivatives called interest rate futures.
We define the {\it domestic} and {\it foreign} risk-free rates (by convention, the AONIA in Australia and the SOFR in the U.S.) and the corresponding backward-looking compound rates, the {\it SOFR Average} and the {\it Realised AONIA}. In the formal definition of SOFR and AONIA accounts, we adopt the general convention that the overnight interest rate is continuously compounded, rather than daily as this is done in practice. We henceforth assume that all stochastic processes are defined on the probability space $(\Omega , \mathcal{F}, \mathbb{P})$, which is endowed with the filtration $\mathbb{F}=(\mathcal{F}_t)_{t\in \mathbb{R}_+}$ satisfying the usual conditions of right-continuity and $\mathbb{P}$-completeness.

\bd\label{def2.1}
Let the $\mathbb{F}$-adapted stochastic processes $\rd$ and $\rf$ represent the instantaneous AONIA and SOFR rates, respectively. The continuously compounded {\it AONIA account} $\Bd$ satisfies, for every $t \in \mathbb{R}^+$,
\begin{equation}
\Bd_t= \exp \left( \int_0^t \rd_u \diff u \right)
\end{equation}
and the {\it Realised AONIA} over $[\Ud,\Td]$ is given by
\begin{equation}
\Rd(\Ud,\Td):= \frac{1}{\delta}\left( \exp \left( \int_{\Ud}^{\Td} \rd_u \diff u \right)-1\right)
=\frac{1}{\delta}\left(\frac{\Bd_{\Td}}{\Bd_{\Ud}}-1\right)
\end{equation}
with $\delta=T-U.$ The continuously compounded {\it SOFR account} $\Bf$ satisfies, for every $t \in \mathbb{R}_+$,
\begin{equation}
\Bf_t= \exp \left( \int_0^t \rf_u \diff u \right)
\end{equation}
and the compound {\it SOFR Average} over $[\Ud,\Td]$ is given by
\begin{equation}
\Rf(\Ud,\Td) := \frac{1}{\delta}\left( \exp \left( \int_{\Ud}^{\Td} \rf_u \diff u \right)-1\right)
= \frac{1}{\delta}\left(\frac{\Bf_{\Td}}{\Bf_{\Ud}}-1\right).
\end{equation}
\ed

Notice that the Realised AONIA and the SOFR Average are backward-looking rates since they can be observed at the end of each period $\Td$. More formally, the random variables $\Rf(\Ud,\Td)$ and $\Rd(\Ud,\Td)$ are $\mathcal{F}_{\Td}$-measurable (but not $\mathcal{F}_{t}$-measurable for $t<\Td$). This should be contrasted with the forward-looking LIBORs, which are known at the beginning of each accrual period. Now we are ready to introduce the model-free definitions of futures product as our main hedging tools.

\bd\label{def2.2}
A {\it SOFR futures} for the period $[\Ud,\Td]$ is defined as a futures contract referencing the SOFR Average over $[\Ud,\Td]$, with the {\it SOFR futures} rate denoted by $\Ff_t(\Ud,\Td)$ for $t \in [0,\Td]$.
\ed

The AONIA futures contracts are defined in an analogous manner and the AONIA futures rates at time $t$ are denoted by $\Fd_t(T, \Td)$.  The spot price of acquiring the futures contract is always zero and the dynamics of AONIA and SOFR futures will be studied in Section \ref{sec4} after we introduce in Assumption \ref{ass4.1} a stochastic multi-factor model.

\subsection{Foreign exchange market and currency futures} \label{sec2.1}

A currency futures contract is an exchange traded agreement to buy or sell a specified amount of a particular currency (the {\it base currency}) relative to a second currency (the {\it quoted currency}) at a future date at a specified exchange rate (the {\it contract price}). The last trading day for each currency futures contract determines when settlement will take place and the instrument will automatically expire. On any trading day during the life of the futures contract, a long (or short) position can be closed by placing a sell (or buy) order in the market. Currency futures serve various purposes, including speculation on exchange rate fluctuations and hedging against other currency-related concerns. They enable market participants to manage their exposure to exchange rate fluctuations, thereby reducing uncertainty and promoting financial stability.

The exchange rate $Q$ is an $\ff$-adapted stochastic processes, which at time $t$ is quoted as
\begin{equation*}
Q_t = \frac{\text{Number of units of the domestic currency (AUD)}}{\text{One unit of the foreign currency (USD)}}
\end{equation*}
where, obviously, the choice of any two currencies was arbitrary. We now give the formal definition of currency futures where USD is the base currency and AUD is the quoted currency. Without loss of generality, we assume that the nominal principal is set to be 1 USD.

\bd \label{def2.3}
A {\it currency futures} initiated at time $S$ with the settlement date $\Td$ is defined as a futures contract referencing a predetermined currency pair (say, USD/AUD) with the futures exchange rate at time $t \in [S,\Td]$ denoted by $\Fq_t(\Sd,\Td)$ (or, briefly, $\Fq_{t}$ if the dates $S$ and $\Td$ are predetermined) in AUD. At the maturity date $\Td$, the holder of the currency futures has an obligation to pay $\Fq_{\Td}=Q_{\Td}$ units of the quoted currency (AUD) to obtain one unit of the base currency (USD).
\ed

If a long position in the futures contract from Definition \ref{def2.3} is entered into at time
$s$ and held until time $u$ where $S\leq s < u \leq \Td$, then the holder receives at time $u$ the amount $\Fq_u-\Fq_s$ in the foreign currency (note that the daily settlement mechanism is purposely ignored here). Again, the price of a currency futures contract will always be zero.

\subsection{Constant notional cross-currency basis swaps}   \label{sec2.2.1}

We first describe the actual cash flows in the constant notional CCBS with the tenor structure $0\leq T_0<T_1<\cdots<T_n$ where $\delta_j:=T_{j}-T_{j-1}$ for $j=1,2,\dots,n$ and with no adjustments to the nominal principal at payments dates. For brevity, a generic $n$-period CCBS with the basis spread $\kappa$ is denoted by $\CCBS \,(\cT_n;\kappa )$ where $\cT_n$ symbolizes the tenor structure $T_0<T_1<\cdots<T_n$. Note that Definition \ref{def2.4} covers both the {\it spot} cross-currency swap where $T_0=0$ and the {\it forward} cross-currency swap for which $T_0>0$. By the usual convention, all quantities computed at time 0 are deterministic and those computed at some date $t>0$ may be random. The domestic and foreign cash flows in Definition \ref{def2.4} are given in respective currencies and we write $[x,y]$ to represent the payoff of $x$ units of AUD combined with the payoff of $y$ units of USD.  We first define a general {\it constant notional} cross-currency basis swap.

\bd \label{def2.4}
Let $P^d$ [AUD] and $P^f$ [USD] denote the notional principals exchanged at the inception date $T_0$ and exchanged back at the maturity date $T_n$. The cash flows at $T_0<T_1<\cdots<T_n$ for the long party in a {\it constant notional} {\rm $\CCBS\,(\cT_n;\kappa )$} are given by
\begin{align*}
&\CF_{T_0} =\big[P^d,-P^f\big],\quad \text{at}\ T_0,\\
&\CF_{T_j}=\big[-(\Rd(T_{j-1},T_{j})+\kappa)\delta_j P^d,\,\Rf(T_{j-1},T_{j})\delta_j P^f \big],
\quad \text{at}\ T_j,\ j = 1,2,\dots , n-1, \\
&\CF_{T_n} = \big[-(\Rd(T_{n-1},T_{n})+\kappa)\delta_n P^d-P^d,\,\Rf(T_{n-1},T_{n})\delta_n P^f + P^f \big],
\quad \text{at}\ T_n,
\end{align*}
where the nominal principal amounts $P^d$ and $P^f$ satisfy $P^d = Q_{T_0} P^f$ and the {\it basis spread} $\kappa$ is set at the swap's inception date $t \leq T_0$.
\ed

Suppose first that the basis spread $\kappa $ is set at time $t=0$. Then $\kappa $ is a constant and, in principle,
its fair value, denoted by $\kappa_0(\cT_n)$, should be chosen to ensure that the arbitrage-free price at time 0 of the (spot or forward) CCBS at time 0 is null, that is, the equality $\CCBS_0\,(\cT_n;\kappa_0(\cT_n))=0$ holds. Of course, an analogous argument applies to the forward CCBS starting at some date $0<t \leq T_0$ but then the fair basis spread $\kappa_t(\cT_n)$ satisfies $\CCBS_t \,(\cT_n;\kappa_t(\cT_n))=0$ and thus it is no longer a deterministic constant since its value depends on the market conditions prevailing at time $t$. We will later define the stochastic process $\kappa_t(\cT_n),\, t \in [0,T_0]$ when studying cross currency swaptions.

In practice, the level of the basis spread is agreed upon by the counterparties at the contract's inception and stays constant during the contract's lifetime. It is clear that the fair level of the basis spread depends on several factors, in particular, a currency pair and a tenor structure. Furthermore, it is also impacted by the credit risk connected to the two reference floating rates, which may be either secured or unsecured, the counterparty credit risk associated with a given trade and the manner in which the swap is collateralized.

We are in a position to state a variant of Definition \ref{def2.4}, which is convenient for the computation of its price and hedge from the perspective of an Australian bank. Hence in Definition \ref{def2.4} the Australian dollar is chosen to be the {\it valuation currency} whereas the U.S. dollar is the {\it reference currency}. Notice that in Definition \ref{def2.5} the domestic nominal value $P^d=Q_{T_0}$ [AUD] is fixed throughout and hence has the same value in the domestic currency at time $T_n$. The foreign nominal value $P^f=1$ [USD] is also fixed throughout but it has the domestic value $Q_{T_j}$ [AUD] at time $T_j$ for $j=1,2,\dots,n$, which usually does not coincide with its initial domestic value $Q_{T_0}$ [AUD] at time $T_0$.

The following definition covers the case of a constant notional CCBS, meaning that the principal nominals, $P^f$ expressed in USD and $P^d$ expressed in AUD, are set at $T_0$ and kept constant during the lifetime of a swap, notwithstanding the fluctuations of the exchange rate $Q$. Recall that we have chosen the Australian dollar as the valuation currency for a CCBS and hence the cash flows are represented as effective net cash flows expressed in the domestic currency.

\bd \label{def2.5}
At every payment date $T_{j}$ for $j=0,1,\dots,n$ the cash flow associated with a {\it constant notional} CCBS with $P^f=1$ and $P^d=Q_{T_0}$ are expressed in AUD and are given by
\begin{align*}
&X_0=0,\quad \text{at}\  T_0,\\
&X_j=\Rf(T_{j-1},T_{j})\delta_j Q_{T_{j}}-(\Rd(T_{j-1},T_{j})+\kappa)\delta_j Q_{T_0},
\quad \text{at}\ \, T_j,\ j=1,2,\dots,n-1, \\
&X_n =\Rf(T_{n-1},T_{n})\delta_n Q_{T_{n}}-(\Rd(T_{n-1},T_{n})+\kappa)\delta_n Q_{T_0}+Q_{T_n}-Q_{T_0},
\quad \text{at}\ \, T_n.
\end{align*}
\ed

For convenience, we will also write, for every $j=1,2,\dots ,n-1$,
\begin{align*}
X^{i}_{T_j}(T_0,T_{j-1},T_j):=\Rf(T_{j-1},T_{j})\delta_j Q_{T_{j}}-(\Rd(T_{j-1},T_{j})+\kappa)\delta_j Q_{T_0}
\end{align*}
and $X^{p}_{T_n}(T_0,T_n):=Q_{T_n}-Q_{T_0}$.

The two legs of the basis swap are backward-looking and thus, unlike in the case of forward-looking LIBOR rates, the hedging strategy can be shown to be dynamic not only before but also during each accrual period.  Our market model introduced in Assumption \ref{ass4.1} is invariant with respect to the choice of the valuation currency, which means that the computations performed by the two counterparties based in Australia and the U.S. are analogous but they may yield different pricing formula at any time $t \leq T_n$, of course, assuming that the two prices are expressed in the same currency (either AUD or USD) using the spot exchange rate $Q_t$. In contrast, the choice of a currency in which the collateral is posted affects the pricing formula
for a CCBS since remuneration rate for collateral depends on a currency.

A foreign exchange swap (FX swap) and a CCBS are both derivative instruments utilized in the hedging of foreign currency exposures, but there are essential differences. In an FX swap, the notional principals are exchanged at the maturity date at the forward rate. This should be contrasted with a constant nominal CCBS where the nominal principals are exchanged at the maturity date at the initial spot rate and there are payments attached to interest rates during the term of the contract, which are absent in FX swaps.

\section{Cross-currency futures trading}  \label{sec3}

We propose a multi-curve approach, which differs from the classical one in many respects. First, it is not postulated that the risk-free rates $\rd$ and $\rf$ can be used for funding of the hedge. It is assumed instead that the futures contracts referencing the compound rates $R^d(\Ud,\Td)$ and $R^f(\Ud,\Td)$ are traded in respective economies. Second, we introduce funding costs for the hedge, which are modeled by the short-term rate $r^h$ but it can be easily extended to differing lending and borrowing rates. Third, we assume that the contract is collateralized with the remuneration rate $\rc$ for the collateral amount proportional to the value of the contract. By convention, the collateral is either posted or received in the domestic currency and is subject to rehypothecation though other conventions regarding collateralization can be accommodated within the present framework.

Of course, our postulates regarding the funding rates and collateralization can be further generalized as it was indeed previously done, for instance, in \cite{BR2015} (see also \cite{Biagini2021unified} and \cite{BCR2018}), but we have deliberately chosen to keep our setup relatively simple in order to derive closed-form pricing and hedging results for constant notional and mark-to-market cross-currency basis swaps, instead of relying on theoretical results for  nonlinear backward stochastic differential equations. In contrast, we can use a solution to a linear backward stochastic differential equation when analyzing cross-currency swaptions but with no explicit analytical formulae available for the price and hedge. Therefore, either numerical methods for backward stochastic differential equations or the Monte Carlo method can be used in the latter case.

We now focus on the hedging strategies involving various futures contracts. The computations presented in this section are model-free since no assumptions about the dynamics of domestic and foreign interest rates are made.
In Section \ref{sec4}, we introduce a specific model consisting of the dynamics of interest rates and exchange rate, which allows for explicit computations of price and hedge.

\subsection{Futures-based trading strategies}    \label{sec3.1}

Most CCBSs are long-term instruments, typically ranging from one to thirty years. On the one hand, the most liquidly traded CCBSs generally have maturities concentrated in between one and three years. On the other hand, financial instruments referencing overnight benchmark rates tend to have shorter maturities: 30-day interbank cash rate futures linked to AONIA are listed up to approximately 18 months, while SOFR futures contracts are typically available for maturities up to five years but with liquidity diminishing beyond three years. Currency futures (e.g., USD/AUD or USD/EUR futures) also have short maturities---generally one to two years. Nevertheless, we assume the existence of a sufficiently rich and regular futures market; specifically, we postulate the existence of futures trades in the market for all $t$ during the lifetime of a CCBS.

In practice, given these market constraints, replication of longer-dated CCBS (particularly, those exceeding five years) using short-dated futures contracts necessitates adopting a rolling hedge strategy, whereby expired futures positions are periodically replaced with newly issued contracts. Although this rolling hedge introduces inherent incompleteness due to the unavailability of long-dated futures at inception of a hedge, it closely aligns with practical hedging methods employed in industry: market participants typically hedge short-term exposures precisely using liquid futures, while monitoring longer-term exposures dynamically. Therefore, despite the apparent market incompleteness for long-dated instruments, our proposed modeling and hedging framework remains applicable.

Let us first introduce a generic notation for futures prices. Note that the currency futures are traded domestically in Australia with USD being the base currency. Let continuous semimartingales $\Fd,\Ff$ and $\Fq$ represent the futures prices referencing the AONIA, SOFR and the currency futures, respectively.  We assume that $\diff B_t^h=r_t^h B_t^h \diff t$ for some $\ff$-adapted process $r^h$ representing the {\it hedge funding rate}.

\bd \label{def3.1}
By a {\it futures trading strategy} we mean an $\rr^4$-valued, $\ff$-adapted process $\phiv=(\phiv^0,\phid,\phif,\phiq)$ where the components $\phid,\phif$ and $\phiq$ represent positions in AONIA, SOFR and currency futures, respectively, and the process $\phiv^0$ represents the hedge funding component. Since the value of any position in a futures contract is zero at any time, the value of a futures trading strategy $\phiv$ at any time $t\in [0, T]$ equals $V_t^p(\phiv)=\phiv^0_t B_t^h$.
\ed

\brem
In practice, futures contracts require the buyer or seller deposit cash in the margin account, a portion of the total value of the specified commodity future being bought or sold. This deposit, which is known as the {\it initial margin} for futures trades, must be made with a registered futures commission merchant before a futures contract is bought or sold in accordance with rules established by each futures exchange. Note that this work does not address the concepts of initial and variation margins for futures contracts and default. For the sake of mathematical simplicity, we discuss the situation where all trades in foreign market contracts are settled on a daily basis. In particular, the existence of the margin account is not considered for futures trading. A more general case can be further studied if we release the restrictions of the margin account in the foreign futures market, but maintain no margin account for the domestic futures market.
Of course, it is also possible to include the initial margin and the maintenance margin on the margin account but this would result in high computational complexity that can be further studied using the theory of backward stochastic differential equations.
\erem

\subsection{Collateralized futures trading strategies}   \label{sec3.2}

After the global financial crisis of 2007-2009, the financial markets worldwide have become more cautious and thus OTC contracts usually require posting of collateral, which we denote as $C$ and which is represented by an $\mathbb{F}$-adapted stochastic process. At any date $t$, the sign of $C_t$ represents the direction of collateralization. It is natural to include the interest paid on a margin account in our multi-curve model with a collateral rate denoted by $\rc$. We do not assume that the collateral is delivered in either AUD or USD; it can be posted in any currency with the associated remuneration rate $\rc$ (see Remark \ref{Rem: choice} and Proposition 5.6 in \cite{Ding2024wp}). Additionally, at this stage we make no assumptions regarding the initial value of the collateral (i.e., the {\it initial margin}).

To define a suitable self-financing condition for any trading strategy in the present multi-curve framework, we first consider the net cash flow in a discrete-time framework by considering the {\it values} $V_t^p(\phiv,C)$ and $V_{t+1}^p(\phiv,C)$ of a  collateralized futures strategy $(\phiv,C)$ for any date $t\in[0,T]$. We adopt here
the assumption of daily settlement and we assume that the hedger either receives or pays the accrued interest on collateral depending on who is holding the collateral amount $C'_t$ at time $t$ and in which currency it is posted (not necessarily AUD or USD). Notice that $C'_t$ is expressed in units of the collateral currency and thus $C_t := C'_t Q'_t$ where $Q'_t$ is the corresponding exchange rate so that $C_t$ is the current value of collateral expressed in the domestic currency. Of course, if the collateral is posted in AUD (resp., USD), then $Q'_t = 1$ (resp., $Q'_t=Q_t$).

As will be shown in what follows, under the assumption that the collateral amount is proportional to the current value of the contract, its arbitrage-free price depends on the choice of the collateral currency only through the remuneration rate, denoted by $\rc$, and thus the dynamics of the exchange rate $Q'$ are not relevant for pricing (though they do matter for hedging).
This important feature is a consequence of our standing assumption of rehypothecation of collateral and thus it is not necessarily true under other collateral conventions.

Within the present framework, it is natural to postulate that the value process of a trading strategy $(\phiv,C)$ where $C_t = C'_t Q'_t$ satisfies $V^p_t (\phiv,C) =\phiv^0_t B^h_t$ for all $t$ and
\begin{align*}
V_{t+\Delta t}^p(\phiv,C)&=V_t^p(\phiv,C)+\phiv^0_t(B_{t+\Delta t}^h-B^h_t)+Q'_{t+\Delta t}C'_{t+\Delta t}-Q'_tC'_t
-r_t^c Q'_tC'_t\nonumber \\
&+ \phid_t(\Fd_{t+\Delta t}-\Fd_t) +\phif_tQ_{t+\Delta t}(\Ff_{t+\Delta t}-\Ff_t)
+ \phiq_t(\Fq_{t+\Delta t}-\Fq_t) \nonumber \\
&=V_t^p(\phiv,C)+\phiv^0_t (B_{t+\Delta t}^h-B^h_t) +C_{t+\Delta t}-C_t- r_t^c C_t \Delta t
+ \phid_t(\Fd_{t+\Delta t}-\Fd_t) \nonumber \\
&+\phif_tQ_{t+\Delta t}(\Ff_{t+\Delta t}-\Ff_t)
+ \phiq_t(\Fq_{t+\Delta t}-\Fq_t). \nonumber
\end{align*}
Therefore,
\begin{align*}
V_{t+\Delta t}^p(\phiv,C)&=V_t^p(\phiv ,C)+\phiv^0_t(B_{t+\Delta t}^h-B^h_t)+
C_{t+\Delta t}-C_t-r_t^c C_t \Delta t+\phid_t(\Fd_{t+\Delta t}-\Fd_t)
\\&+\phif_t\left[(Q_{t+\Delta t}-Q_t)(\Ff_{t+\Delta t}-\Ff_t)+Q_t(\Ff_{t+\Delta t}-\Ff_t)\right]
+ \phiq_t(\Fq_{t+\Delta t}-\Fq_t) \nonumber
\end{align*}
where we have used the algebraic relationship
\begin{align*}
Q_{t+\Delta t}(\Ff_{t+\Delta t}-\Ff_t)=(Q_{t+\Delta t}-Q_t)(\Ff_{t+\Delta t}-\Ff_t)+Q_t(\Ff_{t+\Delta t}- \Ff_t).
\end{align*}
In the dynamics above, we have also implicitly postulated the rehypothecation of collateral (as opposed to its segregation), meaning that the collateral amount is assumed to be available for trading purposes in the domestic market (for a detailed study of alternative conventions regarding collateralization we refer to, e.g., \cite{BCR2018,BR2015}).

It is important to notice that from the hedger's point of view, the collateral amount is not part of their assets and thus we define the {\it hedger's wealth} process by the equality $V_t(\varphi,C) := V_t^p(\varphi,C) - C_t$ for all $t$. Formally, in order to compute the hedger's wealth at any date $t$ for an exogenously given process $C$ and any trading strategy $\varphi$ it suffices to use the self-financing condition to compute $V^p_t(\varphi,C)$ and subsequently deduct the current value of collateral $C_t$.  The wealth process  $V(\varphi,C)$ of a hedging strategy is used to formally describe the current marked-to-market value of a trade, which manifestly depends on the level of collateralization.

Following the exposition of stochastic integration theory in Section 12.1 of \cite{oksendal2003sde}, we claim that the dynamics of the process $V^p(\varphi)$ can be extended from a discrete time case to a continuous time setup by taking limits and using the definition of the It\^o integral with respect to a continuous semimartingale. Then we obtain the following definition of a collateralized futures strategy in a continuous time framework where $\qvar{Y^1,Y^2}{}$ denotes the quadratic covariation process of two continuous semimartingales, $Y^1$ and $Y^2$ (see also \cite{Jamshidian1994} for the case of uncollateralized futures trading strategies).

\bd \label{def3.2}
A collateralized futures strategy $(\phiv,C)=(\phiv^0,\phid,\phif,\phiq,C) $ is {\it self-financing} if the value process
$V^p_t(\phiv,C):=\phiv_t^0 B^h_t$ satisfies, for every $t \in [0, T]$,
\begin{align*}
V_t^p(\phiv,C)&=V_0^p(\phiv,C)+\int_0^t \phiv^0_u\diff B_u^h+C_t-\int_0^t r_u^c C_u \diff u+\int_0^t \phid_u \diff \Fd_u \\
&+\int_0^t\phif_u Q_u \diff \Ff_u+\int_0^t\phif_u\diff\qvar{Q,\Ff}{u}+\int_0^t \phiq_u\diff \Fq_u,
\end{align*}
or, equivalently,
\begin{align*}
\diff V_t^p(\phiv,C)= r_t^h V_t^p(\phiv,C) \diff t +\diff C_t-r_t^c C_t\,dt+\phid_t\diff \Fd_t +\phif_t \diff \Ffq_t
+\phiq_t \diff \Fq_t
\end{align*}
where we denote
\begin{align} \label{fQsi}
\Ffq_t:= \Ff_0 + \int_0^t Q_u \diff \Ff_u + \qvar{Q,\Ff}{t}.
\end{align}
\ed

The auxiliary process $\Ffq$ is used to formally represent the impact of the foreign futures component (e.g., SOFR futures) expressed in the domestic currency (in our case, AUD). It is easy to check that the wealth process $V(\phiv,C) = V^p(\phiv,C) - C$ satisfies
\begin{align*}
\diff V_t(\phiv,C)=r_t^h V_t(\phiv,C)\diff t-(r_t^c-r^h_t)C_t\,dt+\phid_t\diff \Fd_t+\phif_t\diff \Ffq_t
+\phiq_t \diff \Fq_t.
\end{align*}
In particular, if $r^h=\rc$ then, as expected, the collateralization does not have any impact on the wealth process $V(\phiv,C)$, that is, $V(\phiv,C)=V(\phiv,0)$ where zero indicates that we deal with an uncollateralized contract.
We will henceforth work under the key postulate of proportional collateralization, in the sense that we set $C_t := -\beta_t V_t(\phiv,C)$ for all $t$ where $\beta$ is a non-negative, $\mathbb{F}$-adapted stochastic process.

Under this standing assumption of proportional collateralization, the dynamics of the wealth process
$V_t(\phiv,C)$ are governed by the following equation
\begin{align*}
V_t(\phiv,C)=V_0(\phiv,C) + \int_0^t r^\beta_u V_u(\phiv,C)\diff u  +\int_0^t \phid_u \diff \Fd_u +\int_0^t \phif_u\diff \Ffq_u
+\int_0^t \phiq_u\diff \Fq_u
\end{align*}
where the {\it effective hedge funding rate} $r^{\beta}$ for a collateralized contract is given by $r^\beta := (1-\beta)r^h + \beta \rc$ and $\phid,\phif$ and $\phiq$ are arbitrary $\mathbb{F}$-adapted processes for which all integrals above are well defined. To make the dynamics of the discounted wealth easier to handle, we introduce the fictitious bank account $B^\beta$ with the dynamics $\diff B_t^\beta = r_t^\beta B_t^\beta \diff t$. The next result is an easy consequence of the It\^o formula and thus its proof is omitted.

\bp \label{pro3.1}
 Let $(\phiv,C)$ be a self-financing collateralized futures strategy
 with the proportional collateral $C =-\beta V(\phiv,C)$ for some $\ff$-adapted process $\beta$.
Then the discounted wealth process $\wt{V}^\beta(\phiv,C) := (B^\beta)^{-1} V(\phiv,C)$
satisfies, for every $t\in [0,T]$,
\begin{align} \label{er}
\wt{V}_t^\beta(\phiv,C) = \wt{V}_0^\beta(\phiv,C)+\wt{G}^\beta_t(\phiv,C)
\end{align}
where the discounted gains process $\wt{G}^\beta (\phiv,C)$ is given by, for every $t \in [0,T]$,
\begin{align*}
\wt{G}^\beta_t(\phiv,C)=\int_0^t \big(B^{\beta}_u\big)^{-1}\phid_u\diff\Fd_u+\int_0^t \big(B^{\beta}_u\big)^{-1}\phif_u\diff \Ffq_u
+\int_0^t \big(B^{\beta}_u\big)^{-1}\phiq_u\diff \Fq_u.
\end{align*}
\ep

\brem\label{Rem: choice}
A simple but useful observation from the above proposition is that our model can easily accommodate the case of a collateral amount posted in any currency. For instance, consider $C_t = Q'_t C'_t$, where $C'_t$ is the collateral amount in another currency, e.g., the euro, and $Q'_t$ is the corresponding exchange rate. This situation is still covered by the assumption that $C_t = -\beta_t V_t(\phiv, C)$. As was argued in \cite{Fujii2011Choice}, the choice of a collateral currency has a non-negligible impact on the pricing of financial derivatives. We reach the same conclusion here: when collateral is posted in a different currency, the collateral rate will naturally change as well. If the hedger wisely selects the collateral currency (e.g., choosing the one with the highest or lowest value of $\rc$, depending on their strategy), the discounting factor will be adjusted accordingly, which may result in a more favorable derivative price.
\erem

Note that when collateral is posted in a third currency, additional modeling complexity arises. Specifically, such collateral convention would require the introduction of an additional stochastic factor. Although conceptually straightforward, the resulting computations for hedging involving interest rate futures in a third economy would become cumbersome. We briefly outline preliminary steps for this extension but refrain from explicit calculations here for brevity. As shown later in Assumption \ref{ass4.1}, our model already accounts for risks arising from domestic rates, foreign rates, and exchange rates driven by a three-dimensional Brownian motion. Introducing a third collateral currency requires modeling an additional interest rate process, along with the corresponding exchange rates, thereby increasing the dimensionality of the model. While this extension remains fully tractable due to the Gaussian structure, the resulting computations become more involved. A numerical demonstration illustrating the impact of collateral currency is provided in Section \ref{sec7.5}.

\subsection{Discounted wealth and martingale measure}    \label{sec3.3}

In the present setup, the concept of a martingale measure  for the market model with collateralized trading can be introduced through the following definition where the underlying probability space $(\Omega,\cF,\mathbb{F},\pp)$, which is endowed with the statistical probability measure $\pp$ and the filtration $\mathbb{F}$, is assumed to be given.

\bd \label{def3.3}
A probability measure $\wtQ$ is called a \PMM for the date $\Td$ if $\wtQ$ is equivalent on $(\Omega,\cF_{\Td})$ to the statistical probability measure $\pp$ and the process $(\wt{V}_t^\beta(\phiv,C),\, t \in [0,T])$ is a $\wtQ$-local martingale with respect to the reference filtration $\mathbb{F}$ for any self-financing collateralized futures strategy $(\phiv,C)$ with an arbitrary proportional collateralization level $\beta$.
\ed

The local martingale property of the process $\wt{V}_t^\beta(\phiv,C)$ under $\wtQ$ can be used to establish the arbitrage-free property of our market model. As customary, we postulate that only trading strategies with the discounted wealth bounded from below by a constant are admissible in order to ensure that the process $\wt{V}_t^\beta(\phiv,C)$ is in fact a supermartingale under $\wtQ$, which in turn excludes arbitrage opportunities.

It is clear from Definition \ref{def3.3} that a \PMM does not depend on a level of proportional collateralization. To make the process $\wt{V}^\beta(\phiv)$ a local martingale under some probability measure $\wtQ$ equivalent to $\pp$, it suffices to ensure that the processes $\Fd,\Ffq$ and $\Fq$ are $\wtQ$-local martingales on $[0,T]$ and, in fact, the latter property is also a necessary condition given that a trading strategy is arbitrary. Therefore, to establish the existence of a \PMM we need to introduce an arbitrage-free model for foreign and domestic interest rates and exchange rate under some probability measure $\Q$ (see Section \ref{sec4.1}) and then to demonstrate that $\wtQ$ can be obtained from $\Q$ (see Section \ref{sec4.6}).

Before proceeding to an explicit construction of a term structure model for two economies, let us first show how to use the probability measure $\wtQ$ for pricing purposes under the postulate of proportional collateralization. For simplicity of presentation, we focus here on a simple contract of European style with a single payoff but it is clear that more complex contracts (also of an American style) can
be dealt with using $\wtQ$.

\bd \label{def3.4}
We say that a collateralized contract $(X_{\Td},\beta)$ with the terminal payoff $X_{\Td}$ at time $\Td$ and the proportional collateralization at rate $\beta$ is {\it attainable} if there exists a self-financing collateralized futures strategy $(\phiv ,C)$ where $C=-\beta V$ such that $V_{\Td}(\phiv,C)= X_{\Td}$.
\ed

As usual, we need to focus on the class of {\it admissible} trading strategies, that is, all trading strategies for which the relative wealth $\wt{V}^\beta(\phiv,C):= (B^{\beta})^{-1}V^\beta(\phiv,C)$ is a martingale under $\wtQ$. The following proposition is an immediate consequence of Definition \ref{def3.3} combined with the definition of attainability of a collateralized contract.

\bp \label{pro3.2}
Consider a contract $(X_{\Td},\beta)$ with the terminal payoff $X_{\Td}$ at time $\Td$ and the proportional collateralization at rate $\beta$. If the random variable $(B^{\beta}_{\Td})^{-1}X_{\Td}$ is $\wtQ$-integrable, then the arbitrage-free price process for $(X_{\Td},\beta)$ satisfies, for every $t\in[0,T]$,
\begin{equation}  \label{eq3.2}
\pi^{\beta}_t(X_{\Td})=B_t^\beta \,\EwtQ\Big[\big(B^{\beta}_{\Td}\big)^{-1} X_{\Td} \,\Big|\,\cF_t\Big].
\end{equation}
\ep

\begin{proof}
The proof is omitted since it suffices to use the postulated martingale property of the process $\wt{V}^\beta(\phiv,C)$ under $\wtQ$ and the equality $\wt{V}_{\Td}(\phiv,C)=(B^{\beta}_{\Td})^{-1}X_{\Td}$.
\end{proof}

Notice that in our modeling approach, we may consider a variety of contracts with different maturities, terminal payoffs, and various levels of proportional collateralization. As expected, the price process $\pi^{\beta}(X_{\Td})$ will depend on the level of proportional collateralization through the process $B^{\beta}$ if the interest rates $r^h$ and $\rc$ differ and it reduces to the classical price of an uncollateralized contract when $\beta=0$ or $r^h=\rc$. However, as was already mentioned, the probability measure $\wtQ$ does not depend on $\beta$ and thus it can be used to compute prices of contracts with differing levels of proportional collateralization.

For brevity, we will also use the shorthand notation $\Xbeta_t := \pi^{\beta}_t(X_{\Td})$ for every $t\in [0,T]$ so that, in particular, the equality $\Xbeta_{\Td}=X_{\Td}$ is valid.
As an example of application of Proposition \ref{pro3.2} we will state the pricing formula for the multi-period CCBS with the tenor structure $\cT_n$ and basis spread $\kappa$. We start by noticing Proposition \ref{pro3.2} gives the general expression for the arbitrage-free price $\CCBS_t(\cT_n;\kappa )$, for every $t\in [0,\Sd]$,
\begin{align*}
\CCBS_t(\cT_n;\kappa)=\sum_{j=1}^n B_t^\beta \,\EwtQ\Big[\big(B^{\beta}_{T_j}\big)^{-1}X_j \,\Big|\,\cF_t\Big].
\end{align*}
\newpage
\section{Term structure model and futures contracts}  \label{sec4}

In the previous section, we have formalized the concept of a futures trading strategy based on SOFR futures, AONIA futures and currency futures as hedging tools. These futures products are all actively traded short and medium term interest rate derivatives, which provide high liquidity for hedging of cross-currency swaps. After the introduction of a market model in Assumption \ref{ass4.1} we will be able to explicitly compute the dynamics of various futures prices.

\subsection{Cross-currency multi-curve term structure model} \label{sec4.1}

For the sake of concreteness and analytical tractability, we use the classical Vasicek's model to describe the dynamics of the process $\rd$ in the domestic currency and the process $\rf$ in the foreign currency, which is complemented by the Garman-Kohlhagen model for the exchange rate~$Q$. This is a convenient choice in a multi-curve framework since we can also set $\alpha^h:=r^h-\rd$ where $\alpha^h$ represents the spread for the funding rate.

Furthermore, we postulate that all trades in the domestic and foreign futures referencing AONIA and SOFR (as well as trades in the domestic and foreign equities) are funded using the market interest rates denoted by $\xd$ and $\xf$, respectively, which do not necessarily coincide with the benchmark risk-free overnight rates $\rd$ and $\rf$ so we will write $\alpha^d:=\xd-\rd$ and $\alpha^f:=\xf-\rf$ to denote the respective spreads. As a special case of our  market model, one can assume that $\xd=r^h$ and, analogously that $\xf$ represents the hedge funding rate of the foreign counterparty, but this is by no means a necessary assumption and thus it is not made in what follows.

However, to accurately capture the initial term structure of interest rates observed in the market, the Vasicek model can be extended by introducing a deterministic shift function satisfying suitable integrability conditions. Specifically, instead of modeling the short rate $r_t$ directly, one sets $r_t =x_t+ \phi(t)$, where $x_t$ is governed by the standard Vasicek dynamics and $\phi(t)$ is a deterministic function, which is chosen to match the initial yield curve. This approach ensures the model is capable of reproducing market quotes for bond prices at inception, while preserving the analytical tractability for explicit pricing and hedging.

\subsubsection{Domestic martingale measure}

To construct a multi-curve cross-currency term structure model, we start by postulating that the dynamics of the risk-free overnight rates $\rd,\rf$ and the exchange rate $Q$ under the probability measure $\Q$ are, for all $t \in \rr_+$,
\begin{align}
&\diff\rd_t=(a-b\rd_t)\diff t+\sigma\diff\Zone_t,  \nonumber \\
&\diff\rf_t=(\wh{c}-\wh{b}\rf_t)\diff t+\wh{\sigma}\diff\Ztwo_t, \label{eq4.2b} \\
&\diff Q_t=Q_t(\xd_t-\xf_t)\diff t+Q_t\barsig\diff\Zthree_t,  \nonumber
\end{align}
where $a,b,\sigma,c,\wh{b},\wh{\sigma}$ and $\barsig$ are positive constants and the processes $\Zone,\Ztwo$ and $\Zthree$ are one-dimensional Brownian motions under $\Q$ with the correlations $\qvar{Z^i,Z^j}{t}=\rho_{ij} \diff t$.
Notice that the drift term in the dynamics of the exchange rate $Q$ under $\Q$ is due to the interpretation of processes $\xd$ (resp., $\xf$) as the money market interest rate prevailing in the domestic (resp., foreign) fixed-income market (see, e.g., Chapter 14 in \cite{MR2005}). We stress that the equation governing the exchange rate $Q$ does not depend on any particular specification of dynamics of $\xd$ and $\xf$ under $\Q$; it suffices to assume that the associated money market accounts satisfy $d\bar{B}^d_t = \xd_t \bar{B}^d_t\diff t$ and $d\bar{B}^f_t = \xf_t \bar{B}^f_t\diff t$.
Then $\Q$ is the {\it domestic martingale measure} for the arbitrage-free cross-currency model given by the triplet $(\xd,\xf,Q)$,
in the sense that the process $Q_t\bar{B}^f_t (\bar{B}^d_t)^{-1}$ is a martingale under $\Q$. We observe that $\xd$ and $\xf$ are given by the shifted Vasicek's model under $\Q$ if $\alpha^d$ and $\alpha^f$ are assumed to be deterministic functions.

Although our model is fully specified by \eqref{eq4.2b} and the set of correlation coefficients
$\rhodf,\rhod$ and $\rhof$, we find it useful to provide its equivalent representation in terms of a standard Brownian motion
$\Ww=(\Wone, \Wtwo, \Wthree)$, which is defined under $\Q$ through the equalities
\begin{equation*}
\Zone_t=\Wone_t,\ \,
\Ztwo_t=\rhodf\Wone_t+\sqrt{1-\rhodf^2}\,\Wtwo_t,\ \,
\Zthree_t=\alone\Wone_t+\altwo\Wtwo_t+\althree\Wthree_t
\end{equation*}
where we denote
\begin{align} \label{eq4.2}
\alone := \rhod,\quad \altwo:=\frac{\rhof-\rhodf\rhod}{\sqrt{1-\rhodf^2}},\quad \althree:= \sqrt{1-\rhod^2-\altwo^2}.
\end{align}

Then the dynamics of the processes $\rd,\rf$ and $Q$ under $\Q$ can be represented as follows
\begin{align}
&\diff \rd_t=(a- b\rd_t)\diff t+\sigma\diff \Wone_t, \nonumber \\
&\diff \rf_t=(\wh{c}-\wh{b}\rf_t)\diff t+\wh{\sigma}\Big(\rhodf\diff\Wone_t+\sqrt{1-\rhodf^2}\diff\Wtwo_t\Big), \label{eq4.2c} \\
&\diff Q_t=Q_t(\xd_t-\xf_t)\diff t+Q_t\barsig\Big(\alone\diff\Wone_t+\altwo\diff\Wtwo_t+\althree\diff\Wthree_t\Big).\nonumber
\end{align}
More succinctly, the triplet $(\rd,\rf,Q)$ of $\ff$-adapted stochastic processes satisfies
\begin{align}
&\diff \rd_t=(a- b\rd_t)\diff t+\sigmad \diff \Ww_t, \nonumber \\
&\diff \rf_t=(\wh{c}-\wh{b}\rf_t)\diff t+\sigmaf \diff \Ww_t , \label{eq4.2cv} \\
&\diff Q_t=Q_t(\xd_t-\xf_t)\diff t + Q_t \sigmaq \diff \Ww_t, \nonumber
\end{align}
where $\Ww=(\Wone,\Wtwo,\Wthree)$ is a standard Brownian motion under $\Q$ with respect to $\ff = \ff^{\Ww}$ and the volatility
vectors $\sigmad ,\sigmaf ,\sigmaq \in \rr^3$ satisfy
\begin{align} \label{eq4.2cb}
\qvar{\sigmad,\sigmaf}{}=\sigma \wh{\sigma} \rhodf ,\quad
\qvar{\sigmad,\sigmaq}{}=\sigma \barsig \rhod ,\quad
\qvar{\sigmaf,\sigmaq}{}= \wh{\sigma} \barsig \rhof.
\end{align}

\subsubsection{Foreign martingale measure}

Let us denote $R_t:=(Q_t)^{-1}$ for every $t\in \rr_+$. It is easy to check that the dynamics of the process $R$ under $\Q$ are
\begin{align}
\diff R_t=R_t(\xf_t-\xd_t)\diff t-R_t\sigmaq \diff (\Ww_t-\sigmaq t).
\end{align}
We fix $\Td>0$ and, using Girsanov's theorem, we define the probability measure $\whQ$ on $(\Omega,\cF_\Td)$
\begin{align} \label{cEq}
\frac{d\whQ}{d\Q}:=e^{\barsig\Zthree_T -\frac{1}{2}\barsig^2 T}
=  e^{\sigmaq \whW_T -\frac{1}{2}\|\sigmaq\|^2 T}=:\cE^q_T
\end{align}
so that the process $\whW$, which is given by $\whW_t:=\Ww_t - \sigmaq t$ for
every $t\in [0,\Td]$, is a standard Brownian motion under $\whQ$. One can observe that the density process
$\cE^q_t,\, t \in [0,T]$ has the continuous martingale part coinciding with the continuous martingale part
of the process  $(Q_0)^{-1} Q_t,\, t \in [0,T]$, which we denote as $\cE^q \mart (Q_0)^{-1} Q$.
Furthermore, the process $R_t\bar{B}^d_t (\bar{B}^f_t)^{-1}$ is a martingale under $\whQ$ and thus
$\whQ$ is the {\it foreign martingale measure}. 

It is worth noting that the processes $\whZone_t=\Zone_t-\barsig\rhod t$, $\whZtwo_t = \Ztwo_t - \barsig\rhof t$ and $\whZthree_t =\Zthree_t - \barsig t$ for every $t\in [0,\Td]$
are correlated Brownian motions under $\whQ$. Observe that under $\whQ$ the process $\rf$ satisfies, for every $t\in [0,\Td]$,
\begin{align}
\diff \rf_t=\big(\wh{c} +\wh{\sigma}\barsig\rhof -\wh{b}\rf_t\big)\diff t+\sigmaf \diff \whW_t .
\end{align}
Therefore, upon denoting $\wh{a}:=\wh{c}+ \wh{\sigma}\barsig\rhof $ and
we conclude that the dynamics of the triplet $(\rd,\rf,Q)$ under the foreign martingale measure $\whQ$ are given by
\eqref{eqC},
which is the foreign market model formally equivalent to the domestic market model \eqref{eq4.2cv}.
The reader is
referred to \cite{Dam2020inflation} for analogous model specification in the context of inflation-linked derivatives.

\subsection{Standing assumptions}

The above considerations lead to the following standing assumption, which we find to be the most convenient representations of our model for explicit computations under $\Q$ and $\whQ$.
For brevity, we write $\lqtT =\int_t^T \lqu \diff u$ where $\lqt := \alpha^d_t-\alpha^f_t$ and
\begin{align} \label{wtQ3}
\LqtT :=e^{\lqtT }=e^{\int_t^{\Td}\lqu \diff u}=e^{\int_t^{\Td}(\alpha^d_u-\alpha^f_u)\diff u}.
\end{align}

\bhyp  \label{ass4.1} {\rm
The dynamics of the triplet $(\rd,\rf,Q)$ of $\ff$-adapted stochastic processes under the domestic martingale measure $\Q$ are
\begin{align}
&\diff \rd_t=(a- b\rd_t)\diff t+\sigmad \diff \Ww_t, \nonumber \\
&\diff \rf_t=(\wh{a}- \wh{\sigma}\barsig\rhof -\wh{b}\rf_t)\diff t+\sigmaf \diff \Ww_t , \label{eqB} \\
&\diff Q_t=Q_t(\rd_t-\rf_t + \lqt)\diff t + Q_t \sigmaq \diff \Ww_t, \nonumber
\end{align}
and the dynamics of the triplet $(\rd,\rf,R)$ where $R=Q^{-1}$ under the foreign martingale measure $\whQ$ are
\begin{align}
&\diff \rd_t=(a+\sigma\barsig\rhod- b\rd_t)\diff t+\sigmad \diff \whW_t, \nonumber \\
&\diff \rf_t=(\wh{a}-\wh{b}\rf_t)\diff t+\sigmaf \diff \whW_t , \label{eqC} \\
&\diff R_t=R_t(\rf_t-\rd_t -\lqt)\diff t-R_t \sigmaq \diff \whW_t. \nonumber
\end{align}
where $\alpha^h,\alpha^d$ and $\alpha^f$ are deterministic functions, integrable on $[0,\Td]$ for every $\Td>0$.}
\ehyp




\subsection{Auxiliary model processes}    \label{sec4.2}

Let us introduce some notation and recall well known computations for Vasicek's model introduced in \cite{Vasicek1977}. For a fixed $u>0$, we define, for every $t \geq 0$,
\begin{align} \label{BatT}
\Bd(t,u):=\EQ\left[e^{-\int_t^u \rd_v \diff v}\,\Big|\,\cF_t\right], \ \,
\Bf(t,u):=\EwhQ \left[e^{-\int_t^u \rf_v \diff v}\,\Big|\,\cF_t\right]
\end{align}
where the dates $t$ and $u$ are arbitrary so that it is not assumed that $t\leq u$. Notice, in particular, that the equality $\Bd(t,u)=e^{\int_u^t \rd_v\,dv}=(\Bd_u)^{-1} \Bd_t$ holds for every $t \geq u$ where the process $\Bd$ satisfies $d\Bd_t= \rd_t \Bd_t \diff t$ and $\Bd_0=1$.  Let us recall without proof the well-known result for Vasicek's model. An analogous result can be formulated for the process $\Bf(t,u)$ using the process $\rf$ and suitably modified functions $\wh{m}(t,u)$ and $\wh{n}(t,u)$. It should be stressed that the auxiliary processes $\Bd(t,u),\Bf(t,u),\Bdd (t,s,u),$ etc. are not assumed to represent traded assets but they are useful in explicit computations of dynamics of futures prices in Section \ref{sec4.3}. For elementary proofs of Proposition \ref{pro4.1} and Proposition \ref{pro4.2} we refer to \cite{Ding2024wp}.

\bp  \label{pro4.1}
Let the interest rate process $\rd$ be defined as a solution to the stochastic differential equation
\begin{equation} \label{sde1}
d\rd_t=(a-b\rd_t) \diff t + \sigma \diff \Zone_t , \quad \rd_0>0,
\end{equation}
where $a,b$ and $\sigma$ are positive constants and $\Zone$ is a Brownian motion. The unique solution to the stochastic differential
equation \eqref{sde1} satisfies, for every $0 \leq s\leq t$,
\begin{equation} \label{sde2}
\rd_t=\rd_s e^{-b(t-s)} + \frac{a}{b}\left(1-e^{-b(t-s)}\right) + \int_s^t\sigma e^{-b(t-v)} \diff \Zone_v .
\end{equation}
For any fixed $u> 0$, the process $(\Bd(t,u),\,t\leq u)$ equals
\begin{align} \label{eqbx}
\Bd(t,u)=e^{m(t,u)- n(t,u)\rd_t}=:B_u(t,\rd_t)
\end{align}
where the function $B_u:[0,u] \times \rr \to \rr$ is given by $B_u(t,x)=e^{m(t,u)- n(t,u)x}$ and
\begin{align*}
m(t,u)&=\frac{1}{2} \int_t^u \sigma^2 n^2(v,u) \diff v -\int_t^u a n(v,u)\diff v, \\
n(t,u)&=\frac{1}{b}\left( 1-e^{-b(u-t)}\right).
\end{align*}
The dynamics of the process $(B_u(t,\rd_t),\, t\leq u)$ are
\begin{align}  \label{eqbx1}
\diff B_u(t,\rd_t)=B_u(t,\rd_t)\big(\rd_t\diff t + \sigma_u(t)\big)\diff \Zone_t
\end{align}
where $\sigma_u(t)=-\sigma n(t,u)$ for every $t\in [0,u]$.
\ep

We now extend equality \eqref{BatT} by defining, for any fixed $0< s<u$ and every $t\leq s$
\begin{align*}
\Bdd (t,s,u):=\EQ\left[e^{-\int_s^u \rd_v \diff v}\,\Big|\,\cF_t\right]
= \EQ\left[ \EQ\Big[ e^{-\int_s^u \rd_v \diff v}\,\big|\,\cF_s\Big]\,\Big|\,\cF_t\right]
=  \EQ\big[ B^d(s,u) \,\big|\,\cF_t\big].
\end{align*}
Let us denote, for every $0\leq t \leq s \leq u$,
\begin{align*}
\nnN (t,s,u):= 
\int_t^s \sigma_s(v)(\sigma_u(v)-\sigma_s(v))\,dv.
\end{align*}

\bp  \label{pro4.2}
For any $t\leq s<u$, we have that $\Bdd (t,s,u)=\FB_{s,u}(t,\rd_t)$  where
\begin{align} \label{xBx}
\FB_{s,u}(t,\rd_t)= \frac{B_u(t,\rd_t)}{B_s(t,\rd_t)}\, e^{\nnN (t,s,u)}  
\end{align}
The process $(\FB_{s,u}(t,\rd_t),\,t\leq s)$ has the following dynamics under $\Q$
\begin{align} \label{Bx}
\diff \FB_{s,u}(t,\rd_t)=\FB_{s,u}(t,\rd_t)\big( \sigma_u(t)-\sigma_s(t)\big)\diff \Zone_t.
\end{align}
\ep

For any two continuous semimartingales $Y^1$ and $Y^2$ defined on a common probability space, we write $Y^1\mart Y^2$ whenever $Y^1$ and $Y^2$ have the same continuous local martingale part in their respective canonical semimartingale decomposition. In the present framework, if the equality $Y^1 \mart Y^2$ holds under $\Q$ on $(\Omega,\cF_T)$ then, due to the Girsanov theorem for a Brownian motion, it is also satisfied under any probability measure on $(\Omega,\cF_T)$ that is equivalent to $\Q$ and hence, in particular,
under $\whQ$.

\brem \label{rem4.1}
For ease of reference, we will formulate some consequences of Propositions \ref{pro4.1} and \ref{pro4.2}.
First, Proposition \ref{pro4.1} shows that for any fixed $\Td>0$ the process $(B_\Td(t,\rd_t),\, t \leq \Td )$ satisfies
\begin{align*}
\diff B_\Td(t,\rd_t)\mart B_\Td(t,\rd_t)\btT\diff \Zone_t
\end{align*}
where $\btT =-\sigma n(t,T)$ for every $t\in [0,T]$. Similarly, for any fixed $U>0$ the process $(B_\Ud(t,\rd_t),\, t \leq \Ud )$ satisfies
\begin{align*}
\diff B_\Ud(t,\rd_t)\mart B_\Ud(t,\rd_t)\btU \diff \Zone_t
\end{align*}
where $\btU =-\sigma n(t,U)$ for every $t\in [0,U]$.
Second, in view of Proposition \ref{pro4.2}, for any fixed $\Sd<\Td$, the process $(\FB_{\Sd,\Td}(t,\rd_t),\,t\leq\Sd)$ satisfies
\begin{align*}
\diff \FB_{\Sd,\Td}(t,\rd_t)=\FB_{\Sd,\Td}(t,\rd_t)\btST\diff \Zone_t,\quad \btST:=\btT-\btS.
\end{align*}
Similarly, for any fixed $\Sd<\Ud$, the process $(\FB_{\Sd,\Ud}(t,\rd_t),\,t\leq\Sd)$ satisfies
\begin{align*}
\diff \FB_{\Sd,\Ud}(t,\rd_t)=\FB_{\Sd,\Ud}(t,\rd_t)\btSU\diff \Zone_t,\quad \btSU:=\btU-\btS.
\end{align*}
\erem

\brem \label{rem4.1x}
Analogous results can be obtained for the foreign counterparts of processes $B_\Td(t,\rd_t)$
and $\FB_{\Sd,\Td}(t,\rd_t)$. For instance, for any fixed $\Td>0$ the process $(\Bf(t,\Td),\, t \leq \Td)$ satisfies
\begin{align} \label{feqbx}
\Bf(t,\Td)=e^{\wh{m}(t,\Td)- \wh{n}(t,\Td)\rf_t}=:\wh{B}_{\Td}(t,\rf_t)
\end{align}
where
\begin{align*}
\wh{m}(t,\Td)&=\frac{1}{2} \int_t^{\Td} \hsbvT \diff v
-\int_t^{\Td} \wh{a} \wh{n}(v,\Td)\diff v, \\
\wh{n}(t,\Td)&=\frac{1}{\wh{b}}\left( 1-e^{-\wh{b}(\Td-t)}\right).
\end{align*}

Then the dynamics of the process $(\wh{B}_{\Td}(t,\rf_t),\, t\leq \Td)$ under $\whQ$  are
\begin{align}  \label{feqbx1}
\diff \wh{B}_{\Td}(t,\rf_t)=\wh{B}_{\Td}(t,\rf_t)\big(\rf_t\diff t+ \hbtT \big)\diff \whZtwo_t
\end{align}
where $\hbtT := -\wh{\sigma}\wh{n}(t,\Td)$ and thus under $\Q$ we have that, for $t\in [0,\Td],$
\begin{align*}
\diff \wh{B}_\Td(t,\rf_t)\mart \wh{B}_\Td(t,\rf_t)\hbtT \diff \Ztwo_t.
\end{align*}
Finally, we write
\begin{align*}
\hnnN (t,s,u):= 
\int_t^s \wh{\sigma}_s(v)(\wh{\sigma}_u(v)-\wh{\sigma}_s(v))\,dv.
\end{align*}
\erem

\subsection{Futures prices under the term structure model}   \label{sec4.3}

After introducing the model for the domestic and foreign interest rate and the exchange rate, our next goal is to compute the dynamics of futures prices.

\subsubsection{Dynamics of interest rate futures}   \label{sec4.3.1}

Recall that the conventional expressions for the backward-looking Realised AONIA and SOFR Average are given in Definition \ref{def2.1}. It should be stressed again that the price of futures contract with rate $\Fd_t(\Ud,\Td)$ (resp., $\Ff_t(\Ud,\Td)$) is denominated in AUD (resp., USD).

\bd  \label{def4.1}
The {\it AONIA futures price}, denoted by $\Fd_t(\Ud,\Td)$, is defined by the futures contract referencing the Realised AONIA and the {\it SOFR futures price}, denoted by $\Ff_t(\Ud,\Td)$, is determined by the futures contract referencing the SOFR Average. We set, for every $t\in [0,\Td]$,
\begin{align*}
\Fd_t(\Ud,\Td):=\EQ (\Rd(\Ud,\Td)\,|\,\cF_t), \quad \Ff_t(\Ud,\Td):=\EwhQ (\Rf(\Ud,\Td)\,|\,\cF_t).
\end{align*}
\ed

Notice that the SOFR and AONIA futures prices introduced above are martingales under the probability measure $\Q$ and satisfy, for all $t\in[0,\Td]$,
\begin{align*}
1+\delta\Fd_t(\Ud,\Td)=\EQ\Big[e^{\int_{\Ud}^{\Td} \rd_u\diff u}\,\Big|\,\cF_t\Big], \quad
1+\delta\Ff_t(\Ud,\Td)=\EwhQ\Big[e^{\int_{\Ud}^{\Td} \rf_u\diff u}\,\Big|\,\cF_t\Big],
\end{align*}
and, for all $t\in[\Ud,\Td]$,
\begin{align*}
1+\delta\Fd_t(\Ud,\Td)=e^{\int_{\Ud}^t\rd_u\diff u}\,\EQ\Big[e^{\int_t^{\Td}\rd_u\diff u}\,\Big|\,\cF_t\Big], \quad
1+\delta\Ff_t(\Ud,\Td)=e^{\int_{\Ud}^t\rf_u\diff u}\,\EwhQ\Big[e^{\int_t^{\Td}\rf_u\diff u}\,\Big|\,\cF_t\Big].
\end{align*}

\bp \label{pro4.3}
The AONIA futures price referencing the accrual period $[\Ud,\Td]$ satisfies, for every $t\in[0,\Ud]$,
\begin{align} \label{futa}
1 + \delta \Fd_t(\Ud,\Td) 
=\frac{B_U(t,\rd_t)}{B_T(t,\rd_t)}\,e^{\nnN (t,\Ud,\Td)+\int_{\Ud}^{\Td} \sbvT\,dv }
\end{align}
Furthermore, for every $t\in[\Ud,\Td]$,
\begin{align}  \label{futa1}
1 + \delta \Fd_t(\Ud,\Td) 
=\frac{\Rd (U,t)}{B_T(t,\rd_t)}\,e^{\int_t^T \sbvT \,dv}.
\end{align}
The dynamics of AONIA futures price are, for every $t\in[0,\Ud]$,
\begin{align*}
\diff \Fd_t(\Ud,\Td)=\delta^{-1}(1+\delta \Fd_t(\Ud,\Td))(\btU - \btT ) \diff \Zone_t
\end{align*}
and, for every $t\in[\Ud,\Td]$,
\begin{align*}
\diff \Fd_t(\Ud,\Td)=-\delta^{-1}(1+\delta \Fd_t(\Ud,\Td)) \btT \diff \Zone_t.
\end{align*}
\ep

\begin{proof}
Straightforward computations show that, for every $t\leq \Ud < \Td$,
\begin{align*}
1+\delta \Fd_t(\Ud,\Td)=\EQ\Big[e^{\int_\Ud^{\Td}\rd_u\diff u}\,\Big|\,\cF_t\Big]=e^{\mu_{\Ud,\Td}(t,\rd_t)+\frac{1}{2}v^2_{\Ud,\Td}(t)}
\end{align*}
where
\begin{align} \label{eqmu}
\mu_{\Ud,\Td}(t,\rd_t):=(n(t,\Td)-n(t,\Ud))\rd_t+\int_t^{\Td} an(v,\Td)\diff v-\int_t^{\Ud} an(v,\Ud)\diff v
\end{align}
\begin{align}  \label{eqv}
v^2_{\Ud,\Td}(t):=\int_t^{\Ud}(\sigma_{\Td}(v)-\sigma_{\Ud}(v))^2\diff v
 +\int_{\Ud}^{\Td} \sigma^2_{\Td}(v) \diff v.
\end{align}
Furthermore, for every $t\in [\Ud,\Td]$,
\begin{align*}
1+\delta \Fd_t(\Ud,\Td)&
=\Rd(U,t)\,e^{\mu_{\Td}(t,\rd_t)+\frac{1}{2}v^2_{\Td}(t)}
\end{align*}
where
\begin{align*}
\mu_{\Td}(t,\rd_t):=n(t,\Td)\rd_t + \int_t^{\Td} an(u,\Td) \diff u, \quad
v^2_{\Td}(t):=\int_t^{\Td} \sbuT \diff u .
\end{align*}
To complete the proof it suffices to apply the It\^o formula and results from Section \ref{sec4.2}.
\end{proof}


The result for the SOFR futures price is identical to Proposition \ref{pro4.3} but with an appropriately modified notation.

\bp \label{pro4.4}
The SOFR futures price referencing the accrual period $[\Ud,\Td]$ is given by, for all $t\in[0,\Ud]$,
\begin{align} \label{futs}
1+\delta \Ff_t(\Ud,\Td)
= \frac{\whB_U(t,\rf_t)}{\whB_T(t,\rf_t)}\,e^{\hnnN (t,\Ud,\Td) +\int_U^T \hsbvT\,dv }.
\end{align}
and, for all $t\in[\Ud,\Td]$,
\begin{align*}
1+\delta \Ff_t(\Ud,\Td)=\frac{\Rf (U,t)}{\whB_T(t,\rf_t)}\,e^{\int_t^T \hsbvT\,dv}.
\end{align*}
The dynamics of SOFR futures price are, for all $t\in[0,\Ud]$,
\begin{align*}
\diff \Ff_t(\Ud,\Td)&=\delta^{-1}(1+\delta \Ff_t(\Ud,\Td))( \hbtU -\hbtT )\diff\whZtwo_t,
\end{align*}
and, for all $t\in[\Ud,\Td]$,
\begin{align*}
\diff \Ff_t(\Ud,\Td)&=-\delta^{-1}(1+\delta \Ff_t(\Ud,\Td))\hbtT \diff\whZtwo_t.
\end{align*}
\ep

\brem \label{rem4.2}
From Proposition \ref{pro4.3}, the futures price $\Fd =\Fd(\Ud,\Td)$ satisfies $\diff \Fd_t =\nud_t \diff \Zone_t$ where, for all $t\in[0,\Ud]$,
\begin{align*}
\nud_t :=\delta^{-1}(1+\delta\Fd_t(\Ud,\Td))(\btU -\btT)=-\delta^{-1}(1+\delta \Fd_t(\Ud,\Td))\btUT
\end{align*}
and, for all $t\in[\Ud,\Td]$,
\begin{align*}
\nud_t :=-\delta^{-1}(1+\delta\Fd_t(\Ud,\Td))\btT .
\end{align*}
Similarly, from Proposition \ref{pro4.4}, the dynamics of the futures price $\Ff = \Ff(\Ud,\Td)$ are $\diff \Ff_t=\nuf_t \diff \whZtwo_t$ (hence $\diff \Ff_t \mart \nuf_t \diff \Ztwo_t$)  where, for all $t\in[0,\Ud]$,
\begin{align*}
\nuf_t :=\delta^{-1}(1+\delta \Ff_t(\Ud,\Td))(\hbtU -\hbtT) =-\delta^{-1}(1+\delta \Ff_t(\Ud,\Td))\hbtUT
\end{align*}
and, for all $t\in[\Ud,\Td]$,
\begin{align*}
\nuf_t :=-\delta^{-1}(1+\delta \Ff_t(\Ud,\Td))\hbtT.
\end{align*}
\erem

\subsubsection{Dynamics of currency futures}  \label{sec4.3.2}

We now examine the dynamics of currency futures with a fixed maturity $\Td>0$, as given by Definition \ref{def2.3}. Recall that the settlement price at $\Td$ of the currency futures contract equals $Q_{\Td}$ where we assume, without loss of generality, that the contract's nominal principal equals 1 USD. The currency futures price for maturity $\Td$ is expressed in the domestic currency (AUD) and thus it is formally given by the equality $\Fq_t(\Td):=\EQ(Q_{\Td}\,|\,\cF_t)$.

\bp  \label{pro4.5}
The currency futures price $\Fq(\Td)$ equals, for every $t\in [0,\Td]$,
\begin{align} \label{futq}
\Fq_t(\Td) = \frac{\LqtT Q_t\whB_{\Td}(t,\rf_t)}{B_{\Td}(t,\rd_t)}e^{\cqtT}
\end{align}
where
\begin{align*}
\cqtT=\int_t^{\Td}\buT\big(\buT-\hbuT\rhodf-\barsig\rhod\big)\diff u.
\end{align*}
The dynamics of $\Fq(\Td)$ under $\Q$ are
\begin{equation*}
\diff\Fq_t(\Td)=\Fq_t(\Td)\big(-\btT\diff \Zone_t+ \hbtT \diff\Ztwo_t+\barsig\diff\Zthree_t\big).
\end{equation*}
\ep

\proof
In view of Assumption \ref{ass4.1} we have that
\begin{align*}
\Fq_t(\Td)&=\EQ(Q_{T}\,|\,\cF_t)=Q_t\LqtT e^{-\frac{1}{2} \barsig^2(\Td-t)}\,\EQ\Big[ e^{\int_t^{\Td}(\rd_u-\rf_u)\diff u+\barsig(\Zthree_{\Td}-\Zthree_t)}\,\Big|\,\cF_t\Big] \\
&=Q_t\LqtT e^{-\frac{1}{2} \barsig^2(\Td-t)}\,\EQ\big[e^{\Phiq_{t,T}}\,\big|\,\cF_t\big]
\end{align*}
where we denote
\begin{align*}
\Phiq_{t,T}:=\int_t^{\Td} (\rd_u-\rf_u) \diff u + \barsig\big(\Zthree_{\Td}-\Zthree_t\big).
\end{align*}

Using \eqref{sde2} and an analogous equation for $\rf$ we obtain under $\Q$
\begin{align*}
\Phiq_{t,T}&= \rd_t n(t,\Td)-\rf_t\wh{n}(t,\Td)+\int_t^{\Td} \big(an(u,\Td)-\wh{c}\wh{n}(u,\Td)\big)\diff u
  - \int_t^{\Td} \buT\diff \Zone_u\\
&+\int_t^{\Td} \hbuT \diff\Ztwo_u+\barsig\big(\Zthree_{\Td}-\Zthree_t\big)
=\rd_t n(t,\Td)-\rf_t\wh{n}(t,\Td)\\ \,& +\int_t^{\Td} \big(an(u,\Td)-\wh{c}\wh{n}(u, T)\big)\diff u+\int_t^{\Td} \big(
\barsig \alone-\buT+\hbuT \rhodf \big)\diff \Wone_u\\
&+\int_t^{\Td} \Big(\barsig\altwo+\hbuT\sqrt{1-\rhodf^2}\Big)\diff\Wtwo_u
+\int_t^{\Td}\barsig\althree\diff \Wthree_u.
\end{align*}
In view of the independence of Brownian motions $\Wone, \Wtwo$ and $\Wthree$, the $\cF_t$-conditional distribution of $\Phiq_{t,T}$ under $\Q$ is Gaussian with the conditional expectation $\wt{\mu}_{\Td}(t,\rd_t,\rf_t)$ and the conditional variance $\wt{v}^2_{\Td}(t)$
where
\begin{equation*}
\wt{\mu}_{\Td}(t,\rd_t,\rf_t):=\rd_t n(t,\Td)-\rf_t\wh{n}(t,\Td) + \int_t^{\Td} \big(an(u,\Td)-\wh{a}\wh{n}(u,\Td)\big)\diff u
\end{equation*}
and
\begin{equation*}
\wt{v}^2_{\Td}(t):=\int_t^{\Td}\bigg[\Big(\barsig\alone-\buT+\hbuT\rhodf\Big)^2
+ \Big(\barsig\altwo+\hbuT\sqrt{1-\rhodf^2}\Big)^2+\barsig^2\althree^2 \bigg] \diff u.
\end{equation*}
Consequently,
\begin{equation*} 
\Fq_t(\Td)=Q_t \LqtT e^{\,\wt{\mu}_{\Td}(t,\rd_t,\rf_t)+\frac{1}{2}\wt{v}^2_{\Td}(t)-\frac{1}{2} \barsig^2(\Td-t)}
\end{equation*}
and the asserted formula follows by straightforward computations. The dynamics of currency futures can be easily obtained using the It\^o formula.
\endproof

\brem \label{rem4.3}
From Proposition \ref{pro4.5}, the currency futures price $\Fq=\Fq(\Td)$ satisfies
\begin{equation*}
\diff \Fq_t= \nu^{\cq,1}_t\diff \Zone_t+\nu^{\cq,2}_t\diff\Ztwo_t+\nu^{\cq,3}_t\diff\Zthree_t
\end{equation*}
where $\nu^{\cq,1}_t:=-\btT\Fq_t ,\,\nu^{\cq,1}_t:=\hbtT\Fq_t$ and $\nu^{\cq,1}_t:=\barsig\Fq_t$.
Furthermore, $\diff Q_t\mart Q_t\barsig \diff\Zthree_t$.
\erem

\subsection{Pricing martingale measure} \label{sec4.6}

Recall that our model was constructed under the probability measure $\Q$, whereas the problem of hedging  and hence also the arbitrage-free pricing for collateralized contracts is more conveniently formulated and solved under the \PMM $\wtQ$ introduced in Definition \ref{def3.3} and employed in Proposition \ref{pro3.2}. Therefore, our next goal is to establish the existence of the \PMM $\wtQ$ within the present framework. Since this part of the analysis relies critically on the dynamics of foreign futures, it is developed hereafter to ensure we have all necessary components.

\bp \label{pro4.7}
The \PMM $\wtQ$ exists and coincides with the domestic martingale measure $\Q$.
\ep

\begin{proof}
Recall from \eqref{fQsi} that
\begin{align*}
\Ffq_t=\Ffq_0+\int_0^t Q_u\diff \Ff_u+\qvar{Q,\Ff}{t}
\end{align*}
and thus, from Proposition \ref{pro4.4} we have that (recall that $\whZtwo_t =\Ztwo_t - \barsig\rhof t$)
\begin{align*}
d\Ffq_t&=Q_t\nuf_t\big(d\whZtwo_t+\barsig\rhof\diff t\big)=Q_t\nuf_t\diff \Ztwo_t
= Q_t\nuf_t\Big(\rhodf\diff\Wone_t+\sqrt{1-\rhodf^2}\diff\Wtwo_t\Big),
\end{align*}
which shows that $\Ffq$ is a (local) martingale under $\Q$. In view of their definitions, the processes $\Fd$ and $\Fq$
are martingales under $\Q$, which shows that $\Q$ is a pricing martingale measure. Its uniqueness
is a consequence of the model completeness, which will be examined in Proposition \ref{pro4.8}.
\end{proof}

\brem \label{rem5.1}
Note that $d\Ffq_t=\nufq_t\diff \Ztwo_t$ where $\nufq_t = Q_t\nuf_t$ with $\nuf_t$ given in Remark \ref{rem4.2}.
\erem

\subsection{Model completeness}  \label{sec4.7}

We now address the issue of attainability of an arbitrary contingent claim $X_{\Td}$. We will argue that a replicating strategy for $X_{\Td}$ based on futures contracts can be identified by matching terms with different Brownian motions under $\Q$. We have the following result, which allows us to identify the replicating strategy $\phiv$ using the auxiliary process $\psi$ introduced in equation \ref{rep3} and the dynamics \eqref{rep2a} of futures contracts. As expected, the model completeness is a consequence of the predictable representation property of the Brownian motion. We henceforth write $\phiv_t = (\phiv^1_t,\phiv^2_t,\phiv^3_t)=(\phid_t,\phif_t,\phiq_t)$ for every $t\in [0,\Td]$.

\bp \label{pro4.8}
Consider a collateralized contract $(X_{\Td},\beta)$ with the terminal payoff $X_{\Td}$ at time $\Td$ and proportional collateralization at rate $\beta$. If the random variable $X_{\Td}(B^{\beta}_T)^{-1}$ is $\Q$-integrable, then the contract $(X_{\Td},\beta)$ can be replicated by a (unique) futures trading strategy $\phiv$ where
\begin{align*}
&\phid_t=(\nud_t)^{-1}\big(\psi^1_t-\psi^3_t(\nu^{\cq,3}_t)^{-1}\nu^{\cq,1}_t\big),\\
&\phif_t=(\nufq_t)^{-1}\big(\psi^2_t-\psi^3_t(\nu^{\cq,3}_t)^{-1}\nu^{\cq,2}_t \big),\\
&\phiq_t=\psi^3_t(\nu^{\cq,3}_t)^{-1},
\end{align*}
or, equivalently,
\begin{align} \label{hedge1}
\begin{bmatrix}
\phiv^1_t \\
\phiv^2_t \\
\phiv^3_t
\end{bmatrix}=
\begin{bmatrix}
(\nud_t)^{-1} &0& -(\nud_t)^{-1}(\nu^{\cq,3}_t)^{-1}\nu^{\cq,1}_t\\
0&(\nufq_t)^{-1}&-(\nufq_t)^{-1}(\nu^{\cq,3}_t)^{-1}\nu^{\cq,2}_t \\
0&0& (\nu^{\cq,3}_t)^{-1}
\end{bmatrix}
\begin{bmatrix}
\psi^1_t \\
\psi^2_t \\
\psi^3_t
\end{bmatrix}
\end{align}
where $\psi_t =(\psi^1_t,\psi^2_t,\psi^3_t)$ is a unique process satisfying under $\Q$
\begin{align} \label{rep3}
\diff \big( (B^{\beta}_t)^{-1} \pi^{\beta}_t(X_{\Td}) \big)=\big(B^{\beta}_t\big)^{-1}[\psi^1_t,\psi^2_t,\psi^3_t]\diff Z_t.
\end{align}
\ep

\proof
On the one hand, we know from Proposition \ref{pro3.1} that the discounted wealth of a collateralized futures trading strategy $(\phiv,C)$ where $\phiv=[\phid,\phif,\phiq]$ and $C = \beta V$ satisfies
\begin{align} \label{rep1}
\diff \wt{V}_t^\beta(\phiv,C)=\big(B^{\beta}_t\big)^{-1}\big(\phid_t\diff \Fd_t+\phif_t\diff \Ffq_t+\phiq_t\diff \Fq_t\big)=
\big(B^{\beta}_t\big)^{-1}[\phiv^1_t,\phiv^2_t,\phiv^3_t]\diff F_t
\end{align}
where $F:=[\Fd,\Ffq,\Fq]^{\perp}$. Recall that the processes $\Fd,\,\Ffq$ and $\Fq$ are strictly positive, continuous local martingales under $\Q$ and satisfy (see Remarks \ref{rem4.2}, \ref{rem4.3} and \ref{rem5.1})
\begin{align*} 
\diff \Fd_t=\nud_t \diff \Zone_t, \quad
\diff\Ffq_t=\nufq_t\diff \Ztwo_t,\quad
\diff \Fq_t= \nu^{\cq,1}_t\diff \Zone_t+\nu^{\cq,2}_t\diff\Ztwo_t+\nu^{\cq,3}_t\diff\Zthree_t,
\end{align*}
that is,
\begin{align} \label{rep2a}
\diff \Fd_t=[\nud_t,0,0] \diff Z_t, \quad
\diff\Ffq_t=[0,\nufq_t,0]\diff Z_t,\quad
\diff \Fq_t=[\nu^{\cq,1}_t,\nu^{\cq,2}_t,\nu^{\cq,3}_t]\diff Z_t
\end{align}
where $\nud,\nufq,\nu^{\cq,1},\nu^{\cq,2}$ and $\nu^{\cq,3}$ are strictly positive, continuous stochastic processes.
 More explicitly, we have that
 \begin{align*} 
 &\diff \Fd_t=[-\delta^{-1}(1+\delta \Fd_t)\btT,0,0] \diff Z_t, \nonumber \\
 &\diff\Ffq_t=[0,- Q_t\delta^{-1}(1+\delta \Ff_t)\hbtT ,0]\diff Z_t, \\
 &\diff \Fq_t=[-\Fq_t \btT , \Fq_t \hbtT,\Fq_t \barsig]\diff Z_t. \nonumber
 \end{align*}

On the other hand, the discounted price process $\wt{\pi}_t^{\beta}:=(B^{\beta}_t)^{-1}\pi_t^{\beta}$ is also
a continuous local martingale under $\Q$ and thus, from the predictable representation property of the Brownian motion $Z$, it can be uniquely represented as follows
\begin{align*} 
\diff \wt{\pi}_t^{\beta}=\big(B^{\beta}_t\big)^{-1}\big(\psi^1_t\diff \Zone_t+\psi^2_t\diff\Ztwo_t+\psi^3_t\diff \Zthree_t\big)
=\big(B^{\beta}_t\big)^{-1}[\psi^1_t,\psi^2_t,\psi^3_t]\diff Z_t
\end{align*}
where the processes $\psi^1,\psi^2$ and $\psi^3$ can be computed using the It\^o formula provided that the closed-form solution for the price $\wt{\pi}_t^{\beta}$ is available.

We observe that the processes $\nud,\nufq$ and $\nu^{\cq,3}$ are strictly positive and thus we obtain from \eqref{rep2a}
\begin{align*}
\diff \Zone_t&=(\nud_t)^{-1}\, \diff \Fd_t, \quad
\diff \Ztwo_t=(\nufq_t)^{-1}\diff\Ffq_t,\\
\diff\Zthree_t&=(\nu^{\cq,3}_t)^{-1}\big(\diff \Fq_t-\nu^{\cq,1}_t(\nud_t)^{-1}\,\diff \Fd_t-\nu^{\cq,2}_t(\nufq_t)^{-1}\diff\Ffq_t \big).
\end{align*}
Then $J:=[\psi^1_t,\psi^2_t,\psi^3_t]\diff Z_t$ can be represented
in terms of $\Fd,\,\Ffq$ and $\Fq$
\begin{align*}
&J=\psi^1_t(\nud_t)^{-1} d\Fd_t+\psi^2_t(\nufq_t)^{-1}d\Ffq_t+\psi^3_t(\nu^{\cq,3}_t)^{-1}
(\diff\Fq_t-\nu^{\cq,1}_t(\nud_t)^{-1}d \Fd_t-\nu^{\cq,2}_t (\nufq_t)^{-1}d\Ffq_t)\\
&=(\nud_t)^{-1}\big( \psi^1_t-\psi^3_t(\nu^{\cq,3}_t)^{-1}\nu^{\cq,1}_t\big)\diff \Fd_t
+(\nufq_t)^{-1}\big(\psi^2_t-\psi^3_t(\nu^{\cq,3}_t)^{-1}\nu^{\cq,2}_t \big)\diff\Ffq_t
+\psi^3_t(\nu^{\cq,3}_t)^{-1}\diff \Fq_t.
\end{align*}
The equalities from the statement of Proposition \ref{pro4.8} now follow by comparing the last equality with \eqref{rep1}.
\endproof

\section{Arbitrage-Free Pricing of Collateralized Cross-Currency Basis Swaps} \label{sec5}

Collateralization is used to mitigate the counterparty credit risk and hence increase trading volumes; it is also enforced by regulators of financial markets. Therefore, in our further calculations, we take collateral as the default setting with the rate $\beta$ indicating the level of proportional collateralization. Needless to say, the pricing and hedging results for uncollateralized contracts can be obtained upon setting $\beta=0$.

We first derive in Proposition \ref{pro5.1} and Proposition \ref{pro5.2} explicit pricing formulae for a CCBS with domestic collateralization. Next, in
Proposition \ref{pro5.1c} we obtain the corresponding results for the case of a collateral posted in the foreign currency.
The tentative pricing results obtained using the \PMM will be supported in Section \ref{sec6} by further computations demonstrating the existence of a replicating strategy, although we only present here expressions for the hedging strategy in the case of the domestic collateralization.

\subsection{Convexity corrections}  \label{sec5.1}

The date $t \geq 0$ is arbitrary but it can be assumed that it is fixed throughout this subsection. We consider a family of dates $T_0,T_1,\dots ,T_n$ such that $T_0 \geq t$ and $T_i>t$ for every $i=1,2,\dots,n$ but we do not assume that these dates are ordered.  Let $Z^1, Z^2, \dots , Z^m$ be correlated Brownian motions on $\mathbb{R}_+$ under $\Q$ with constant correlations $\rho_{k,l}$ for $k,l=1,2,\dots,m$. We fix $l\leq n$ and we define the processes $X^1,X^2,\dots ,X^l$ satisfying
\begin{align*}
\int_t^{T_i} X^i_u\diff u=X^i_t+\int_t^{T_i}\mu^i_u\diff u+\int_t^{T_i}\sigma^i_u\diff Z^{k_i}_u
\end{align*}
for some $k_i=0,1,\dots,m$, as well as the processes $\wh{X}^{l+1},\wh{X}^{l+2},\dots ,\wh{X}^n$ satisfying
the equality
\begin{align*}
\int_t^{T_i}\wh{X}^i_u\diff u=\wh{X}^i_t+\int_t^{T_i}\wh{\mu}^i_u\diff u+\int_t^{T_i}\wh{\sigma}^i_u\diff Z^{k_i}_u
\end{align*}
for some $k_i=0,1,\dots,m$. By assumption, the integrands $\mu^i,\sigma^i, \wh{\mu}^i$ and $\wh{\sigma}^i$ are assumed
to be deterministic functions on respective intervals. We define, for every $i=1,2,\dots,l$,
\begin{align} \label{eqD}
D(t,T_i):=\EQ\left[e^{-\int_t^{T_i} X^i_u\diff u}\,\Big|\,\cF_t\right]
\end{align}
and, for every $i=l+1,\dots, n$,
\begin{align}  \label{eqhD}
\wh{D}(t,T_i):=\EQ\left[\mathcal{E}_{t,T_i}\,e^{-\int_t^{T_i}\wh{X}^i_u\diff u}\,\Big|\,\cF_t\right].
\end{align}
Finally, the stochastic exponential $\mathcal{E}_{t,s}$ is given by the equality, for every $s \geq t$,
\begin{align} \label{eqE}
\mathcal{E}_{t,s}=e^{\int_t^{s}\sigma^0_u\diff Z^0_u-\frac{1}{2}\int_t^s (\sigma^0_u)^2\diff u}
= e^{X^0_s}
\end{align}
where $\sigma^0$ is a deterministic function and the process $X^0$ satisfies, for every $s \geq t$,
\begin{align*}
X^0_s=\int_t^{s}\sigma^0_u \diff Z^0_u-\frac{1}{2}\int_t^s(\sigma^0_u)^2\diff u.
\end{align*}
Then we have the following elementary lemma, which will allow us to effectively compute prices of various cash flows
associated with  cross-currency basis swaps under the domestic and foreign collateralization.

\bl \label{newlm}
Let $J$ be given by the equality
\begin{align} \label{eqJ}
J:=\EQ\Big[\,e^{X^0_{T_0}-\alpha_1\int_t^{T_1}X^1_u\diff u-\ldots -\alpha_l\int_t^{T_l}X^l_u\diff u
-\alpha_{l+1}\int_t^{T_{l+1}}\wh{X}^{l+1}_u\diff u - \ldots - \alpha_n\int_t^{T_n}\wh{X}^n_u\diff u}\,\Big|\,\cF_t \Big]
\end{align}
where $\alpha_i\in \mathbb{R}$ for every $i=1,2,\dots,n$. Then $J$ has the representation
\begin{align}  \label{repJ}
J=[D(t,T_1)]^{\alpha_1}\dots [D(t,T_l)]^{\alpha_l}[\wh{D}(t,T_{l+1})]^{\alpha_{l+1}}
 \dots [\wh{D}(t,T_n)]^{\alpha_n}\Gamma (t,T_0,T_1,\dots,T_n)
\end{align}
where the total adjustment function $\Gamma (t,T_0,T_1,\dots,T_l)$ equals
\begin{align*}
\Gamma (t,T_0,T_1,\dots,T_n)=\Pi_{i=1}^n\Gamma_{i}(t,T_i)\Pi_{i,j=0,i<j}^n\Gamma_{i,j}(t,T_i,T_j)
\end{align*}
where, for every $i,j=0,1,\dots,n$, the self-adjustment $\Gamma_{i}(t,T_i)=\Gamma_{i}(t\wedge T_i,T_i)$ satisfies,
for $t\in [0,T_i]$,
\begin{align*}
&\EQ\Big[e^{-\alpha_i\int_t^{T_i} X^i_u\diff u}\,\Big|\,\cF_t\Big]= [D(t,T_i)]^{\alpha_i}\Gamma_{i}(t,T_i),\\
&\EQ\Big[e^{-\alpha_i\int_t^{T_i}\wh{X}^i_u\diff u}\,\Big|\,\cF_t\Big]= [\wh{D}(t,T_i)]^{\alpha_i}\Gamma_{i}(t,T_i),
\end{align*}
and, for every $i,j=0,1,\dots,n$ such that $i<j$, the cross-adjustment $\Gamma_{i,j}(t,T_i,T_j)= \Gamma_{i,j}(t \wedge T_i \wedge T_j,T_i,T_j) $ is such that, for $t\in [0,T_i \wedge T_j]$,
\begin{align*}
&\EQ\Big[e^{-\alpha_i\int_t^{T_i} X^i_u\diff u -\alpha_j\int_t^{T_j} X^j_u\diff u}\,\Big|\,\cF_t\Big]
       = [D(t,T_i)]^{\alpha_i} [D(t,T_j)]^{\alpha_i}\Gamma_{i,j}(t,T_i,T_j),\\
&\EQ\Big[e^{-\alpha_i\int_t^{T_i} X^i_u\diff u -\alpha_j\int_t^{T_j} \wh{X}^j_u\diff u}\,\Big|\,\cF_t\Big]
       = [D(t,T_i)]^{\alpha_i} [\wh{D}(t,T_j)]^{\alpha_j}\Gamma_{i,j}(t,T_i,T_j),\\
&\EQ\Big[e^{-\alpha_i\int_t^{T_i} \wh{X}^i_u\diff u -\alpha_j\int_t^{T_j} \wh{X}^j_u\diff u}\,\Big|\,\cF_t\Big]
       = [\wh{D}(t,T_i)]^{\alpha_i} [\wh{D}(t,T_j)]^{\alpha_j}\Gamma_{i,j}(t,T_i,T_j).
\end{align*}
\el

\proof
None of the terms in the sum in the exponent in \eqref{eqJ} should be repeated so we deal with a unique representation
of the conditional expectation $J$. The proof hinges on elementary properties of exponential moments of a lognormal distribution, which are applied directly to \eqref{eqD}, \eqref{eqhD} and \eqref{eqJ}, and thus it is omitted.
\endproof

In fact, representation \eqref{repJ} can be made explicit since closed-form expressions
for all adjustments are available. We find it convenient to classify adjustments according to their origin and expression
by examining all combinations of $i$ and $j$ and every possible situation.

First, for any $i=1,2,\dots,n$, we have the {\it self adjustment} $\Gamma^s$, which occurs whenever $\alpha_i \ne 1$. It is easy to handle since it can be checked that, for every $\alpha_i\in\mathbb{R}$,
\begin{align} \label{Gams}
\Gamma^s_{i}(t,T_i)=e^{\frac{1}{2}\int_t^{T_i}\alpha_i(\alpha_i-1)(\sigma_u^i)^2\diff u}=:\Gamma^s_t(T_i,\alpha_i,\sigma^i)
\end{align}
where $\sigma_u^i$ should be replaced by $\wh{\sigma}_u^i$ for $i=l,l+1,\dots,n$.

Second, if we assume that $X^i$ and $X^j$ for some $i,j=1,2,\dots,n$ such that $i < j$ are driven by a common Brownian motion $Z^{k_i}=Z^{k_j}$ (i.e., $\rho_{k_i,k_j}=1$) but the maturities $T_i$ and $T_j$ are different, that is, $T_i \ne T_j$. Then we deal with the {\it maturity adjustment}  $\Gamma^m$, which equals
\begin{align}  \label{Gamm}
\Gamma^m_{i,j}(t,T_i,T_j)=e^{\int_t^{T_i \wedge T_j}\alpha_i  \sigma_u^i\alpha_j \sigma_u^j  \diff u}=:
\Gamma^m_t(T_i\wedge T_j,\alpha_i\sigma^i,\alpha_j\sigma^j)
\end{align}
where $\sigma_u^i$ (resp. $\sigma_u^j)$ should be replaced by $\wh{\sigma}_u^i$ (resp., $\wh{\sigma}_u^j)$
if $i\geq l+1$ (resp., $j\geq l+1$).

Third, we assume that $X^i$ and $X^j$ for some $i,j=1,2,\dots,n$ such that $i < j$ are driven by different Brownian
motions $Z^{k_i}$ and $Z^{k_j}$ with $\rho_{k_i,k_j}\ne 1$ (note that the maturities $T_i$ and $T_j$ may differ as well),
we obtain the {\it correlation adjustment}  $\Gamma^c$, which satisfies
\begin{align}  \label{Gamc}
\Gamma^c_{i,j}(t,T_i,T_j)=e^{\int_t^{T_i \wedge T_j}\alpha_i  \sigma_u^i\alpha_j \sigma_u^j \rho_{k_i,k_j} du}
=:\Gamma^c_t(T_i\wedge T_j,\alpha_i\sigma^i,\alpha_j\sigma^j,\rho_{k_i,k_j})
\end{align}
where $\sigma_u^i$ (resp., $\sigma_u^j)$ should be replaced by $\wh{\sigma}_u^i$ (resp., $\wh{\sigma}_u^j)$
if $i\geq l+1$ (resp., $j\geq l+1$).

Finally, it remains to consider the {\it drift adjustment} $\Gamma^d$ for every $i=1,2,\dots,n$, which arises when considering, for instance,
\begin{align*}
\EQ\Big[ \mathcal{E}_{t,T_0}\,e^{-\alpha_i\int_t^{T_i} X^i_u\diff u}\,\Big|\,\cF_t\Big].
\end{align*}
Recall that we denote, for every $t\leq s$,
\begin{align*}
\mathcal{E}_{t,s}=e^{\int_t^s\sigma^0_u\diff Z^0_u-\frac{1}{2}\int_t^s(\sigma^0_u)^2\diff u}.
\end{align*}
Hence the name {\it drift adjustment} is justified since $\mathcal{E}_{t,T_0}$ can be interpreted as a change of a probability measure and thus, by the Girsanov theorem, as a change of drift in Brownian motions $Z^0,Z^1,\dots ,Z^l$.
For $i=1,2,\dots,l$ the drift-adjustments have the same form as correlation adjustments. In contrast, for $i=l+1,l+2,\dots,n$ they should be computed due to the definition of $\wh{D}_i(t,T_i)$, which includes $\mathcal{E}_{t,T_i}$.

For $i=1,2,\dots,l$, the drift adjustment equals
\begin{align}   \label{Gamdx}
\Gamma^d_{0,i}(t,T_0,T_i)=e^{-\int_t^{T_0\wedge T_i}\sigma^0_u\alpha_i\sigma_u^i\rho_{0,k_i} du}
=: \Gamma^d_t(T_0\wedge T_i,\sigma^0,\alpha_i\sigma^i,\rho_{0,k_i})
\end{align}
so it occurs whenever $T_0>t$ and $\rho_{0,k_i}\ne 0$. For $i=l+1,l+2,\dots,n$, we consider two cases: (a) $t< T_i \leq T_0$ and (b) $0\leq T_0< T_i$.  Observe that there is no drift adjustment in case (a) since
\begin{align*}
\wh{D}(t,T_i):=\EQ\Big[\mathcal{E}_{t,T_i}\,e^{-\alpha_i\int_t^{T_i}\wh{X}^i_u\diff u}\,\Big|\,\cF_t\Big]
=\EQ\Big[\mathcal{E}_{t,T_0}\,e^{-\alpha_i\int_t^{T_i}\wh{X}^i_u\diff u}\,\Big|\,\cF_t\Big].
\end{align*}
due to the equality $\EQ [\mathcal{E}_{t,T_0}\,|\,\cF_{T_i}]=\mathcal{E}_{t,T_i}$ when $t< T_i \leq T_0$.
In case (b), simple computations show that
\begin{align*}
\EQ\Big[\mathcal{E}_{t,T_0}\, e^{-\alpha_i\int_t^{T_i}\wh{X}^i_u\diff u}\,\Big|\,\cF_t \Big]
&=\EQ\Big[\mathcal{E}_{t,T_i}\, e^{-\alpha_i\int_t^{T_i}\wh{X}^i_u\diff u}\,\Big|\,\cF_t \Big]
e^{\int_{T_0}^{T_i}\sigma^0_u \alpha_i\wh{\sigma}_u^i\rho_{0,k_i} du} \\
&=[\wh{D}(t,T_i)]^{\alpha_i}\wh{\Gamma}^d_{0,i}(t,T_0,T_i)
\end{align*}
where the drift adjustment equals
 \begin{align} \label{Gamdy}
\wh{\Gamma}^d_{0,i}(t,T_0,T_i)=e^{\int_{t \vee T_0}^{T_i}\sigma^0_u\alpha_i\wh{\sigma}_u^i\rho_{0,k_i}\diff u} =:
\wh{\Gamma}^d_t(t \vee T_0,T_i,\sigma^0,\alpha_i\wh{\sigma}^i,\rho_{0,k_i}).
\end{align}
Due to their exponential form, all adjustments specified above can be also collectively referred to as {\it convexity corrections} (see, e.g., \cite{Henrard2018}, \cite{Jamshidian1994}, Chapter 11 in \cite{Pelsser2000} and \cite{Pelsser2003}).

\subsection{Single-period forward CCBS}  \label{sec5.02}

We will argue that to compute the price and hedging strategy for an $n$-period CCBS, it suffices to examine specific cash flows in a single-period setup and consequently generalize the single-period pricing formulae to the multi-period case using Proposition \ref{pro5.1} and Proposition \ref{pro5.2} (resp., Proposition \ref{pro5.1c}) for domestic (resp., foreign) collateralization combined with the linearity of arbitrage-free pricing operator from Proposition \ref{pro3.2}.

Therefore, we find it convenient to separately examine the two cases: the exchange of interest payments at the end of each accrual period and the exchange of nominal principals at the maturity date. The building blocks of a constant notional $\CCBS \,(\cT_n;\kappa )$ of Definition \ref{def2.5} are single-period cross-currency basis swaps given by the following definition where we write $X_{T_j}(T_0,T_{j-1},T_j)$ to represent $X_j$ for $j=1,2,\dots ,n$.

\bd  \label{def5.1}
A {\it single-period CCBS}  with no exchange of nominal principals, the inception date $0\leq T_0 \leq T_{j-1}$, and the accrual period $[T_{j-1},T_{j}]$, is defined by the net interest rate cash flow at payment date $T_{j}$ given by, for any fixed $j = 1,2,\dots,n$,
\begin{align*}
X_{T_j}(T_0,T_{j-1},T_j)& =\Rf(T_{j-1},T_{j})\delta_j Q_{T_{j}}-(\Rd(T_{j-1},T_{j})+\kappa)\delta_j Q_{T_0}\\
&= \Xf_{T_j}(T_0,T_{j-1},T_j)-\Xd_{T_j}(T_0,T_{j-1},T_j)=\Xf_j-\Xd_j
\end{align*}
where $\Xf_j=\Xf_{T_j}(T_0,T_{j-1},T_j)$ and $\Xd_j= \Xd_{T_j}(T_0,T_{j-1},T_j)$ represent the cash flows from the foreign and domestic interest rate leg, respectively.
\ed

To find the price a constant notional $\CCBS \,(\cT_n;\kappa )$, we also need to examine the exchange of nominal principals, which is represented by the net cash flow  $X^p_{T_n}(T_0,T_n)$  (also denoted by $X_n^{p}$) at maturity date $T_n$
equal to
\begin{equation*}
X^p_{T_n}(T_0,T_n)= X_n^{p}=Q_{T_n}-Q_{T_0} =Z_n-X_n.
\end{equation*}

The arbitrage-free pricing operator $\pi^{\beta}$, which is introduced in Proposition \ref{pro3.2} and applied to the market model from Assumption \ref{ass4.1}, is additive and homogeneous with respect to nominal principal amounts. Therefore, it suffices to fix $j$, denote $\Sd=T_0,\Ud= T_{j-1}, \Td = T_{j}$ and consider arbitrary three dates $0\leq \Sd \leq \Ud < \Td$ with the length of the accrual period $[\Ud,\Td]$ denoted by $\delta :=\Td-\Ud>0$. Then we will separately examine two payoffs: the contingent claim $X_{\Td} (\Sd,\Ud,\Td)$ at time $\Td$, which corresponds to interest rate cash flows, and is given by
\begin{equation} \label{ccbsa}
X^{i}_{\Td} (\Sd,\Ud,\Td):=\Xf_{\Td} (\Sd,\Ud,\Td)- \Xd_{\Td}(\Sd,\Ud,\Td)=
\Rf(\Ud,\Td)\delta Q_{\Td}-(\Rd(\Ud,\Td)+\kappa)\delta Q_{\Sd}
\end{equation}
where $\Xf_{\Td} (\Sd,\Ud,\Td)$ and $\Xd_{\Td}(\Sd,\Ud,\Td)$ represent the foreign and domestic legs, respectively,
and the contingent claim $X^{p}_{\Td}(\Sd,\Td)$ at time $\Td$ associated with the exchange of notional principals
\begin{equation}  \label{ccbsb}
X^{p}_{\Td}(\Sd,\Td):=Q_{\Td}-Q_{\Sd}.
\end{equation}
Notice that it suffices to find the arbitrage-free price and replicating strategy for a single-period CCBS given be \eqref{ccbsa} and \eqref{ccbsb}, which is denoted by $\CCBS \,(\Sd,\Ud,\Td,\kappa )$. Then the expressions for the price and hedge at any time $t\leq T_n$ for a multi-period $\CCBS \,(\cT_n;\kappa )$ will be obtained by summation with respect to all periods outstanding and an analogous comment applies to hedging strategies based on futures contracts. Recall from \eqref{ccbsb} that in a single-period CCBS the net cash flow $X_{\Td}(\Sd,\Ud,\Td)$ at time $\Td$ associated with interest
rate payments equals
\begin{align*}
X^{i}_{\Td} (\Sd,\Ud,\Td)= \Xf_{\Td}-\Xd_{\Td} &=\Rf(\Ud,\Td)\delta Q_{\Td}
-(\Rd(\Ud,\Td)+\kappa)\delta Q_{\Sd} \\
&=\Big(e^{\int_{\Ud}^{\Td} \rf_u\, du}-1\Big) Q_{\Td}-\Big(e^{\int_{\Ud}^{\Td}\rd_u\,du}-\kpde\Big)Q_{\Sd}
\end{align*}
where we denote $\Xf_{\Td}=\Xf_{\Td}(\Sd,\Ud,\Td)$ and $\Xd_{\Td}=\Xd_{\Td}(\Sd,\Ud,\Td)$ and, for brevity, we write $\kpde:= (1-\kappa\delta)$. From \eqref{ccbsb} we have that $X^{p}_{\Td}=X^{p}_{\Td}(\Sd,\Td)=Q_{\Td}-Q_{\Sd}$. Then, in view of Proposition \ref{pro3.2}, the arbitrage-free price of a single-period CCBS equals, for every $t\in [0,\Td]$,
\begin{align*}
\CCBS^{\beta}_t(\Sd,\Ud,\Td;\kappa )=\pi^{\beta}_t\big(\Xf_{\Td}\big)-\pi^{\beta}_t\big(\Xd_{\Td}\big)
+\pi^{\beta}_t\big(X^{p}_{\Td}\big)=\Xbetaf_t-\Xbetad_t+\Xbetap_t.
\end{align*}

\subsection{Single-period forward CCBS with domestic collateralization}  \label{sec5.2}

Our goal is to compute the arbitrage-free price $\pi^{\beta}_t(X_{\Td}(\Sd,\Ud,\Td))$ for every $t \leq \Td$.
We first consider in Proposition \ref{pro5.1} and Proposition \ref{pro5.2} a single-period swap contract with proportional collateralization at rate $\beta$, which is posted in domestic currency so that we set $\rc := \rd +\alpha^c$. Then the effective hedge rate equals
$r^\beta=\beta \rc+(1-\beta)r^h=\rd + \alpha^{\beta,d}$ where $\alpha^{\beta,d} = \beta \alpha^c+(1-\beta)\alpha^h$.
We introduce the discount function $\Abst := e^{-\int_s^t \alpha_u^{\beta,d} \diff u}$ associated with the domestic collateralization and we recall that $\LqtT$ is given by \eqref{wtQ3}.

Recall also that $B_{u}(t,\rd_t)$ (resp., $\whB_{u}(t,\rf_t)$) is given
by \eqref{eqbx} in Proposition \ref{pro4.2} (resp., \eqref{feqbx} in Remark \ref{rem4.1}) and $B_{s,u}(t,\rd_t)$ is given by \eqref{xBx} in Proposition \ref{pro4.2}.Furthermore, we define the combined correlation and drift adjustments $\GSU$ and $\GST$ by setting
\begin{align} \label{gsut}
\GSU:=\exp \Big[ \int_t^{\Sd} \big( (\buU-\buS)(\barsig\rhod+\hbuS\rhodf)\big)\diff u\Big]
\end{align}
and
\begin{align}  \label{gstt}
\GST:=\exp \Big[ \int_t^{\Sd} \big((\buT-\buS)(\barsig\rhod+\hbuS\rhodf)\big)\diff u\Big].
\end{align}
It should be observed that suitable self-adjustments $\Gamma^s$ and maturity adjustments $\Gamma^m$ are implicit in definitions of $\FB_{\Sd,\Ud}$ and $\FB_{\Sd,\Td}$ but they are also computed explicitly in the proof of Proposition \ref{pro5.1}.

We are ready to establish two key pricing results for the forward-start single-period CCBS, which constitute central contributions of this paper. The first result, Proposition \ref{pro5.1}, deals with pricing of the exchange of interest payments at the maturity date $\Td$. The second result, Proposition \ref{pro5.2}, complements the former by deriving the arbitrage-free price for the exchange of nominal principals at time $\Td$, represented by the net cash flow $X^{p}_{\Td}(\Sd,\Td) = Q_{\Td} - Q_{\Sd}$.


\bp  \label{pro5.1}
Let $0 \leq \Sd \leq \Ud < \Td$ be arbitrary fixed dates. The cash flow $X_{\Td}(\Sd,\Ud,\Td)$ representing the exchange of interest payments at time $\Td$ can be replicated and its arbitrage-free price satisfies
$\pi^{\beta}_t(X_{\Td}(\Sd,\Ud,\Td))=\Xbeta_t = \Xbetaf_t-\Xbetad_t$ where the price processes $\Xbetaf_t$ and $\Xbetad_t$ for the foreign and domestic legs are given by the pricing functions, for every $t\in[0,\Sd]$,
\begin{align*}
&\Xbetaf(t,\rf_t,Q_t)=\AbtT \LqtT Q_t \big[\whB_{\Ud}(t,\rf_t)-\whB_{\Td}(t,\rf_t) \big],\\
&\Xbetad(t,\rd_t,\rf_t,Q_t)=\AbtT\LqtS   Q_t\whB_{\Sd}(t,\rf_t)\big[\GSU\FB_{\Sd,\Ud}(t,\rd_t)-\kpde \GST\FB_{\Sd,\Td}(t,\rd_t)\big],
\end{align*}
for every $t\in[\Sd,\Ud]$,
\begin{align*}
&\Xbetaf(t,\rf_t,Q_t)=\AbtT \LqtT  Q_t \big[\whB_{\Ud}(t,\rf_t)-\whB_{\Td}(t,\rf_t) \big],\\
&\Xbetad(t,\rd_t,Q_{\Sd})=\AbtT Q_{\Sd}\big[B_{\Ud}(t,\rd_t)-\kpde B_{\Td}(t,\rd_t)\big],
\end{align*}
and, for every $t \in [\Ud,\Td]$,
\begin{align*}
&\Xbetaf(t,\rf_t,\Bf_t,Q_t)=\AbtT \LqtT  Q_t \big[(\Bf_{\Ud})^{-1} \Bf_t-\whB_{\Td}(t,\rf_t)\big],\\
&\Xbetad(t,\rd_t,\Bd_t,Q_{\Sd})=\AbtT Q_{\Sd}\big[(\Bd_{\Ud})^{-1}\Bd_t-\kpde B_{\Td}(t,\rd_t)\big].
\end{align*}
\ep


\begin{proof}
It suffices to apply the pricing formula established in Proposition \ref{pro3.2} to the contingent claim $X_{\Td}(\Sd,\Ud,\Td)=\Xf_{\Td}-\Xd_{\Td}$. We obtain
\begin{align*}
\pi^{\beta}_t(X_{\Td}(\Sd,\Ud,\Td))=\,&\EQ\left[e^{-\int_t^{\Td} r_u^\beta \diff u} e^{\int_{\Ud}^{\Td} \rf_u \diff u}\, Q_{\Td}\,\Big|\, \cF_t \right] -\EQ\left[e^{-\int_t^{\Td}r_u^\beta\diff u}\,Q_{\Td}\,\Big|\,\cF_t \right]
\\&-\EQ\left[ e^{-\int_t^{\Td} r_u^\beta \diff u}e^{\int_{\Ud}^{\Td} \rd_u \diff u}\,Q_{\Sd} \,\Big|\,\cF_t \right]+\kpde \,\EQ\left[e^{-\int_t^{\Td} r_u^\beta\diff u}\,Q_{\Sd}\,\Big|\,\cF_t\right]
\\&=I_t^1-I_t^2-I_t^3+\kpde I_t^4 =\pi_t^{\beta}(\Xf_{\Td})- \pi_t^{\beta}(\Xd_{\Td})=\Xbetaf_t-\Xbetad_t
\end{align*}
where $\Xbetaf_t=\pi_t^{\beta}(\Xf_{\Td})=I_t^1-I_t^2$ (resp., $\Xbetad_t=\pi_t^{\beta}(\Xd_{\Td})=I_t^3-\kpde I_t^4$) is the price of the long position in the foreign leg $\Xf_{\Td}=\Xf_{\Td} (\Sd,\Ud,\Td)$ (resp., the domestic leg $\Xd_{\Td}=\Xd_{\Td} (\Sd,\Ud,\Td)$) of the swap.

 We know that the dynamics of the exchange rate $Q$ under $\Q$ are
\begin{align*}
\diff Q_t=Q_t\big( \rd_t-\rf_t+\lqt\big)\diff t+Q_t\barsig\diff\Zthree_t
\end{align*}
and thus, for every $t\in [0,\Td]$,
\begin{equation*}
Q_{\Td}= \LqtT Q_t\,\cE^q_{t,\Td}\,e^{\int_t^{\Td} (\rd_u-\rf_u) \diff u}
\end{equation*}
and, for every $t\in [0,\Sd]$,
\begin{equation*}
Q_{\Sd}=\LqtS Q_t\,\cE^q_{t,\Sd}\,e^{\int_t^{\Sd} (\rd_u-\rf_u) \diff u}.
\end{equation*}
Notice that the pricing formulae for the components of the domestic leg involve convexity corrections, which were
introduced in Section \ref{sec5.1}.

{\it Foreign leg.} We first consider pricing of the foreign leg using the equality $\Xbetaf_t=I_t^1-I_t^2$.
We obtain, for every $t \in [0,\Ud]$,
\begin{align*}
I_t^1&=\EQ\left[e^{-\int_t^{\Td} r_u^\beta \diff u}\,e^{\int_{\Ud}^{\Td} \rf_u \diff u}\, Q_{\Td}\,\Big|\, \cF_t \right]\\
&=\EQ\left[e^{-\int_t^{\Td}(\rd_u+ \alpha_u^\beta)\diff u}\,e^{\int_{\Ud}^{\Td} \rf_u \diff u}\,\LqtT
Q_t\, \cE^q_{t,\Td} \, e^{\int_t^{\Td} (\rd_u-\rf_u) \diff u} \,\Big|\, \cF_t \right]\\
&= \AbtT \LqtT Q_t\,
\EQ \left[ \cE^q_{t,\Ud}\, e^{-\int_{t}^{\Ud}\rf_u\diff u}\,\Big|\, \cF_t \right]\\
&= \AbtT \LqtT \,Q_t \whB_{\Ud}(t,\rf_t)
\end{align*}
and, for every $t \in [\Ud,\Td]$,
\begin{align*}
I_t^1&=\EQ\left[e^{-\int_t^{\Td} r_u^\beta \diff u}\,e^{\int_{\Ud}^{\Td} \rf_u \diff u}\, Q_{\Td}\,\Big|\, \cF_t \right]\\
&=\EQ\left[e^{-\int_t^{\Td}(\rd_u+ \alpha_u^\beta)\diff u}\,e^{\int_{\Ud}^{\Td} \rf_u \diff u}\,\LqtT
Q_t\, \cE^q_{t,\Td} \, e^{\int_t^{\Td} (\rd_u-\rf_u) \diff u}\,\Big|\, \cF_t \right]\\
&= \AbtT\LqtT Q_t\,
\EQ \left[ \cE^q_{t,\Td}\, e^{\int_{\Ud}^{\Td}\rf_u\diff u -\int_{t}^{\Td}\rf_u\diff u} \,\Big|\, \cF_t \right]\\
&= \AbtT \LqtT Q_t e^{\int_{\Ud}^t \rf_u\diff u}\,
\EQ \big[\cE^q_{t,\Td}\,|\,\cF_t \big]\\
&= \AbtT \LqtT Q_t(\Bf_{\Ud})^{-1} \Bf_t.
\end{align*}

Next, the term $I_t^2$ satisfies, for every $t \in [0,\Td]$,
\begin{align*}
I_t^2&= \EQ\left[ e^{-\int_t^{\Td} r_u^\beta \diff u}\,Q_{\Td}\,\Big|\,\cF_t\right]
=\AbtT \LqtT Q_t\, \EQ \left[ \cE^q_{t,\Td}\, e^{-\int_t^{\Td} \rf_u \diff u}\,\Big|\,\cF_t \right]\\
&=\AbtT \LqtT Q_t\whB_{\Td}(t,\rf_t).
\end{align*}

{\it Domestic leg.} We now compute the price of the domestic leg $\Xbetad_t=I_t^3-\kpde I_t^4$.
Let us first consider any date $t \in [0,\Sd]$. Then for the term $I_t^3$, we obtain
\begin{align*}
I_t^3&=\EQ\left[e^{-\int_t^{\Td} r_u^\beta \diff u}\, e^{\int_{\Ud}^{\Td} \rd_u \diff u}\,Q_{\Sd} \,\Big|\,\cF_t \right]\\
&=\EQ\left[e^{-\int_t^{\Td} (\rd_u+\alpha^\beta_u) \diff u}\, e^{\int_{\Ud}^{\Td} \rd_u \diff u}\LqtS  Q_t\, \cE^q_{t,\Sd}\, e^{\int_t^{\Sd}(\rd_u-\rf_u) \diff u} \,\Big|\,\cF_t \right] \\
&=\AbtT \LqtS Q_t\,\EQ\left[ \cE^q_{t,\Sd}\, e^{\int_t^{\Sd}\rd_u\diff u-\int_t^{\Ud}\rd_u\diff u
-\int_t^{\Sd}\rf_u\diff u}\,\Big|\,\cF_t \right]\\
&= \AbtT \LqtS Q_t[B_{\Sd}(t,\rd_t)]^{-1}B_{\Ud}(t,\rd_t)\whB_{\Sd}(t,\rf_t)
\Gamma^s_t(\Sd,-1,\bS)\Gamma^m_t(\Sd,\bU,-\bS) \\
&\quad \ \Gamma^c_t(\Sd,-\bS,\hbS,\rhodf)\Gamma^c_t(\Sd,\bU,\hbS,\rhodf)\Gamma^d_t(\Sd,\barsig,-\bS,\rhod)
\Gamma^d_t(\Sd,\barsig,\bU,\rhod) \\
&=\AbtT\LqtS Q_t\FB_{\Sd,\Ud}(t,\rd_t)\whB_{\Sd}(t,\rf_t)\GSU
\end{align*}
where the penultimate equality follows by an application of Lemma \ref{newlm} and the last one is obtained using \eqref{xBx} and \eqref{gsut}. The arguments for $I_t^4$ are almost identical, although with $\Ud$ replaced by $\Td$ and $\GST$ given by \eqref{gstt}. We obtain, for every $t\in [0,\Sd]$,
\begin{align*}
I_t^4&=\EQ\left[e^{-\int_t^{\Td}r_u^\beta \diff u}\,Q_{\Sd}\,\Big|\,\cF_t \right]\\
&=\EQ\left[e^{-\int_t^{\Td}(\rd_u+\alpha^\beta_u)\diff u}\LqtS  Q_t\,\cE^q_{t,\Sd}\,e^{\int_t^{\Sd}(\rd_u-\rf_u)\diff u}\,\Big|\,\cF_t\right] \\
&=\AbtT \LqtS  Q_t\,\EQ\left[\cE^q_{t,\Sd}\,e^{\int_t^{\Sd}\rd_u\diff u-\int_t^{\Td}\rd_u\diff u
  -\int_t^{\Sd}\rf_u\diff u}\,\Big|\,\cF_t\right]\\
&= \AbtT \LqtS Q_t [B_{\Sd}(t,\rd_t)]^{-1} B_{\Td}(t,\rd_t)\whB_{\Sd}(t,\rf_t)
\Gamma^s_t(\Sd,-1,\bS)\Gamma^m_t(\Sd,\bT,-\bS) \\
&\quad \ \Gamma^c_t(\Sd,-\bS,\hbS,\rhodf) \Gamma^c_t(\Sd,\bT,\hbS,\rhodf) \Gamma^d_t(\Sd,\barsig,\bT,\rhod)\Gamma^d_t(\Sd,\barsig,-\bS,\rhod)\\
&=\AbtT\LqtS Q_t\FB_{\Sd,\Td}(t,\rd_t)\whB_{\Sd}(t,\rf_t)\GST
\end{align*}
where in the last equality can be deduced from \eqref{xBx} and \eqref{gstt}.

It remains to consider the case of $t\in [\Sd,\Td]$. Since $Q_{\Sd}$ is $\cF_t$-measurable when $t\geq \Sd$ we obtain
\begin{align*}
I_t^3&=Q_{\Sd}\,\EQ\left[e^{-\int_t^{\Td}r_u^\beta \diff u}e^{\int_{\Ud}^{\Td}\rd_u \diff u}\,\Big|\,\cF_t\right]
= \AbtT Q_{\Sd}\Bd(t,\Ud),\\
I_t^4&=Q_{\Sd}\,\EQ\left[e^{-\int_t^{\Td} r_u^\beta \diff u}\,\Big|\, \cF_t \right]=\AbtT Q_{\Sd}B_{\Td}(t,\rd_t),
\end{align*}
where we recall that $\Bd(t,\Ud)=B_{\Ud}(t,\rd_t)$ is given by \eqref{eqbx} for every $t \in [\Sd,\Ud]$
and by the equalities $\Bd(t,\Ud) = e^{\int_{\Ud}^t \rd_u\, du}=(\Bd_{\Ud})^{-1} \Bd_t$ for every $t \in [\Ud,\Td]$.
\end{proof}

In the second pricing result, which complements Proposition \ref{pro5.1}, we derive the arbitrage-free price for the exchange of nominal principals at time $\Td$, which is given by the net cash flow at time $\Td$ equal to $X^{p}_{\Td}(\Sd,\Td)=Q_{\Td}-Q_{\Sd}.$

\bp  \label{pro5.2}
Let $0 \leq \Sd \leq \Ud < \Td$ be arbitrary dates. The cash flow $X^{p}_{\Td}(\Sd,\Td)$ representing the exchange of nominal principals at time $\Td$ can be replicated and its arbitrage-free price satisfies {\rm $\pi^{\beta}_t( X^{p}_{\Td}(\Sd,\Td))=\Xbetap_t$} where, for every $t\in[0,\Sd]$,
\begin{align*}
\Xbetap_t=\Xbetap(t,\rd_t,\rf_t,Q_t)=\AbtT \big[\LqtT Q_t \whB_{\Td}(t,\rf_t)
-\LqtS  Q_t\FB_{\Sd,\Td}(t,\rd_t)\whB_{\Sd}(t,\rf_t)\GST\big]
\end{align*}
and, for every $t\in[\Sd,\Td]$,
\begin{align*}
\Xbetap_t=\Xbetap(t,\rd_t,\rf_t,Q_t,Q_{\Sd})=\AbtT \big[\LqtT Q_{t}\whB_{\Td}(t,\rf_t)-Q_{\Sd}B_{\Td}(t,\rd_t)\big].
\end{align*}
\ep

\begin{proof}
We now apply the pricing formula of Proposition \ref{pro3.2} to the contingent claim $X^{p}_{\Td}(\Sd,\Td):=Q_{\Td}-Q_{\Sd}$
\begin{align*}
\pi^{\beta}_t(X^{p}_{\Td}(\Sd,\Td))=\EQ\left[e^{-\int_t^{\Td}r_u^\beta \diff u}\,Q_{\Td}\,\Big|\,\cF_t\right]
-\EQ\left[e^{-\int_t^{\Td}r_u^\beta\diff u}\,Q_{\Sd}\,\Big|\,\cF_t\right]=I^2_t-I^4_t.
\end{align*}
For any date $t \in [\Sd,\Td]$, the random variable $Q_{\Sd}$ is $\cF_t$-measurable. From the computations for the processes $I^2$ and $I^4$ in the proof of  Proposition \ref{pro5.1} we obtain, for every $t \in [0,\Sd]$
\begin{align*}
\pi^{\beta}_t(X^{p}_{\Td}(\Sd,\Td))=\AbtT\LqtT Q_t\whB_{\Td}(t,\rf_t)
-\AbtT\LqtS  Q_t\FB_{\Sd,\Td}(t,\rd_t)\whB_{\Sd}(t,\rf_t)\GST
\end{align*}
and for every $t \in [\Sd,\Td]$
\begin{align*}
\pi^{\beta}_t(X^{p}_{\Td}(\Sd,\Td))=I^2_t-I^4_t=\AbtT\LqtT Q_t\whB_{\Td}(t,\rf_t)-\AbtT Q_{\Sd} B_{\Td}(t,\rd_t)
\end{align*}
as was required to show.
\end{proof}

\subsection{Single-period forward CCBS with foreign collateralization}  \label{sec5.6.3}

In Section \ref{sec5.2}, we worked under the convention of collateral posted in domestic currency and remunerated at the domestic collateral rate $\rc = \rd + \alpha^c$. In practice, the collateral remuneration rate $\rc$ should be aligned with the currency in which the collateral is posted and thus the price and hedging strategy for a CCBS depends on a choice of collateral currency.

In this section, we consider a contract with a proportional collateralization at a rate $\beta$, which is posted in the foreign currency. We define the foreign collateral remuneration rate $\rc = \rf +\alpha^{c,f}$ so that the effective hedge rate equals $r^\beta=\beta \rf + (1-\beta)\rd + \alpha^{\beta,f}$ where $\rf$ and $\rd$ are given by \eqref{eq4.2} and $\alpha^{\beta,f}:=\beta\alpha^{c,f} + (1-\beta)\alpha^h$. The discount function $\Abfst$ is given by $\Abfst:= e^{-\int_s^t \alpha_u^{\beta,f} \diff u}$.
The following result is a counterpart of Propositions \ref{pro5.1} and \ref{pro5.2} for the case of a foreign collateralization remunerated at a rate $\rc = \rf + \alpha^{c,f}$. It is worth noting that for an arbitrary real number $\beta$
\begin{align*}
\Gamma^s_t(\Td,\beta,\bT)=e^{\frac{1}{2}\int_t^{\Td}\beta(\beta-1)\sbuT \diff u}=\Gamma^s_t(\Td,1-\beta,\bT)
\end{align*}
and thus
\begin{align*}
\Gamma^s_t(\Td,-\beta,\bT)=e^{\frac{1}{2}\int_t^{\Td}\beta(\beta+1)\sbuT \diff u}.
\end{align*}
For the reader's convenience the convexity corrections are presented explicitly.


\bp  \label{pro5.1c}
The arbitrage-free price of the exchange of interest payments at time $\Td$ satisfies $\pi^{\beta}_t(X_{\Td}(\Sd,\Ud,\Td))=\Xbeta_t = \Xbetaf_t-\Xbetad_t =I_t^1-I_t^2-I_t^3+\kpde I_t^4 $
and the arbitrage-free price of the exchange of nominal principals satisfies $\pi^{\beta}_t(X^{p}_{\Td}(\Sd,\Td))=I^2_t-I^4_t$. \\
(i) The term $I^1_t$ equals, for every $t \in [0,\Ud]$,
\begin{align*}
I^1_t=\AbftT \LqtT Q_t [B_T(t,\rd_t)]^{-\beta}\wh{B}_U(t,\rf_t)[\wh{B}_T(t,\rf_t)]^{\beta}\fbGUT
\end{align*}
where the convexity correction satisfies
\begin{align*}
\fbGUT &=\Gamma^s_t(\Td,-\beta,\bT)\Gamma^s_t(\Td,\beta,\hbT)\Gamma^m_t(\Ud,\hbU,\beta \hbT)
\Gamma^c_t(\Ud,\bU,\beta \hbT,\rhodf) \Gamma^c_t(\Td,-\beta \bT,\beta \hbT,\rhodf)\\
&\quad \ \Gamma^d_t(\Td,\barsig,-\beta\bT,\rhod)
\end{align*}
and, for every $t \in [\Ud,\Td]$,
\begin{align*}
I^1_t&=\AbftT\LqtT Q_t(\Bf_{\Ud})^{-1} \Bf_t [B_T(t,\rd_t)]^{-\beta}[\wh{B}_T(t,\rf_t)]^{\beta}\fbGT
\end{align*}
where
\begin{align*}
\fbGT &=\Gamma^s_t(\Td,-\beta,\bT)\Gamma^s_t(\Td,\beta,\hbT)\Gamma^c_t(\Td,-\beta \bT,\beta \hbT,\rhodf) \Gamma^d_t(\Td,\barsig,-\beta\bT,\rhod).
\end{align*}
(ii) The term $I^2_t$ equals, for every $t \in [0,\Td]$,
\begin{align*}
I^2_t=\AbftT\LqtT Q_t[B_T(t,\rd_t)]^{-\beta}[\wh{B}_T(t,\rf_t)]^{\beta+1}\fbdGT
\end{align*}
where
\begin{align*}
\fbdGT =\Gamma^s_t(\Td,-\beta,\bT)\Gamma^s_t(\Td,\beta+1,\hbT)\Gamma^c_t(\Td,-\beta \bT, (\beta+1)\hbT,\rhodf) \Gamma^d_t(\Td,\barsig,-\beta\bT,\rhod).
\end{align*}
(iii) The term $I^3_t$ equals, for every $t \in [0,\Sd]$,
\begin{align*}
I^3_t&=\AbftT\LqtS Q_t[B_S(t,r^d_t)]^{-1}B_{\Ud}(t,\rd_t)[B_T(t,\rd_t)]^{-\beta}\wh{B}_{\Sd}(t,\rf_t)[\wh{B}_{\Td}(t,\rf_t)]^{\beta}
 \fbtGSUT
\end{align*}
where
\begin{align*}
\fbtGSUT &=\Gamma^s_t(\Sd,-1,\bS)\Gamma^s_t(\Td,-\beta,\bT)\Gamma^s_t(\Td,\beta,\hbT)\Gamma^m_t(\Sd,-\bS,\bU)
\Gamma^m_t(\Sd,-\bS,-\beta \bT)\\
&\quad \ \Gamma^m_t(\Ud,\bU,-\beta \bT) \Gamma^m_t(\Sd,\hbS,\beta\hbT) \Gamma^c_t(\Sd,-\bS,\hbS,\rhodf)
\Gamma^c_t(\Sd,-\bS,\beta \hbT,\rhodf)\\
&\quad \  \Gamma^c_t(\Sd,\bU,\hbS,\rhodf) \Gamma^c_t(\Ud,\bU,\beta \hbT,\rhodf)\Gamma^c_t(\Sd,-\beta \bT,\hbS,\rhodf) \Gamma^c_t(\Td,-\beta \bT,\beta \hbT,\rhodf) \\
&\quad \ \Gamma^d_t(\Sd,\barsig,-\bS,\rhod)\Gamma^d_t(\Sd,\barsig,\bU,\rhod)\Gamma^d_t(\Sd,\barsig,-\beta \bT,\rhod) \wh{\Gamma}^d_t(\Sd,\Td,\barsig,\beta \hbT,\rhof),
\end{align*}
for every $t \in [\Sd,\Ud]$,
\begin{align*}
I^3_t=\AbftT Q_{\Sd}B_{\Ud}(t,\rd_t)[B_T(t,\rd_t)]^{-\beta}[\wh{B}_T(t,\rf_t)]^{\beta}\fbtGUT
\end{align*}
where
\begin{align*}
\fbtGUT &=\Gamma^s_t(\Td,-\beta,\bT)\Gamma^s_t(\Td,\beta,\hbT)\Gamma^m_t(\Ud,\bU,-\beta\bT)
\Gamma^c_t(\Ud,\bU,\beta \hbT,\rhodf)\\
&\quad \ \Gamma^c_t(\Td,-\beta \bT,\beta \hbT,\rhodf)\wh{\Gamma}^d_t(t,\Td,\barsig,\beta \hbT,\rhof),
\end{align*}
and, for every $t \in [\Ud,\Td]$,
\begin{align*}
I^3_t=\AbftT Q_{\Sd}(B^d_{\Ud})^{-1}B^d_t[B_T(t,\rd_t)]^{-\beta}[\wh{B}_T(t,\rf_t)]^{\beta}\fbtGT
\end{align*}
where
\begin{align*}
\fbtGT =\Gamma^s_t(\Td,-\beta,\bT)\Gamma^s_t(\Td,\beta,\hbT)\Gamma^c_t(\Td,-\beta \bT,\beta \hbT,\rhodf)\wh{\Gamma}^d_t(t,\Td,\barsig,\beta \hbT,\rhof).
\end{align*}
(iv) The term $I^4_t$ equals, for every $t \in [0,\Sd]$,
\begin{align*}
I^4_t&=\AbftT \LqtS Q_t[B_{\Sd}(t,r^d_t)]^{-1}[B_{\Td}(t,\rd_t)]^{1-\beta}\wh{B}_{\Sd}(t,\rf_t)[\wh{B}_{\Td}(t,\rf_t)]^{\beta}
\fbcGST
\end{align*}
where
\begin{align*}
&\fbcGST = \Gamma^s_t(\Sd,-1,\bS)\Gamma^s_t(\Td,1-\beta,\bT)\Gamma^s_t(\Td,\beta,\hbT)\Gamma^m_t(\Sd,-\bS,(1-\beta)\hbT) \Gamma^m_t(\Sd,\hbS,\beta \hbT)\\
&\quad  \Gamma^c_t(\Sd,-\bS,\hbS,\rhodf)\Gamma^c_t(\Sd,-\bS,\beta \hbT,\rhodf)\Gamma^c_t(\Sd,(1-\beta)\bT,\hbS,\rhodf)
\Gamma^c_t(\Td,(1-\beta)\bT,\beta \hbT,\rhodf)\\
&\quad \Gamma^d_t(\Sd,\barsig,-\bS,\rhod)\Gamma^d_t(\Sd,\barsig,(1-\beta)\bT,\rhod)
\wh{\Gamma}^d_t(\Sd,\Td,\barsig,\beta\hbT,\rhof)
\end{align*}
and, for every $t \in [\Sd,\Td]$,
\begin{align*}
I^4_t=\AbftT Q_{\Sd}[B_T(t,\rd_t)]^{1-\beta}[\wh{B}_{\Td}(t,\rf_t)]^{\beta}\fbcGT
\end{align*}
where
\begin{align*}
\fbcGT &=\Gamma^s_t(\Td,1-\beta,\bT)\Gamma^s_t(\Td,\beta,\hbT)
   \Gamma^c_t(\Td,(1-\beta)\bT,\beta \hbT,\rhodf)\Gamma^d_t(\Td,\barsig,(1-\beta)\bT,\rhod)\\
&\quad \  \wh{\Gamma}^d_t(t,\Td,\barsig,\beta\hbT,\rhof).
\end{align*}
\ep


\begin{proof} The proof is analogous to the proofs of Proposition \ref{pro5.1}. Recall that, for every $t\in [0,\Td]$,
\begin{equation*}
Q_{\Td}= \LqtT Q_t\,\cE^q_{t,\Td}\,e^{\int_t^{\Td} (\rd_u-\rf_u) \diff u}
\end{equation*}
and, for every $t\in [0,\Sd]$,
\begin{equation*}
Q_{\Sd}=\LqtS Q_t\,\cE^q_{t,\Sd}\,e^{\int_t^{\Sd} (\rd_u-\rf_u) \diff u}.
\end{equation*}
The effective hedge rate equals $r^\beta = \beta \rf + (1-\beta)\rd +\alpha^{\beta,f}$ and
$\AbftT:= e^{-\int_t^{\Td} \alpha_u^{\beta,f} \diff u}$.

\vskip 5 pt \noindent {\it Foreign leg.}
We first consider pricing of the foreign leg using the equality $\Xbetaf_t=I_t^1-I_t^2$. \\
(i) For any date $t \in [0,\Ud]$, we obtain
\begin{align*}
I_t^1&=\EQ\left[e^{-\int_t^{\Td}r_u^\beta\diff u}\,e^{\int_{\Ud}^{\Td}\rf_u\diff u}\,Q_{\Td}\,\Big|\,\cF_t \right]\\
&=\AbftT\,\EQ\left[e^{-\int_t^{\Td}(\beta\rf_u+(1-\beta)\rd_u)\diff u}\,e^{\int_{\Ud}^{\Td}\rf_u\diff u}
\LqtT Q_t\cE^q_{t,\Td}\,e^{\int_t^{\Td}(\rd_u-\rf_u) \diff u}\,\Big|\,\cF_t\right]\\
&=\AbftT\LqtT Q_t\,\EQ \left[ \cE^q_{t,\Td}\,e^{\int_t^{\Td}\beta\rd_u\diff u-\int_{t}^{\Ud}\rf_u\diff u
-\int_t^{\Td}\beta\rf_u\diff u}\,\Big|\,\cF_t\right]\\
&=\AbftT\LqtT Q_t[B_T(t,\rd_t)]^{-\beta}\wh{B}_U(t,\rf_t)[\wh{B}_T(t,\rf_t)]^{\beta}
\Gamma^s_t(\Td,-\beta,\bT)\Gamma^s_t(\Td,\beta,\hbT)\Gamma^m_t(\Ud,\hbU,\beta \hbT)\\
&\quad \ \Gamma^c_t(\Ud,\bU,\beta \hbT,\rhodf) \Gamma^c_t(\Td,-\beta \bT,\beta \hbT,\rhodf)
\Gamma^d_t(\Td,\barsig,-\beta\bT,\rhod)
\end{align*}
and, for every $t \in [\Ud,\Td]$,
\begin{align*}
I_t^1&=\EQ\left[e^{-\int_t^{\Td}r_u^\beta \diff u}\,e^{\int_{\Ud}^{\Td}\rf_u\diff u}\,Q_{\Td}\,\Big|\,\cF_t\right]\\
&=\AbftT\,\EQ \left[e^{-\int_t^{\Td}(\beta \rf_u +(1-\beta)\rd_u)\diff u}\,e^{\int_{\Ud}^{\Td}\rf_u\diff u}
\LqtT Q_t\cE^q_{t,\Td}\,e^{\int_t^{\Td}(\rd_u-\rf_u)\diff u}\,\Big|\,\cF_t\right]\\
&=\AbftT\LqtT Q_t e^{\int_{\Ud}^t \rf_u\diff u}\,
\EQ \Big[\cE^q_{t,\Td}e^{\int_t^{\Td}\beta \rd_u\diff u-\int_t^{\Td}\beta\rf_u\diff u}\,\Big|\,\cF_t\Big]\\
&=\AbftT\LqtT Q_t\fbGT (\Bf_{\Ud})^{-1} \Bf_t [B_T(t,\rd_t)]^{-\beta}[\wh{B}_T(t,\rf_t)]^{\beta}
\Gamma^s_t(\Td,-\beta,\bT)\Gamma^s_t(\Td,\beta,\hbT)\\
&\quad \ \Gamma^c_t(\Td,-\beta \bT,\beta \hbT,\rhodf)\Gamma^d_t(\Td,\barsig,-\beta\bT,\rhod).
\end{align*}
(ii) The term $I_t^2$ satisfies, for every $t \in [0,\Td]$,
\begin{align*}
I_t^2&=\EQ\left[e^{-\int_t^{\Td} r_u^\beta\diff u}\,Q_{\Td}\,\Big|\,\cF_t\right]\\
&=\AbftT\,\EQ\left[e^{-\int_t^{\Td} (\beta \rf_u + (1-\beta)\rd_u) \diff u}\,\LqtT
Q_t\,\cE^q_{t,\Td}\,e^{\int_t^{\Td} (\rd_u-\rf_u) \diff u}\,\Big|\, \cF_t \right]\\
&=\AbftT \LqtT Q_t\, \EQ \left[ \cE^q_{t,\Td}\, e^{\int_t^{\Td}\beta \rd_u\diff u
-\int_t^{\Td}(\beta+1)\rf_u\diff u}\,\Big|\,\cF_t \right]\\
&=\AbftT \LqtT Q_t[B_T(t,\rd_t)]^{-\beta}[\wh{B}_T(t,\rf_t)]^{\beta+1}
\Gamma^s_t(\Td,-\beta,\bT)\Gamma^s_t(\Td,\beta+1,\hbT)\\
&\quad \ \Gamma^c_t(\Td,-\beta \bT, (\beta+1)\hbT,\rhodf) \Gamma^d_t(\Td,\barsig,-\beta\bT,\rhod).
\end{align*}
\noindent {\it Domestic leg.} For the domestic leg we use the equality $\Xbetad_t=I_t^3-\kpde I_t^4$.\\
(iii) The term $I^3_t$ equals, for every $t\in [0,\Sd]$,
\begin{align*}
I_t^3&=\EQ\left[ e^{-\int_t^{\Td} r_u^\beta \diff u}e^{\int_{\Ud}^{\Td} \rd_u \diff u}\,Q_{\Sd} \,\Big|\,\cF_t \right]\\
&=\AbftT\,\EQ\left[ e^{-\int_t^{\Td} (\beta \rf_u + (1-\beta)\rd_u) \diff u}\,e^{\int_{\Ud}^{\Td} \rd_u \diff u}\LqtS  Q_t\, \cE^q_{t,\Sd}\,e^{\int_t^{\Sd} (\rd_u-\rf_u) \diff u} \,\Big|\,\cF_t \right] \\
&=\AbftT\LqtS Q_t\,\EQ\left[\cE^q_{t,\Sd}\,e^{-\int_t^{\Td}(\beta\rf_u+(1-\beta)\rd_u) \diff u
+\int_{\Ud}^{\Td}\rd_u\diff u + \int_t^{\Sd}\rd_u\diff u-\int_t^{\Sd}\rf_u\diff u} \,\Big|\,\cF_t \right] \\
&=\AbftT\LqtS Q_t\,\EQ\left[\cE^q_{t,\Sd}\,e^{\int_t^{\Sd} \rd_u \diff u-\int_t^{\Ud}\rd_u \diff u
+\int_t^{\Td}\beta\rd_u\diff u-\int_t^{\Sd}\rf_u \diff u-\int_t^{\Td}\beta\rf_u\diff u}\,\Big|\,\cF_t \right],
\end{align*}
which yields
\begin{align*}
I_t^3&=\AbftT\LqtS Q_t[B_S(t,r^d_t)]^{-1}B_{\Ud}(t,\rd_t)[B_T(t,\rd_t)]^{-\beta}\wh{B}_{\Sd}(t,\rf_t) [\wh{B}_{\Td}(t,\rf_t)]^{\beta}
\Gamma^s_t(\Sd,-1,\bS)\\
&\quad \ \Gamma^s_t(\Td,-\beta,\bT)\Gamma^s_t(\Td,\beta,\hbT)\Gamma^m_t(\Sd,-\bS,\bU)\Gamma^m_t(\Sd,-\bS,-\beta \bT)\Gamma^m_t(\Sd,\hbS,\beta\hbT)\\
&\quad \ \Gamma^c_t(\Sd,-\bS,\hbS,\rhodf)\Gamma^c_t(\Sd,-\bS,\beta \hbT,\rhodf)\Gamma^c_t(\Sd,\bU,\hbS,\rhodf) \Gamma^c_t(\Ud,\bU,\beta \hbT,\rhodf)\\
&\quad \ \Gamma^c_t(\Sd,-\beta \bT,\hbS,\rhodf)\Gamma^c_t(\Td,-\beta \bT,\beta \hbT,\rhodf)
\Gamma^d_t(\Sd,\barsig,-\bS,\rhod)\Gamma^d_t(\Sd,\barsig,\bU,\rhod)\\
&\quad \ \Gamma^d_t(\Sd,\barsig,-\beta \bT,\rhod)\wh{\Gamma}^d_t(\Sd,\Td,\barsig,\beta \hbT,\rhof).
\end{align*}

Furthermore, for every $t\in [\Sd,\Ud]$,
\begin{align*}
I_t^3&=Q_{\Sd}\,\EQ\left[ e^{-\int_t^{\Td} r_u^\beta \diff u}\,e^{\int_{\Ud}^{\Td} \rd_u \diff u}\,\Big|\,\cF_t \right]\\
&=\AbftT Q_{\Sd}\,\EQ\left[ e^{\int_t^{\Td}\beta\rd_u\diff u - \int_t^{\Ud}\rd_u \diff u
-\int_t^{\Td} \beta\rf_u \diff u}\,\Big|\,\cF_t \right]\\
&=\AbftT Q_{\Sd}B_{\Ud}(t,\rd_t)[B_T(t,\rd_t)]^{-\beta}[\wh{B}_T(t,\rf_t)]^{\beta}
\Gamma^s_t(\Td,-\beta,\bT)\Gamma^s_t(\Td,\beta,\hbT)\\
&\quad \ \Gamma^m_t(\Ud,\bU,-\beta\bT)\Gamma^c_t(\Ud,\bU,\beta \hbT,\rhodf)\Gamma^c_t(\Td,-\beta \bT,\beta \hbT,\rhodf) \wh{\Gamma}^d_t(t,\Td,\barsig,\beta \hbT,\rhof).
\end{align*}
Finally, for every $t\in [\Ud,\Td]$,
\begin{align*}
I_t^3&= Q_{\Sd}\,\EQ\left[ e^{-\int_t^{\Td} r_u^\beta \diff u}\,e^{\int_{\Ud}^{\Td} \rd_u \diff u}\,\Big|\,\cF_t \right]\\
&=\AbftT Q_{\Sd}e^{- \int_t^{\Ud}\rd_u \diff u}\,\EQ\left[ e^{\int_t^{\Td}\beta\rd_u\diff u
-\int_t^{\Td} \beta\rf_u \diff u}\,\Big|\,\cF_t \right]\\
&= \AbftT Q_{\Sd} (B^d_{\Ud})^{-1} B^d_t [B_T(t,\rd_t)]^{-\beta}[\wh{B}_T(t,\rf_t)]^{\beta}
\Gamma^s_t(\Td,-\beta,\bT)\\ &\quad \ \Gamma^s_t(\Td,\beta,\hbT)\Gamma^c_t(\Td,-\beta \bT,\beta \hbT,\rhodf)\wh{\Gamma}^d_t(t,\Td,\barsig,\beta \hbT,\rhof).
\end{align*}
(iv) For the term $I_t^4$ we have, for every $t\in [0,\Sd]$,
\begin{align*}
I_t^4&=\EQ\left[e^{-\int_t^{\Td}r_u^\beta \diff u}\,Q_{\Sd}\,\Big|\,\cF_t \right]\\
&=\AbftT\,\EQ\left[e^{-\int_t^{\Td} (\beta \rf_u + (1-\beta)\rd_u) \diff u}\,\LqtS  Q_t\,\cE^q_{t,\Sd}\,e^{\int_t^{\Sd}(\rd_u-\rf_u)\diff u}\,\Big|\,\cF_t\right] \\
&=\AbftT\LqtS Q_t\,\EQ\left[\cE^q_{t,\Sd}\,e^{\int_t^{\Sd}\rd_u\diff u-\int_t^{\Td} (1-\beta)\rd_u\diff u
 -\int_t^{\Sd}\rf_u\diff u -\int_t^{\Td} \beta \rf_u  \diff u}\,\Big|\,\cF_t\right]
 \end{align*}
and thus we obtain
 \begin{align*}
I_t^4 &=\AbftT\LqtS Q_t[B_{\Sd}(t,r^d_t)]^{-1}[B_{\Td}(t,\rd_t)]^{1-\beta}\wh{B}_{\Sd}(t,\rf_t)
 [\wh{B}_{\Td}(t,\rf_t)]^{\beta} \Gamma^s_t(\Sd,-1,\bS)\\
&\quad \ \Gamma^s_t(\Td,1-\beta,\bT)\Gamma^s_t(\Td,\beta,\hbT) \Gamma^m_t(\Sd,-\bS,(1-\beta)\hbT)
\Gamma^m_t(\Sd,\hbS,\beta \hbT) \\
&\quad \ \Gamma^c_t(\Sd,-\bS,\hbS,\rhodf)\Gamma^c_t(\Sd,-\bS,\beta \hbT,\rhodf)\Gamma^c_t(\Sd,(1-\beta)\bT,\hbS,\rhodf)
\\ &\quad \ \Gamma^c_t(\Td,(1-\beta)\bT,\beta \hbT,\rhodf)\Gamma^d_t(\Sd,\barsig,-\bS,\rhod)
\Gamma^d_t(\Sd,\barsig,(1-\beta)\bT,\rhod) \\
&\quad \ \wh{\Gamma}^d_t(\Sd,\Td,\barsig,\beta\hbT,\rhof).
\end{align*}
Finally, for every $t\in [\Sd,\Td]$,
\begin{align*}
I_t^4&=Q_{\Sd}\,\EQ\left[e^{-\int_t^{\Td} r_u^\beta \diff u}\,\Big|\, \cF_t \right]\\
&=\AbftT Q_{\Sd}\,\EQ\left[e^{-\int_t^{\Td} (1-\beta)\rd_u\diff u
 -\int_t^{\Sd}\rf_u\diff u -\int_t^{\Td} \beta \rf_u  \diff u}\,\Big|\, \cF_t \right]\\
&= \AbftT Q_{\Sd}[B_T(t,\rd_t)]^{1-\beta}[\wh{B}_{\Td}(t,\rf_t)]^{\beta}\Gamma^s_t(\Td,1-\beta,\bT)
\Gamma^s_t(\Td,\beta,\hbT)\\
&\quad \  \Gamma^c_t(\Td,(1-\beta)\bT,\beta \hbT,\rhodf)\Gamma^d_t(\Td,\barsig,(1-\beta)\bT,\rhod)\wh{\Gamma}^d_t(t,\Td,\barsig,\beta\hbT,\rhof),
\end{align*}
which ends the proof of the proposition.
\end{proof}

\subsection{Pricing of a constant notional CCBS with domestic collateralization} \label{sec5.6.1}

We henceforth focus exclusively on contracts with domestic collateralization but, in view of  Proposition \ref{pro5.1c}, an extension of all foregoing pricing and hedging results to the case of swaps with foreign collateralization is straightforward. We are ready to state the pricing formula for the multi-period CCBS with tenor structure $\cT_n$ and basis spread $\kappa$. The arbitrage-free price $\CCBS_t(\cT_n;\kappa )$ satisfies, for every $t\in [0,T_0]$,
\begin{align*}
\CCBS_t(\cT_n;\kappa)=\sum_{j=1}^n\pi^{\beta}_t\big(X_{T_j}(T_0,T_{j-1},T_j)\big)+\pi^{\beta}_t\big(X^{p}_{T_n}(T_0,T_n)\big)
\end{align*}
and thus, using also Propositions \ref{pro5.1} and \ref{pro5.2}, we obtain an explicit pricing formula, which is stated here for $t\in [0,T_0]$. Of course, one can also formulate without any difficulties the pricing result for
the multi-period CCBS for any $t \in [T_0,T_n]$. Let us denote $\wt{\kappa}_j = 1-\kappa\delta_j $.

\bp \label{pro5.6}
The arbitrage-free price {\rm $\CCBS_t(\cT_n;\kappa )$} equals, for every $t\in [0,T_0]$,
\begin{align} \label{neq6.18}
{\rm \CCBS}_t(\cT_n;\kappa)= \sum_{j=1}^n \big( X_j^{f,\beta}(t,\rf_t,Q_t)-X_j^{d,\beta}(t,\rd_t,\rf_t,Q_t)\big)+X^{p,\beta}_n(t,\rd_t,\rf_t,Q_t)
\end{align}
where
\begin{align*}
X_j^{f,\beta}:=X_j^{f,\beta}(t,\rf_t,Q_t),\
X_j^{d,\beta}:=X_j^{d,\beta}(t,\rd_t,\rf_t,Q_t), \
X_n^{p,\beta}:=X^{p,\beta}(t,\rd_t,\rf_t,Q_t)
\end{align*}
are given by, for every $t\in [0,T_0]$,
\begin{align*}
&X_j^{f,\beta}=\AbtTj\LqtTj Q_t \big[\whB_{T_{j-1}}(t,\rf_t)-\whB_{T_j}(t,\rf_t) \big],\\
&X_j^{d,\beta}=\AbtTj\LqtTo Q_t\whB_{T_0}(t,\rf_t)
\big[\Gamma_{t,T_0,T_{j-1}}\FB_{T_0,T_{j-1}}(t,\rd_t)-\wt{\kappa}_j \Gamma_{t,T_0,T_{j}}\FB_{T_0,T_j}(t,\rd_t)\big],\\
&X_n^{p,\beta}=\AbtTn Q_t \big[\LqtTn \whB_{T_n}(t,\rf_t)
-\LqtTo \Gamma_{t,T_0,T_n}\FB_{T_0,T_n}(t,\rd_t)\whB_{T_0}(t,\rf_t)\big].
\end{align*}
\ep

Observe that only the term $X_j^{d,\beta}$ depends on the basis spread $\kappa$ and the arbitrage-free price $\CCBS(\cT_n;\kappa)$ is a linear function of $\kappa$.

\bd \label{def5.3}
The constant notional CCBS with tenor structure $\cT_n$ and $T_0=0$ is said to be {\it fair} if {\rm $\CCBS_{0}(\cT_n;0)=0$}.
We say that the forward constant notional CCBS with tenor $\cT_n$ and $T_0>0$ is {\it fair} if {\rm $\CCBS_{0}(\cT_n;0)=0$} and it is {\it strongly fair} if {\rm $\CCBS_{T_0}(\cT_n;0)=0$} and thus also {\rm $\CCBS_t(\cT_n;0)=0$} for every $t\in [0,T_0]$.
\ed

We introduce the following definition of the \textit{fair basis spread}, which makes the cross-currency basis
swap valueless at a given date $t \in [0,T_0]$.

\bd \label{def5.4}
For any fixed $t \in [0,T_0]$, the \textit{fair basis spread} at time $t$ for the constant notional cross-currency basis swap
is a unique $\cF_{t}$-measurable random variable $\kappa_t(\cT_n)$, which satisfies the equality {\rm $\CCBS_t(\cT_n;\kappa_t(\cT_n) )=0$}.
\ed

If $0 \leq t<T_0$ then $\kappa_t(\cT_n)$ can also be called the \textit{forward} fair basis spread.
Observe that if the fair basis spread $\kappa_t(\cT_n)=0$ for some $t \in [0,T_0]$, then $\kappa_s(\cT_n)=0$ for any date $s \in [0,t]$ since the pricing operator is clearly time-consistent. In general, by solving for $\kappa $ the linear equation $\CCBS_t(\cT_n;\kappa)=0$ where the price $\CCBS_t(\cT_n;\kappa)$ is given by \eqref{neq6.18}, we obtain the following lemma.

\bl \label{lem5.3}
At any time $t \in [0,T_0]$, the fair basis spread in the multi-period CCBS with tenor structure $\cT_n$ equals
$\kappa_t(\cT_n) =\big(\Ifb_t-\Idb_t+ \Ipb_t\big)\big(\Kdb_t\big)^{-1}$ where
\begin{align*}
&\Ifb_t:=\sum_{j=1}^n \AbtTj\LqtTj \big[\whB_{T_{j-1}}(t,\rf_t)-\whB_{T_j}(t,\rf_t) \big],\\
&\Idb_t:=\sum_{j=1}^n \AbtTj \LqtTo Q_t\whB_{T_0}(t,\rf_t)
        \big[\Gamma_{t,T_0,T_{j-1}}\FB_{T_0,T_{j-1}}(t,\rd_t)-\Gamma_{t,T_0,T_{j}}\FB_{T_0,T_j}(t,\rd_t)\big],\\
&\Kdb_t:=\sum_{j=1}^n \delta_j \AbtTj\LqtTo Q_t
         \Gamma_{t,T_0,T_j}\FB_{T_0,T_j}(t,\rd_t)\whB_{T_0}(t,\rf_t),\\
&\Ipb_t:=\sum_{j=1}^n \AbtTn Q_t \big[\LqtTn \whB_{T_n}(t,\rf_t)-\LqtTo\Gamma_{t,T_0,T_n} \FB_{T_0,T_n}(t,\rd_t)\whB_{T_0}(t,\rf_t) \big].
\end{align*}
\el

\subsection{Cross-currency swaptions}   \label{sec5.6.4}

The notion of a cross-currency basis swaption is a natural extension of the classical concept of an interest rate swaption
in a single economy, that is, an option on the value of a floating-for-floating interest rate swap referencing a single currency. The underlying asset for a cross-currency basis swaption is the constant notional swap $\CCBS (\cT_n;\kappa)$ where $\kappa$ is a real number.

\bd \label{def5.5}
The \textit{payer cross-currency swaption} with strike $\kappa $ and maturity $T_0$ is a call option of European style with
the terminal payoff {\rm $\PSwn_{T_0}(\kappa):=(\CCBS_{T_0}(\cT_n;\kappa ))^{+}$} at the option's maturity date $T_0$.
\ed

The payoff $\PSwn_{T_0}(\kappa)$ can be represented as $\PSwn_{T_0}(\kappa)=\Kdb_{T_0}\big(\kappa_{T_0}(\cT_n)-\kappa\big)^+$
and thus it can be seen as a call option written on the fair basis spread with strike $\kappa $ and nominal value $\Kdb_{T_0}$. Similarly, the \textit{receiver cross-currency swaption} with strike $\kappa $ and maturity $T_0$ is a put option with the payoff at time $T_0$ equal to $\RSwn_{T_0}(\kappa):=(-\CCBS_{T_0}(\cT_n;\kappa ))^{+}$ or, equivalently,
$\RSwn_{T_0}(\kappa)=\Kdb_{T_0}\big(\kappa-\kappa_{T_0}(\cT_n)\big)^+$.
The payoffs of a cross-currency swaption satisfy at time $T_0$
\begin{align*}
\PSwn_{T_0}(\kappa)- \RSwn_{T_0}(\kappa)=\Kdb_{T_0}\big(\kappa_{T_0}(\cT_n)-\kappa\big)=
\Ifb_{T_0}-\Idb_{T_0}+ \Ipb_{T_0}-\kappa \Kdb_{T_0}
\end{align*}
and thus the put-call parity for swaptions reads, for every $t \in [0,T_0]$,
\begin{align*}
\PSwn_{t}(\kappa)- \RSwn_{t}(\kappa)=\Ifb_t-\Idb_t+ \Ipb_t-\kappa \Kdb_t.
\end{align*}

In contrast to the case of a multi-period CCBS, which is a relatively simple portfolio of payoffs for which explicit pricing and hedging results are available, the arbitrage-free pricing and hedging of a cross-currency basis swaption is a computationally challenging task. To describe the exercise set, one needs to compute the arbitrage-free price of $\CCBS_{T_0}(\cT_n;\kappa )$ or, equivalently, the fair basis spread $\kappa_{T_0}(\cT_n)$. However, these random variables are given as rather complex functions of the model variables $\rd_{T_0},\rf_{T_0}$ and $Q_{T_0}$ and the same comment applies to their representations in terms of the market variables. Therefore, the Monte Carlo simulation seems to be the most appropriate method for pricing of cross-currency basis swaptions within the present framework, although explicit pricing formula for options on backward-looking swaps was obtained in \cite{Bickersteth2025SOFR} where a single currency case was studied.

\section{Hedging of a forward-start single-period CCBS}  \label{sec6}

In Section \ref{sec6}, we deal with a single-period CCBS with collateral posted in the domestic currency, which was studied in
Proposition \ref{pro5.1} and Proposition \ref{pro5.2}. For clarity of presentation, the derivation of the replicating strategy for a CCBS is split into several steps. Recall that we denote $X_{\Td}(\Sd,\Ud,\Td) =\Xf_{\Td}-\Xd_{\Td}$  where $\Xf_{\Td}$ and $\Xd_{\Td}$ represent the payoff from the foreign and domestic leg, respectively. Consequently, we write $\phiv = \phiv^f-\phiv^d$ where $\phiv^f$ and $\phiv^d$ denote the replicating strategies for the contingent claims $\Xf_{\Td}$ and $\Xd_{\Td}$, respectively. They are studied in Section \ref{sec6.2} and Section \ref{sec6.3} on each time interval between the dates $0,\Sd,\Ud$ and $\Td$.
The replicating strategy $\phiv$ on $[0,\Td]$ for a forward-start single-period CCBS can be obtained by a concatenation of these results. 

We observe that in all questions studied in this section it suffices to identify the processes $\phid ,\phif $ and $\phiq$ corresponding to futures contracts since, in view of Definition \ref{def3.2}, the cash component $\varphi^0$ can always be found from the equality $(1-\beta_t)V(\varphi)=\varphi^0_t B^h_t$ where $r^h_t = \rd_t+\alpha^d_t$ for every $t\in [0,\Td]$. Hence in the case of the domestic leg of a single-period CCBS we have that $\varphi^{d,0}_t = (1-\beta_t)(B^h_t)^{-1}\Xbetad_t$ and for its foreign leg the equality $\varphi^{f,0}_t=(1-\beta_t)(B^h_t)^{-1}\Xbetaf_t$ holds. Obviously, in the case of the full collateralization (that is, when $\beta_t=1$ for every $t\in [0,\Td]$) the cash component $\varphi^0$ vanishes since a replicating strategy is fully funded through the collateral rate $r^c$, that is, the remuneration of collateral amount pledged or received.

\subsection{General representation of hedging strategies}  \label{sec6.1}

Our goal is to apply the first hedging method introduced in Section \ref{sec4.7} to find explicit expressions for the processes $\psi^1,\psi^2$ and $\psi^3$ for the case of a single-period CCBS without exchange of notional principals. To this end, it suffices to apply the It\^o product rule to the pricing formulae established in Proposition \ref{pro5.1} and Proposition \ref{pro5.2}
(or Proposition \ref{pro5.1c}) and make use of the previously established dynamics of relevant stochastic processes (see Remark \ref{rem4.1} and Remark \ref{rem4.1x}). We also aim to express the hedging strategies $[\phid, \phif, \phiq]$ in terms of market observables, namely futures contracts, by applying the linear transformation described in Proposition \ref{pro4.8}. This mathematical step is essential for practical implementation of hedging strategies.

We first introduce the shorthand notation for the vector of processes, which appear in the pricing formula for a CCBS, specifically, we write $[Y^1_t,Y^2_t,\dots, Y^8_t]$ to denote the vector
\begin{align*}
[Q_t,\whB_{\Ud}(t,\rf_t),\whB_{\Td}(t,\rf_t),\whB_{\Sd}(t,\rf_t),\FB_{\Sd,\Ud}(t,\rd_t),\FB_{\Sd,\Td}(t,\rd_t),B_\Ud(t,\rd_t),B_\Td(t,\rd_t)]
\end{align*}
and $[\sigma^1_t,\sigma^2_t,\dots,\sigma^8_t]$ stands for the vector of the associated deterministic volatilities
\begin{align*}
[\barsig ,\hbtU,\hbtT,\hbtS,\btSU ,\btST,\btU,\btT].
\end{align*}
Using the dynamics of the processes $Y^1,Y^2,\dots, Y^8$, we obtain
\begin{align*}
\begin{bmatrix}
\diff Y^1_t/Y^1_t \\
\diff Y^2_t/Y^2_t \\
\diff Y^3_t/Y^3_t \\
\diff Y^4_t/Y^4_t \\
\diff Y^5_t/Y^5_t \\
\diff Y^6_t/Y^6_t \\
\diff Y^7_t/Y^7_t \\
\diff Y^8_t/Y^8_t
\end{bmatrix}
\mart
\begin{bmatrix}
0&0&\sigma^1_t\\
0&\sigma^2_t&0\\
0&\sigma^3_t&0\\
0&\sigma^4_t&0\\
\sigma^5_t&0&0\\
\sigma^6_t&0&0\\
\sigma^7_t&0&0\\
\sigma^8_t&0&0\\
\end{bmatrix}
\begin{bmatrix}
\diff \Zone_t\\
\diff \Ztwo_t\\
\diff \Zthree_t
\end{bmatrix}
=
\begin{bmatrix}
\zeta^1_t\\
\zeta^2_t \\
\zeta^3_t\\
\zeta^4_t\\
\zeta^5_t\\
\zeta^6_t\\
\zeta^7_t\\
\zeta^8_t\\
\end{bmatrix}
\begin{bmatrix}
\diff \Zone_t\\
\diff \Ztwo_t\\
\diff \Zthree_t
\end{bmatrix}
\end{align*}
where the rows of the deterministic volatility matrix are $\zeta^i_t =[\zeta^{i,1}_t,\zeta^{i,2}_t,\zeta^{i,3}_t]$ for $i=1,2,\dots,8$ so that $\diff Y^i_t = Y^i_t\zeta^i_t \diff Z_t$ for $i=1,2,\dots ,8$.
We denote $\gm(t):=A^{\beta}_{t,\Td}$ and, as before, we set $\kpde= (1-\kappa\delta)$. From Proposition \ref{pro5.1}, we know that
the arbitrage-free price $\Xbeta_t$ of a CCBS satisfies
\begin{align*}
\Xbeta_t &=
\begin{cases}
\gm(t)\big[ \LqtT (Y^1_t Y^2_t-Y^1_t Y^3_t)- \LqtS (\GSU   Y^1_t Y^4_t Y^5_t-\kpde\GST  Y^1_t Y^4_t Y^6_t) \big], & t\in [0,\Sd], \\
\gm(t)\big[ \LqtT (Y^1_t Y^2_t- Y^1_t Y^3_t)- Q_{\Sd} Y^7_t+ \kpde Q_{\Sd} Y^8_t \big], & t\in [\Sd,\Ud], \\
\gm(t)\big[ \LqtT (Y^1_t(\Bf_{\Ud})^{-1} \Bf_t- Y^1_t Y^3_t) -Q_{\Sd}(\Bd_{\Ud})^{-1} \Bd_t+\kpde Q_{\Sd} Y^8_t\big], & t\in [\Ud,\Td]
\end{cases}
\end{align*}

We are now ready to give the second main contribution of this work, which is an explicit representation for the hedging strategy of a forward-start single-period $\CCBS(\Sd,\Ud,\Td;\kappa)$. By applying Proposition \ref{pro4.8}, one first obtains
the intermediate triplet $[\psi^1,\psi^2,\psi^3]$ and subsequently the triplet $[\phid,\phif,\phiq]$, which represents
the futures hedging strategy, for the exchange of interest payments and the exchange of nominal principals.
It is important to note that both the dynamics of the futures contracts and the pricing formulae  vary across time intervals. Specifically, futures dynamics differ on the intervals $[0, \Ud]$ and $[\Ud, \Td]$, while pricing expressions may change across $[0, \Sd]$, $[\Sd, \Ud]$, and $[\Ud, \Td]$.

\subsection{Hedging strategy for the domestic leg}\label{sec6.2}

We first present the hedging strategy of the domestic leg over the
three intervals: $[0, \Sd]$, $[\Sd, \Ud]$, and $[\Ud, \Td]$.  For brevity, we denote
\begin{align*}
\whzetad (t,\Sd,\Ud,\Td):=\frac{\btS}{\btUT},\quad
\wtzetad (t,\Ud,\Td):=\frac{\btU}{\btUT},\quad
\zetad (t,\Ud,\Td):=\frac{\btT}{\btUT},
\end{align*}
and, similarly,
\begin{align*}
\whzetaf (t,\Sd,\Ud,\Td):=\frac{\hbtS}{\hbtUT},\quad
\wtzetaf (t,\Ud,\Td):=\frac{\hbtU}{\hbtUT},\quad
\zetaf (t,\Ud,\Td):=\frac{\hbtT}{\hbtUT}.
\end{align*}
Since the dates $\Sd,\Ud$ and $\Td$ are fixed we will also write $\whzetad _t =\whzetad (t,\Sd,\Ud,\Td),\wtzetad _t= \wtzetad (t,\Ud,\Td)$, $
\zetad _t= \zetad (t,\Ud,\Td),$ etc.. The derivation of the hedging strategy for the domestic leg is given
in Appendix \ref{AppD1}.

\medskip

\textbf{Pricing formula.} Let us denote $\Upsilon_t:=\AbtT \LqtS  Q_t\whB_{\Sd}(t,\rf_t)$. The price process
of the domestic leg equals $\Xbetad_t =  \Xbetadx_t - \Xbetady_t$ where the terms $\Xbetadx_t$ and $\Xbetady_t$ satisfy:
\begin{align*}
\Xbetadx_t &=
\begin{cases}
\Upsilon_t \GSU   \FB_{\Sd,\Ud}(t,\rd_t), & t\in [0,\Sd], \\
\AbtT Q_{\Sd} B_{\Ud}(t,\rd_t), & t\in [\Sd,\Ud], \\
\AbtT Q_{\Sd}(\Bd_{\Ud})^{-1}\Bd_t, & t\in [\Ud,\Td],
\end{cases} \quad
\Xbetady_t =
\begin{cases}
\Upsilon_t\kpde\GST  \FB_{\Sd,\Td}(t,\rd_t), & t\in [0,\Sd], \\
\AbtT Q_{\Sd}\kpde B_{\Td}(t,\rd_t), & t\in [\Sd,\Ud], \\
\AbtT Q_{\Sd}\kpde B_{\Td}(t,\rd_t), & t\in [\Ud,\Td].
\end{cases}
\end{align*}

\medskip
\textbf{Hedging strategy.} The hedging strategy $(\phiv^{d,1}_t,\phiv^{d,2}_t,\phiv^{d,3}_t)$ for the domestic leg satisfies
\begin{align*}
\Xbetad_t=\Xbetad_0+\int_0^t r^{\beta}_u\Xbetad_u\diff u+\int_0^t\phiv^{d,1}_u\diff \Fd_u
+\int_0^t\phiv^{d,2}_u\diff \Ffq_u+\int_0^t\phiv^{d,3}_u\diff \Fq_u
\end{align*}
where the hedging components are given by the following equalities:

\medskip

\noindent\textit{Domestic futures strategy:}
\begin{align*}
\phiv^{d,1}_t =
\begin{cases}
\big[\zeta^*_t \Xbetadx_t+(2\zetad _t-\whzetad_t)\Xbetady_t \big]\delta (1 + \delta \Fd_t)^{-1}, & t\in [0,\Sd], \\
\big[-\wtzetad _t\Xbetadx_t+\zetad_t\Xbetady_t\big]\delta(1+\delta \Fd_t)^{-1}, & t\in [\Sd,\Ud],\\
-\Xbetady_t \delta (1+\delta \Fd_t)^{-1}, & t \in [\Ud,\Td],
\end{cases}
\end{align*}
\noindent\textit{Foreign futures strategy:}
\begin{align*}
\phiv^{d,2}_t =
\begin{cases}
\Xbetad_t (Q_t)^{-1}(\zetaf _t-\whzetaf _t)\delta (1 + \delta \Ff_t)^{-1}, & t\in [0,\Sd], \\
0, & \text{otherwise},
\end{cases}
\end{align*}

\noindent\textit{Currency futures strategy:}
\begin{align*}
\phiv^{d,3}_t =
\begin{cases}
\Xbetad_t (\Fq_t)^{-1}, & t\in [0,\Sd], \\
0, & \text{otherwise},
\end{cases}
\end{align*}
with the auxiliary notation $\zeta^*_t := \whzetad _t-\wtzetad _t-\zetad _t$.

\medskip

\brem Notice that $\phiv^{d,1}_t$ can also be represented as follows, for every $t \in [\Ud,\Td]$,
\begin{align*}
\phiv^{d,1}_t=\gm (t)Q_{\Sd}\kpde \, e^{\int_{\Ud}^t\rd_u\diff u+\int_t^\Td \sbuT \diff u}\, \delta(1+\delta \Fd_t)^{-2}
\end{align*}
or, equivalently, \begin{align*}
\phiv^{d,1}_t= (B^{\beta}_t)^{-1} \gm (t)Q_{\Sd}\kpde\delta \, e^{\int_t^\Td\sbuT\diff u}\,(B^d_{\Ud})^{-1}B^d_t (1+\delta \Fd_t)^{-2}.
\end{align*}
\erem

\subsection{Hedging strategy for the foreign leg}\label{sec6.3}

We now turn to the hedging strategy for the foreign leg, which is defined over two distinct intervals: \( [0, \Ud] \) and \( [\Ud, \Td] \). We provide a unified expression for the foreign leg's price process and the associated hedging strategy.
For the derivation of the hedging strategy for the foreign leg, see Appendix \ref{AppD2}.

\medskip

\textbf{Pricing formula.} The price process for the foreign leg can be succinctly expressed as
$\Xbetaf_t = \Xbetafx_t - \Xbetafy_t$ where the terms $\Xbetafx_t$ and $\Xbetafy_t$ are defined by:
\begin{align*}
\Xbetafx_t &=
\begin{cases}
\AbtT \LqtT  Q_t \whB_{\Ud}(t,\rf_t), & t\in [0,\Ud],\\[4pt]
\AbtT \LqtT  Q_t (\Bf_{\Ud})^{-1}\Bf_t, & t\in [\Ud,\Td],
\end{cases} \\
\Xbetafy_t &= \AbtT \LqtT  Q_t \whB_{\Td}(t,\rf_t), \quad \quad \ t\in [0,\Td].
\end{align*}

\medskip

\textbf{Hedging strategy.} The hedging strategy $(\phiv^{f,1}_t,\phiv^{f,2}_t,\phiv^{f,3}_t)$ for the foreign leg satisfies
\begin{align*}
\Xbetaf_t = \Xbetaf_0 +\int_0^t r^{\beta}_u\,\Xbetaf_u\diff u+\int_0^t\phiv^{f,1}_u\diff \Fd_u
+\int_0^t\phiv^{f,2}_u\diff \Ffq_u+\int_0^t\phiv^{f,3}_u\diff \Fq_u
\end{align*}
with the hedging components given by:

\medskip

\noindent\textit{Domestic futures strategy:}
\begin{align*}
\phiv^{f,1}_t =
\begin{cases}
-\Xbetaf_t \zetad_t\delta(1+\delta\Fd_t)^{-1}, & t\in [0,\Ud],\\[4pt]
-\Xbetaf_t \delta(1+\delta \Fd_t)^{-1}, & t\in [\Ud,\Td],
\end{cases}
\end{align*}

\noindent\textit{Foreign futures strategy:}
\begin{align*}
\phiv^{f,2}_t = \Xbetafx_t (Q_t)^{-1}\delta(1+\delta \Ff_t)^{-1}, \quad t\in [0,\Td],
\end{align*}
\noindent\textit{Currency futures strategy:}
\begin{align*}
\phiv^{f,3}_t = \Xbetaf_t (\Fq_t)^{-1}, \quad t\in [0,\Td].
\end{align*}

\medskip

\subsection{Hedging strategy for the exchange of nominal principals}\label{sec6.4}

Finally, we examine the hedging strategy for the exchange of nominal principals. As in previous sections, we partition the analysis into the subintervals \( [0, \Sd] \), \( [\Sd, \Ud] \), and \( [\Ud, \Td] \) and we postpone the
computations to Appendix \ref{AppD3}.

\medskip

\textbf{Pricing formula.} The price process of the exchange of nominal principals is represented as
$\Xbetap_t = \Xbetapx_t - \Xbetapy_t$ where the components $\Xbetapx_t$ and $\Xbetapy_t$ are given explicitly by:
\begin{align*}
\Xbetapx_t &= \AbtT \LqtT  Q_{t} \whB_{\Td}(t,\rf_t), \quad  t\in [0,\Td],
\\[6pt]
\Xbetapy_t &=
\begin{cases}
\AbtT Q_t \LqtS \GST  \FB_{\Sd,\Td}(t,\rd_t)\whB_{\Sd}(t,\rf_t), & t\in [0,\Sd], \\[4pt]
\AbtT Q_{\Sd} B_{\Td}(t,\rd_t), & t\in [\Sd,\Td].
\end{cases}
\end{align*}

\medskip

\textbf{Hedging strategy.} The hedging strategy $(\phiv^{1}_t,\phiv^{2}_t,\phiv^{3}_t)$ satisfies
\begin{align*}
\Xbetap_t = \Xbetap_0 +\int_0^t r^{\beta}_u \Xbetap_u\diff u+\int_0^t\phiv^{1}_u\diff \Fd_u+\int_0^t\phiv^{2}_u\diff \Ffq_u
+\int_0^t\phiv^{3}_u\diff \Fq_u
\end{align*}
where the hedging components satisfy:

\medskip

\noindent\textit{Domestic futures strategy:}
\begin{align*}
\phiv^{1}_t =
\begin{cases}
\big[(\whzetad_t-2\zetad_t)\Xbetapy_t - \zetad_t\Xbetapx_t\big]\delta(1+\delta \Fd_t)^{-1}, & t\in [0,\Sd],\\[4pt]
-\Xbetap_t \zetad_t\delta(1+\delta \Fd_t)^{-1}, & t\in [\Sd,\Ud],\\[4pt]
-\Xbetap_t \delta(1+\delta \Fd_t)^{-1}, & t\in [\Ud,\Td],
\end{cases}
\end{align*}

\noindent\textit{Foreign futures strategy:}
\begin{align*}
\phiv^{2}_t =
\begin{cases}
-\Xbetapy_t\delta(\whzetaf_t-\zetaf_t)(1+\delta \Ff_t)^{-1}, & t\in [0,\Sd],\\[4pt]
0, & t\in [\Sd,\Td],
\end{cases}
\end{align*}

\noindent\textit{Currency futures strategy:}
\begin{align*}
\phiv^{3}_t =
\begin{cases}
\Xbetap_t(\Fq_t)^{-1}, & t\in [0,\Sd], \\[4pt]
\Xbetapx_t (\Fq_t)^{-1}, & t\in [\Sd,\Td].
\end{cases}
\end{align*}

\newpage
\section{Numerical studies}  \label{sec7}

We conclude this work by presenting a numerical study for various CCBS classes. While in Section \ref{sec5} and Section \ref{sec6} we focused on single-period CCBSs, it is clear this can be extended to multi-period CCBSs discussed in Section \ref{sec5.6.1}.

As an example, we take a standard 3-year multi-period CCBS with the domestic notional principal USD 10 million, traded on a non-mark-to-market basis. This CCBS is a floating-for-floating swap involving interest rate payments and notional amounts in USD and AUD. The floating reference rate for each leg is based on the respective six-month backward-looking compound rate calculated using SOFR/AONIA in the corresponding currency. To be more specific, we consider the AUD/USD basis swap in which an investor receives the 6-month USD SOFR and pays the 6-month AUD AONIA plus a spread with interest payments exchanged semiannually. Recall that, unlike other interest rate swaps, CCBS also includes exchanges of notional principals at inception and maturity dates. From Definition \ref{def2.5}, the structure of CCBS
can be described as follows: let $P^d$ [AUD] and $P^f$ [USD] be the equivalent notional principal amounts exchanged at the inception date $T_0$ and exchanged back at the maturity date $T_n$ with $n = 6$. The tenor structure $\cT_n$ is given by $0 < T_1 < \cdots < T_6$, where $T_{j} = 0.5 j$ for $j = 1, \dots, 6$, and $\delta_j := T_{j} - T_{j-1} = 0.5$ for $j = 1, \dots, 6$. At each payment date $T_{j}$ for $j = 1, \dots, 6$, the net cash flow associated with a cross-currency basis swap with $P^f = 10$ mm [USD] and $P^d =Q_0 P^f$ mm [AUD] is given by
\begin{equation*}
\CCBS \,(\cT_{6};\kappa ) =\begin{cases}
0, \quad \text{at}\ \, T_0,\\
0.5 \left[ \Rf(T_{j-1},T_{j}) Q_{T_{j}}-(\Rd(T_{j-1},T_{j})+\kappa) Q_0\right] ,\quad \text{at}\ \, T_j,\ j=1,2,\dots,5, \\
0.5 \left[ \Rf(T_{5},T_{6}) Q_{T_{n}}-(\Rd(T_{5},T_{6})+\kappa) Q_0  \right]+ Q_{T_{6}}- Q_{0} ,\quad \text{at}\ \, T_6.\\
\end{cases}
\end{equation*}
For simplicity, we assume $T_0 = 0$, meaning the contract is initiated at time $0$, rather than considering a forward CCBS entered into at time $0$ but starting at a future date $T_0 > 0$. This assumption is made for clarity in presentation but can be easily relaxed.

\subsection{Model parameters and validation of pricing-hedging framework}  \label{sec7.1}

In our numerical simulations, we utilized the following parameters, which are all given in annualized terms:
\begin{align*}
& a = 0.15,\ \wh{c} = 0.05,\  b = \wh{b} = 5, \\
& \sigma = \wh{\sigma} = 1\%,\ \barsig = 10 \%, \\
& \rho_{23} = \rho_{13} = 0.1,\ \rho_{12} = 0.3, \\
& \rd_0 = \rf_0 = 2 \%,\ Q_0 = 1.5, \\
& \alpha^\beta = 2 \%,\ \alpha^d = \alpha^f.
\end{align*}

\begin{figure}
\centering
\includegraphics[width=10cm]{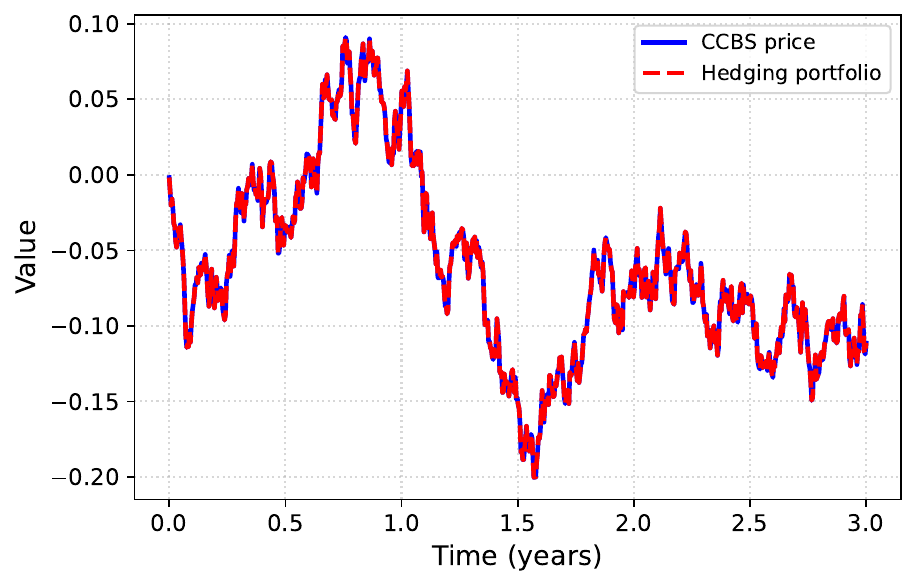}
\caption{Comparison of the simulated CCBS price and the hedging portfolio value}\label{figure: Hedging}
\end{figure}

For the purpose of simulations, these parameters can be random processes but, for the sake of clarity, we have chosen to keep them constant. While we are using artificial parameters here, empirical pricing and hedging results can also be obtained by a model calibration to the current market data. To validate the theoretical pricing framework introduced in Assumption \ref{ass4.1}, we conduct a Monte Carlo simulation of the  multi-period CCBS.  We first simulate sample paths of interest rates and exchange rate based on their dynamics given by Assumption \ref{ass4.1}. Based on these dynamics, we then simulate the corresponding futures prices using results from Section \ref{sec4.3}. The simulated price, computed as the expected discounted cash flow, is shown to match closely with the theoretical price derived in Proposition \ref{pro5.6}, with negligible numerical error. Furthermore, we evaluate the consistency between pricing and hedging by comparing the simulated price process of a multi-period CCBS with the corresponding wealth process generated by the proposed hedging strategy. As shown in Figure \ref{figure: Hedging}, the two trajectories are virtually indistinguishable, providing strong evidence that the derived hedging strategy effectively replicates the contract under the proposed market dynamics.

\subsection{Sensitivity analysis and comparative statics}  \label{sec7.2}

After a brief validation of theoretical pricing and hedging results within the setup of a model from Assumption \ref{ass4.1},
our next objective is to illustrate the pricing outcomes discussed in Section \ref{sec5} and to analyze the effect of varying model parameters, as introduced in Section \ref{sec4}, alongside initial market data, on arbitrage-free pricing of a multi-period CCBS obtained in Proposition \ref{pro5.6}. In the following two tables, we present the fair prices of a standard CCBS contract, adjusted to reflect a typical contract size of $10$ million. The contract, denoted as $\CCBS \,(\cT_{6};\kappa )$, is split into two components: the exchange of interest payments, $X^{\beta, j}_0 = \pi^{\beta}_0(X_{T_j}(T_0,T_{j-1},T_j))$ for $j = 1, \dots, n$, and the exchange of nominal principals, $\Xbetap_0 = \pi^{\beta}_0(X^p_{T_n}(T_0,T_n))$. Additionally, we define $\Xbeta_0 = \sum_{j=1}^{n} X^{\beta, j}_0$ as the total price of the interest payments at time 0. The notation used here is consistent with that introduced in Section \ref{sec5.2}.

For the purpose of this section, we introduce the new notation for the model parameters. Specifically, dynamics in Assumption \ref{ass4.1} is now rewritten as
\begin{align*}
&\diff \rd_t = b(\theta^d - \rd_t)\diff t + \sigma \diff Z^1_t, \\
&\diff \rf_t = \wh{b}(\theta^f - \rf_t)\diff t + \wh{\sigma} \diff Z^2_t, \\
&\diff Q_t = Q_t(\rd_t - \rf_t + \Delta \alpha) \diff t + Q_t \barsig \diff Z^3_t,
\end{align*}
where $\Delta \alpha := \alpha^d - \alpha^f,\theta^d := a/b$ and $\theta^f := \wh{c}/\wh{b}$ represent the long term mean level of the domestic and foreign risk-free rate, respectively. We also denote $\Delta \theta := \theta^d - \theta^f$, which captures the difference in the mean levels of the domestic and foreign interest rates. The quantity $\theta^q := \Delta \theta + \Delta \alpha$ represents the long term drift of the exchange rate $Q_t$ and it encapsulates the combined effects of both the interest rate differential and the drift terms.

The default sample, chosen to encapsulate the swap buyer's requirements, sets the drift of the exchange rate $\theta^q :=\Delta \theta + \Delta\alpha $ at $3\%$ where $\Delta\alpha$ represents the difference in indicative funding costs (market rates) in two currencies, while the term $\Delta \theta$ reflects the difference in long term interest rates in two economies. This choice highlights their critical role in determining the CCBS price. To demonstrate the role of these parameters, we will study the impact of $\Delta\alpha$ and $\theta^q$ separately in Table \ref{Table: parameters study alpha} and Table \ref{Table: parameters study multiperiod}, respectively. We will proceed in this subsection as follows.

First, in Step 1, we examine the impact of $\Delta \alpha$. As shown in Table \ref{Table: parameters study alpha}, it mainly influences the price of the exchange of nominal principals, while its effect on pricing of interest payments is minimal. In Step 2, we focus on the role of $\Delta \theta$. In contrast to $\Delta \alpha$, the parameter $\Delta \theta$ has a significant impact on prices of principal exchanges and interest payments since changes in $\Delta \theta$ not only affect the price of interest payments, but also influence the drift term $\theta^q$ in the exchange rate, which affects the pricing of the exchange of nominal principals.

\textbf{Step 1}: We first fix the difference $\Delta \theta$ at 2\% and we present in Table \ref{Table: parameters study alpha} the effects of the difference in long term market rates, which is given by $\Delta \alpha = \alpha^d - \alpha^f$. It is important to note that only the difference $\Delta \alpha$ matters, as opposed to individual changes in $\alpha^d$ or $\alpha^f$, which is evidenced by the definition of the function $\LqtT$ in Equation \eqref{wtQ3}.

\begin{table}[htbp]
\centering
\caption{Impact of $\Delta \alpha$ on arbitrage-free price and fair basis spread}
\label{Table: parameters study alpha}
\begin{tabular}{cccccc}
\toprule
No. & $\Delta\alpha$ & $\Xbeta_{0} \times 10^7$ & $\Xbetap_{0} \times 10^7$ & $\CCBS_0^\beta \times 10^7$ & Basis spread $\kappa$\\
\midrule
1 & $-0.5\%$ & $-766,922$ & $541,421$ & $-225,501$ & $-54.54$ bps \\
2 & $-0.1\%$ & $-763,923$ & $704,131$ & $-59,792$ & $-14.46$ bps \\
3 & $0\%$ & $-763,169$ & $745,115$ & $-18,055$ & $-4.37$ bps \\
4 & $0.1\%$ & $-762,414$ & $786,221$ & $23,807$ & $5.76$ bps \\
5 & $0.5\%$ & $-759,377$ & $951,886$ & $192,509$ & $46.56$ bps \\
\bottomrule
\end{tabular}
\end{table}

From Table \ref{Table: parameters study alpha}, we observe that changes in $\Delta \alpha$ have a very limited effect on the price of interest payments, $\Xbeta_{0}$. This was expected because, although there is a correlation between the exchange rate and interest rates, the difference in mean interest rates, $\Delta \theta$ remains unchanged. The primary impact is observed on the price of the exchange of principals, which is positively correlated with $\Delta \alpha$, resulting in a positive correlation in the total price.

Specifically, when $\Delta \alpha$ increases from -0.5\% to 0.5\%, the basis spread grows from -54.54 basis points to 46.56 basis points. This indicates that the funding cost in the domestic economy has become more expensive relative to the foreign economy, thereby increasing the overall interest rate cost. The increase of the basis spread reflects this higher cost and is intuitively consistent with the model.

\textbf{Step 2:} We continue the main comparative statics by fixing $\Delta \alpha = 0$. This assumption implies that, from the perspective of funding costs, the two economies are effectively equivalent. In Table \ref{Table: parameters study multiperiod}, we display the fair prices for varying speed of reversion parameters $b$ and~$\wh{b}$.

It is important to note that, by the definition of $\theta^q$, changing only the value of the speed of reversion will also change $\theta^q$. Therefore, when adjusting the speed of reversion parameters $b$ and $\wh{b}$, we ensure that the long term means $\theta^d$ and $\theta^f$ (and hence $\theta^q$) remain unchanged by making suitable adjustments to parameters $a$ and $\hat{a}$.

\begin{table}[htbp]
\centering
\caption{Effect of the speed of reversion on arbitrage-free price and fair basis spread}
\label{Table: parameters study multiperiod}
\begin{tabular}{ccccccc}
\toprule
No. & $b = \wh{b}$ & $\theta^q$ & $\Xbeta_{0} \times 10^7$ & $\Xbetap_{0} \times 10^7$ & $\CCBS_0^\beta \times 10^7$ & Basis spread $\kappa$\\
\midrule
1 & $1$ & $2\%$ & $-556,284$ & $545,481$ & $-10,803$ & $-2.60$ bps \\
2 & $2.5$ & $2\%$ & $-707,603$ & $691,948$ & $-15,655$ & $-3.78$ bps \\
3 & $5$ & $2\%$ & $-763,169$ & $745,115$ & $-18,055 $ & $-4.37$ bps \\
4 & $7.5$ & $2\%$ & $-781,808$ & $762,858$ & $-18,950$ & $-4.59$ bps \\
5 & $10$ & $2\%$ & $-791,141$ & $771,730$ & $-19,411$ & $-4.70$ bps \\
\bottomrule
\end{tabular}
\end{table}

Let us delve into some fundamental features of hedging costs for a CCBS. Note that the sign of the pricing results only indicates the direction of payments. Thus, when we mention an ``increase," we refer to an increase in magnitude.
We observe that a higher speed of reversion leads to higher prices for both interest rate payments, $\Xbeta_{0}$, and the exchange of nominal principals, $\Xbetap_{0}$s. The overall increase in the total value can be attributed to the movement of interest rates; higher speeds of reversion imply a longer dominance of one interest rate over the other (in our case, the domestic interest rate dominates the foreign one). This results in a more pronounced difference between the two economies and, consequently, higher levels for the price of a CCBS.

\begin{table}[htbp]
\centering
\caption{Effect of long term drift $\theta^q$ on arbitrage-free price and fair basis spread}
\label{Table: parameters study multiperiod1}
\begin{tabular}{ccccccc}
\toprule
No. & $b = \wh{b}$ & $\theta^q$ & $\Xbeta_{0} \times 10^7$ & $\Xbetap_{0} \times 10^7$ & $\CCBS_0^\beta \times 10^7$ & Basis spread $\kappa$\\
\midrule
1 & $5$ & $5\%$ & $-1,909,214$ & $1,864,064$ & $-45,150$ & $-10.35$ bps \\
2 & $5$ & $2\%$ & $-763,169$ & $745,115$ & $-18,055 $ & $-4.37$ bps \\
3 & $5$ & $-0.1\%$ & $76,307$ & $-74,502$ & $1,805$ & $0.43$ bps \\
4 & $5$ & $-2\%$ & $784,631$ & $-766,273$ & $18,359$ & $4.24$ bps \\
5 & $5$ & $-5\%$ & $2,046,301$ & $-1,999,224$ & $47,077$ & $10.38$ bps \\
\bottomrule
\end{tabular}
\end{table}

In Table \ref{Table: parameters study multiperiod1}, we give the fair prices for multi-period CCBSs for various levels of the drift of the exchange rate $\theta^q$. We aim here to examine the impact the relative difference in interest rates between the two economies, as represented by $\Delta \theta$. This is the dominant driver of prices for both interest payments and the exchange of principal nominals, unlike $\Delta\alpha$ that have only a strong effect on the exchange of principal nominals. The sign of both prices also depends on the sign of $\theta^q$, which indicates which interest rate dominates the other and thus determines the direction of payments. When the sign of $\theta^q$ is fixed, a larger magnitude of $\theta^q$ (i.e., a wider gap between the two economies) leads to higher contract prices. This is due to the fact that the buyer must pay/receive more to compensate for the disparity between the economies.

\subsection{Performance evaluation of hedging strategies}  \label{sec7.3}

Our next objective is to provide a preliminary study of the hedging results from Section \ref{sec6} by analyzing discretized hedging strategies based on the dynamics of futures prices obtained in Section \ref{sec4.3}. To this end, we simulate the futures price dynamics and subsequently derive the wealth process dynamics for a discretized replicating strategy for a CCBS.

Specifically, aim to assess the impact of various frequencies of the hedge on the distribution of the P\&L profile for a hedging strategy. To this end, we examine the effects of hedging performed on a weekly, monthly, quarterly, and biannual basis. Our goal is to provide an idea of the asset's riskiness by analyzing the quantiles of the P\&L, which can be interpreted as a measure of the hedger's risk exposure.

\begin{figure}[htbp]
\centering
\includegraphics[width=0.6\linewidth]{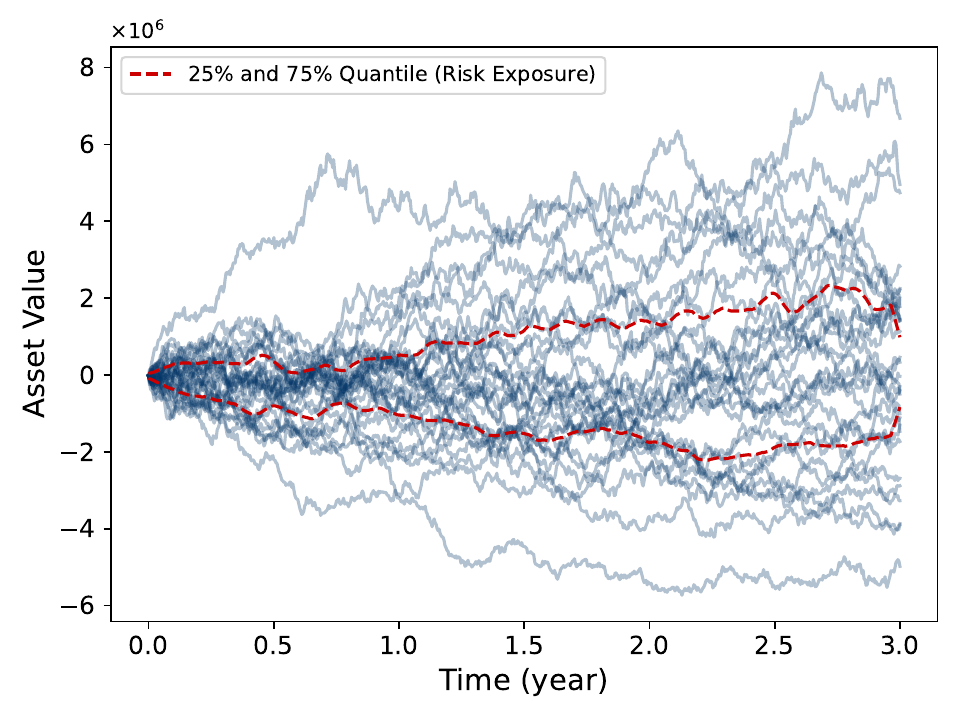}
\caption{Unhedged P\&L plot}
\label{fig: unhedged PnL}
\end{figure}

\begin{figure}[tbhp]
\centering
\subfloat[Weekly hedged P\&L plot]{\label{fig: week hedged PnL}\includegraphics[width=0.45\textwidth]{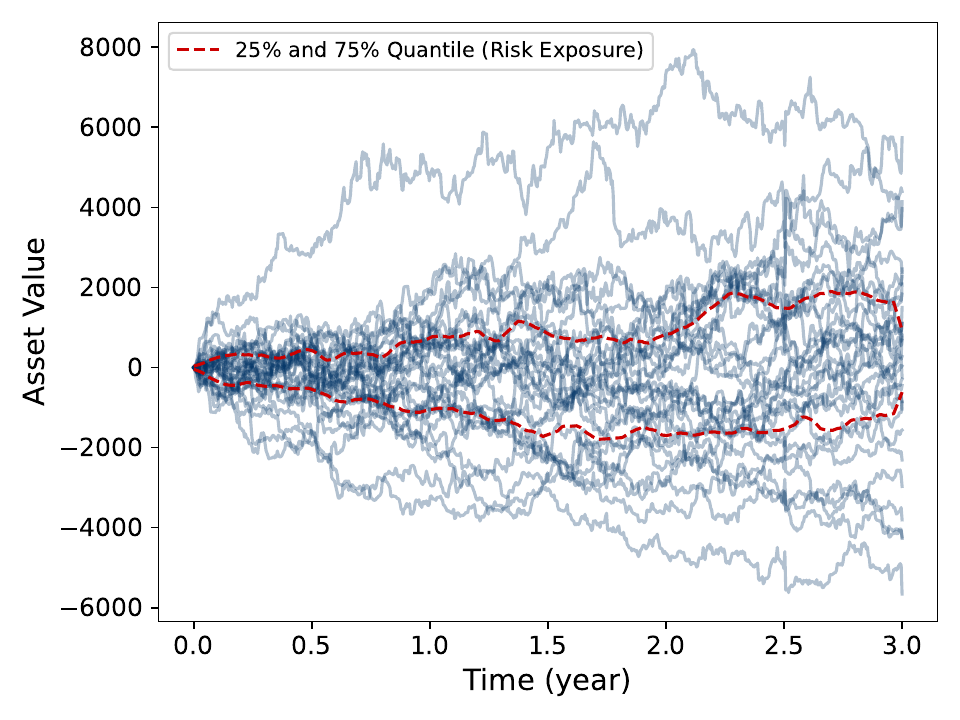}}
\hfill
\subfloat[Monthly hedged P\&L plot]{\label{fig: month hedged PnL}\includegraphics[width=0.45\textwidth]{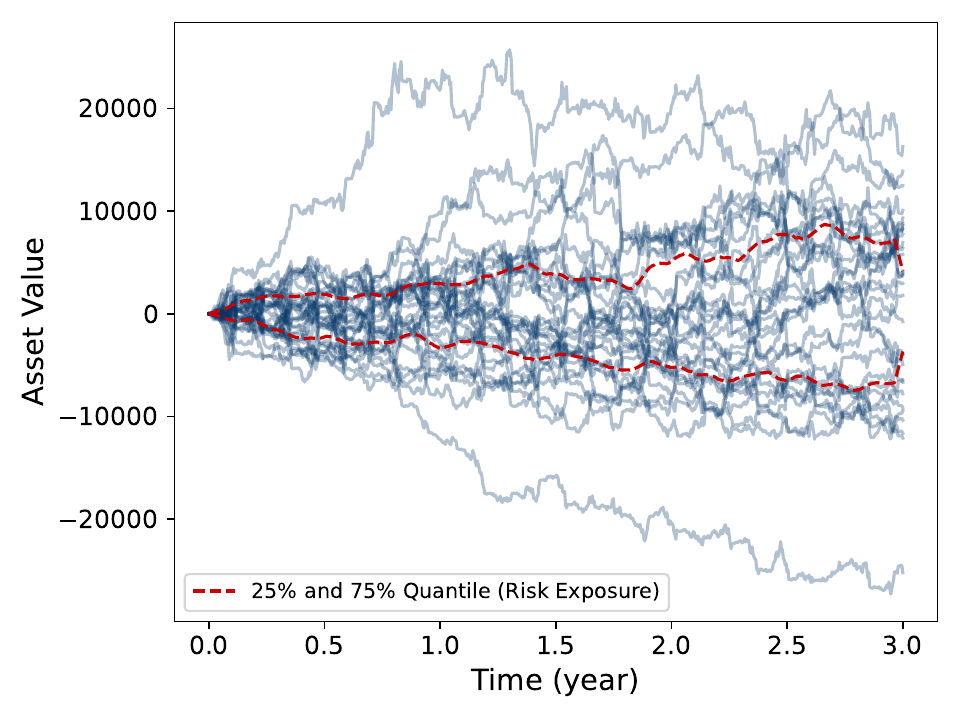}}

\medskip

\subfloat[Quarterly hedged P\&L plot]{\label{fig: quarter hedged PnL}\includegraphics[width=0.45\textwidth]{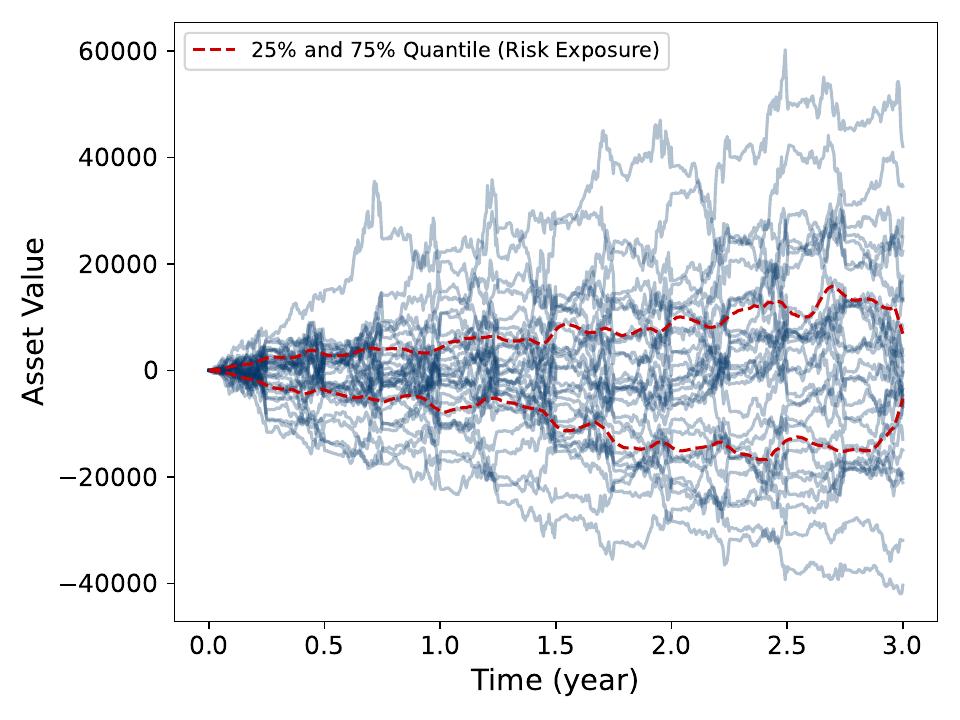}}
\hfill
\subfloat[Biannually hedged P\&L plot]{\label{fig: biannual hedged PnL}\includegraphics[width=0.45\textwidth]{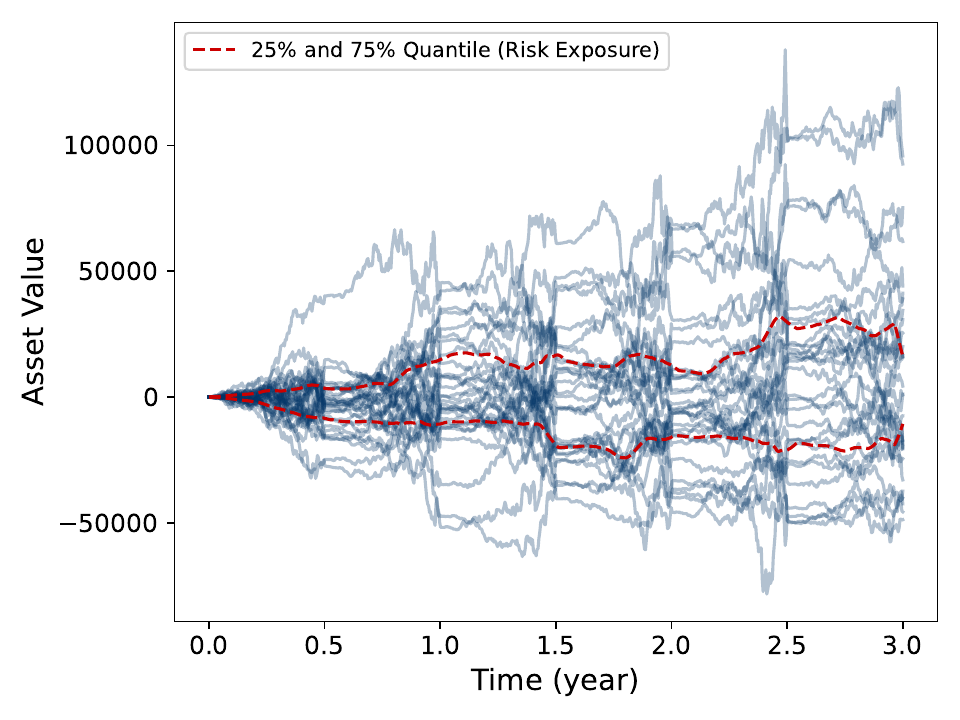}}

\caption{Comparison of P\&L plots for different hedging frequencies.}
\label{fig:different_hedge}
\end{figure}

To illustrate the risk profile, we first simulate unhedged and hedged sample paths and calculate the corresponding 25\% and 75\% quantiles to represent indicative risk exposure levels. For the sake of illustration, we present in
Figure \ref{fig: unhedged PnL} and Figure \ref{fig:different_hedge} thirty sample paths of portfolio's value over the contract's lifetime. Note that the sign of portfolio's value and hence the price of the CCBS only indicates the direction of payments, making both quantiles significant for the analysis of risk exposure.

\begin{figure}[htbp]
\centering
\includegraphics[width=0.8\linewidth]{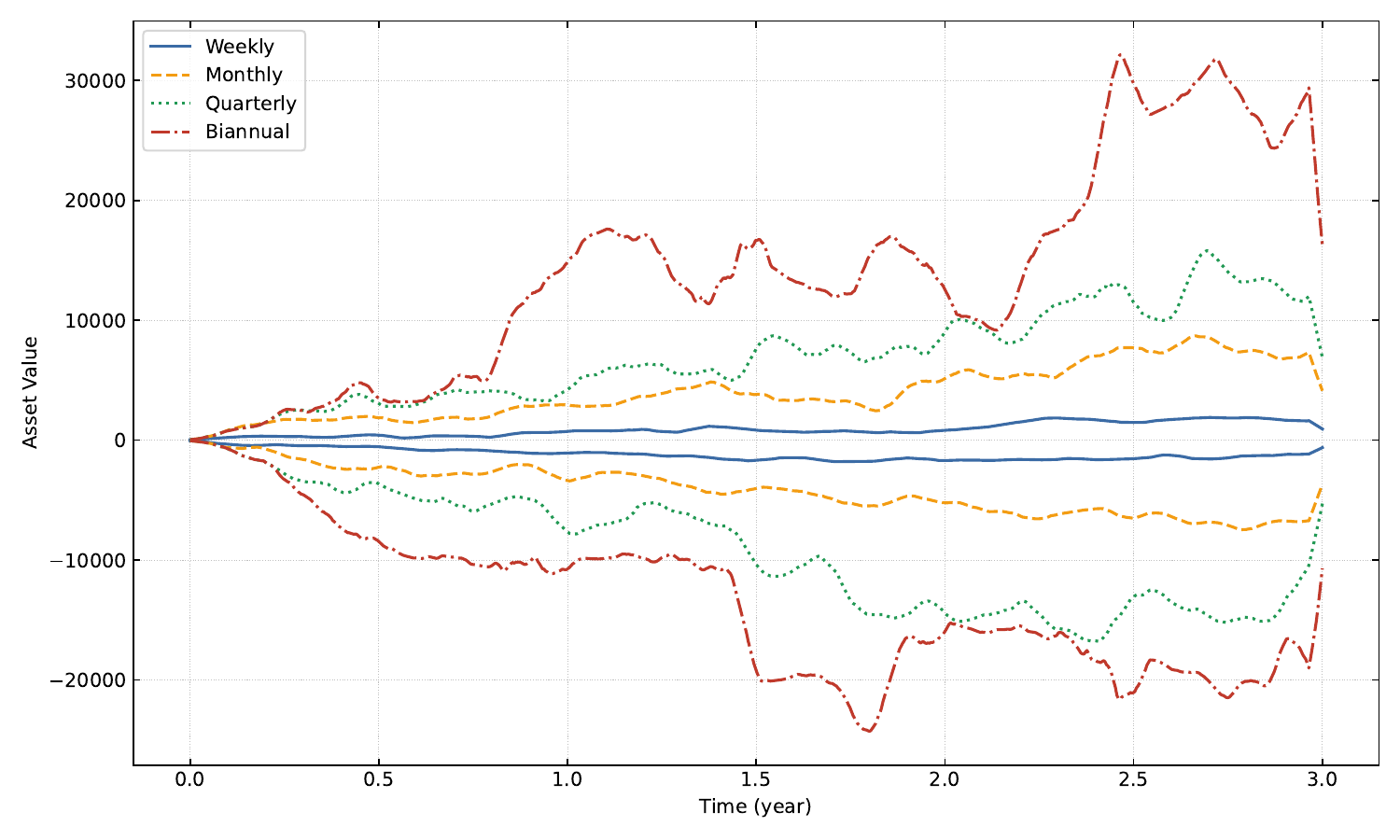}
\caption{P\&L analysis across various hedging frequencies: 25\% and 75\% quantiles}
\label{fig: different quantiles}
\end{figure}

In all five graphs, 
the distance between the 25\% and 75\% quantile tends to increase over time, which leads to a natural conjecture that the risk exposure grows as time progresses. Furthermore, in plots of the hedged positions (see Figure \ref{fig:different_hedge}) we observe that the risk exposure falls dramatically at each time when the hedging action is executed.

 As expected, between the moments of rebalancement of the hedge, the risk exposure increases, which is particularly evident in Figure \ref{fig: biannual hedged PnL} where the behavior of the risk exposure in the case of a semiannual hedging is presented. Additionally, as the hedging frequency decreases -- from weekly to biannual rebalancement of the hedging portfolio -- the risk exposure systematically increases, can be observed in Figure \ref{fig: different quantiles}.

\subsection{Numerical illustration of cross-currency swaptions} \label{sec7.4}

As emphasized in Section \ref{sec5.6.4}, the pricing a cross-currency basis swaption is a computationally demanding task. In this example, we consider a one-year European payer swaption. The \textit{trade date} is today ($t = 0$), and the \textit{exercise date} is one year from now ($T_0 = 1$). At that point, the holder may choose whether to enter into the underlying CCBS. The structure of the underlying swap remains the same as described earlier but with all dates postponed by one year so that the inception date is $T_0 = 1$ and the maturity date is $T_6 = 4$.

The swaption grants the holder the right to lock in a specific spread ($\kappa$) on the swap, which reflects the differential between the domestic and foreign interest rates. The payoff is determined by the prevailing market conditions at the exercise date. If the market spread at time $T_0$ exceeds the strike $\kappa$, the holder exercises the swaption and hence enters into the CCBS with the agreed spread. To numerically determine the price of the swaption, we first calculate the fair value of $\kappa$ at the inception of the contract, ensuring the present value of the CCBS is zero. This strike is then used as the input parameter to compute the swaption's price. Our analysis shows that the fair strike equals $-4.37$ basis points  and the price of the payer swaption is $\PSwn_{0}(\kappa) = \$918,911$, while the price of the receiver swaption is slightly higher at $\RSwn_{0}(\kappa) = \$920,676$. The funding differentials between the two currencies introduce asymmetry in the expected future values of the swap. As a result, the payer and receiver swaptions are exposed to different levels of risk leading to a slight divergence in their prices.


\subsection{Impact of collateral currency choice on pricing} \label{sec7.5}

In the last step of a numerical study, we focus on a collateral currency.
As discussed in Remark \ref{Rem: choice}, the choice of a collateral currency and hence the collateral rate $r^c$ plays a non-negligible role in derivative pricing, particularly when the collateralization level $\beta$ is relatively high.
Recall that the effective hedging rate is defined as $r^\beta := (1 - \beta) r^h + \beta r^c$.
To illustrate this sensitivity, we compute CCBS prices under different collateral currencies in the fully collateralized case $\beta = 1$. Notice that the use of a third-party collateral currency, which is not directly involved in the swap, would necessarily introduce additional modeling complexity.

To maintain tractability, we assume that the third-party benchmark rate follows the same Vasicek-type dynamics as in Assumption \ref{ass4.1}, with the same mean reversion speed but a different long-term mean, chosen to match the corresponding benchmark rate. We simulate this extended setting using a Monte Carlo method.

For illustrative purposes, we consider collateral rates linked to  benchmark overnight rates for five major currencies: Australian dollar (AUD), Euro (EUR), pound sterling (GBP), Canadian dollar (CAD), and Swiss franc (CHF) with the domestic currency (i.e., AUD) serving as a reference case. It is important to emphasize that the rates used in this section are indicative only and do not necessarily reflect current market conditions.

Arbitrage-free prices and fair basis spreads for various collateral currencies are reported in Table \ref{Table: collateral currency impact}. As expected, higher benchmark rates lead to basis spreads of a greater magnitude.

\begin{table}[htbp]
\centering
\caption{Impact of collateral currency on arbitrage-free price and fair basis spread (\(\beta=1\))}
\label{Table: collateral currency impact}
\begin{tabular}{ccccc}
\toprule
No. & Collateral Currency & Benchmark rate & \(\CCBS_0^\beta\times 10^7\) & Basis spread \(\kappa\) \\
\midrule
1 & AUD  & 3\% & $-18,055$ & $-4.37$ bps \\
2 & EUR  & 2.4\% & $-12,801$ & $-3.06$ bps \\
3 & GBP  & 4.5\% & $-30,608$ & $-7.59$ bps \\
4 & CAD  & 2.75\% & $-15,882$ & $-3.82$ bps \\
5 & CHF  & 1.5\% & $-4,662$ & $-1.10$ bps \\
\bottomrule
\end{tabular}
\end{table}

\appendix

\section{Proofs of hedging strategies using futures} \label{AppD}

\subsection{Proof of replication of the domestic leg}     \label{AppD1}

We proceed in two steps: we first find the auxiliary process $(\psi^1,\psi^2,\psi^3)$ satisfying
$$
\diff ( (B^{\beta}_t)^{-1} \pi^{\beta}_t(X_{\Td}))=(B^{\beta}_t)^{-1}[\psi^1_t,\psi^2_t,\psi^3_t]\diff Z_t
$$
using \eqref{rep3} and the It\^o formula. In the second step, we will compute the futures hedging strategy $\phiv^d=(\phiv^{d,1},\phiv^{d,2},\phiv^{d,3})$, which is described in \eqref{hedge1}.

\textit{Step 1.} Let $y^i,\, i=1,2,\dots,n$ be strictly positive processes such that $\diff y^i_t \mart y^i_t \zeta^i_t\diff Z_t$. Then the process $y_t :=\gm (t) \Pi_{i=1}^n y^i_t$   where $\gm(t)$ is a smooth deterministic function satisfies $\diff y_t \mart \gm(t) y_t\Sigma_{i=1}^n \zeta^i_t\diff Z_t$. Therefore, for every $t\in [0,\Sd]$,
\begin{align*}
\diff \wt{X}^{d,\beta}_t = \diff \big((B^{\beta}_t)^{-1} \Xbetad_{t} \big)&=(B^{\beta}_t)^{-1}\gm(t)\LqtS \Big[\GSU Y^1_t Y^4_t Y^5_t \big(\zeta^1_t+\zeta^4_t+\zeta^5_t\big) \\
&\quad\quad\quad - \kpde \GST Y^1_t Y^4_t Y^6_t \big(\zeta^1_t+\zeta^4_t+\zeta^6_t\big)\Big] \diff Z_t.
\end{align*}
Next, for every $t\in [\Sd,\Ud]$,
\begin{align*}
d\wt{X}^{d,\beta}_t=(B^{\beta}_t)^{-1}\gm (t)\big( Q_{\Sd} Y^7_t \zeta^7_t-\kpde  Q_{\Sd} Y^8_t \zeta^8_t\big)\diff Z_t
\end{align*}
and, finally, for every $t\in [\Ud,\Td]$,
\begin{align*}
d\wt{X}^{d,\beta}_t=-(B^{\beta}_t)^{-1}\gm(t)  \kpde  Q_{\Sd} Y^8_t\zeta^8_t \diff Z_t.
\end{align*}
More explicitly, for every $t\in [0,\Sd]$,
\begin{small}
\begin{align*}
&d\wt{X}^{d,\beta}_t=(B^{\beta}_t)^{-1}\gm (t)\LqtS \Big(\GSU Y^1_t Y^4_t Y^5_t[\sigma^5_t,\sigma^4_t,\sigma^1_t]-\kpde
\GST Y^1_t Y^4_t Y^6_t[\sigma^6_t,\sigma^4_t,\sigma^1_t]\Big)\diff Z_t ,
\end{align*}
\end{small}
which implies the first asserted equality. Next, for every $t\in [\Sd,\Ud]$,
\begin{align*}
d\wt{X}^{d,\beta}_t=(B^{\beta}_t)^{-1}\gm (t)\big( Q_{\Sd} Y^7_t[\sigma^7_t,0,0]-\kpde Q_{\Sd} Y^8_t [\sigma^8_t,0,0]\big)\diff Z_t.
\end{align*}
Finally, for every $t\in [\Ud,\Td]$,
\begin{align*}
d\wt{X}^{d,\beta}_t=-(B^{\beta}_t)^{-1}\gm(t)\big(\kpde Q_{\Sd}Y^8_t[\sigma^8_t,0,0]\big)\diff Z_t.
\end{align*}

We conclude that the auxiliary process $\psi$ satisfies, for every $t\in [0,\Sd]$,
\begin{align*}
\begin{bmatrix}
\psi^{d,1}_t \\
\psi^{d,2}_t \\
\psi^{d,3}_t
\end{bmatrix}
&= \gm(t)\LqtS  Y^1_t Y^4_t
\begin{bmatrix}
\GSU Y^5_t \sigma^5_t-\kpde\GST  Y^6_t\sigma^6_t \\
\GSU Y^5_t \sigma^4_t-\kpde \GST Y^6_t\sigma^4_t \\
\GSU Y^5_t\sigma^1_t-\kpde \GST  Y^6_t\sigma^1_t
\end{bmatrix} =
\begin{bmatrix}
\sigma^5_t \\
\sigma^4_t\\
\sigma^1_t
\end{bmatrix} \Xbetadx_t -
\begin{bmatrix}
\sigma^6_t \\
\sigma^4_t\\
\sigma^1_t
\end{bmatrix} \Xbetady_t
\end{align*}
for every $t\in [\Sd, \Ud]$,
\begin{align*}
\begin{bmatrix} \psi^{d,1}_t\\
\psi^{d,2}_t\\
\psi^{d,3}_t
\end{bmatrix}
= \gm(t) Q_{\Sd}
\begin{bmatrix}
Y^7_t\sigma^7_t-\kpde Y^8_t\sigma^8_t \\
0 \\
0
\end{bmatrix} =
\begin{bmatrix}
\sigma^7_t \\
0\\
0
\end{bmatrix} \Xbetadx_t -
\begin{bmatrix}
\sigma^8_t \\
0\\
0
\end{bmatrix} \Xbetady_t
\end{align*}
and, for every $t\in [\Ud, \Td]$,
\begin{align*}
\begin{bmatrix}
\psi^{d,1}_t\\
\psi^{d,2}_t\\
\psi^{d,3}_t
\end{bmatrix}
= \gm(t)Q_{\Sd}
\begin{bmatrix}
-\kpde Y^8_t\sigma^8_t \\
0 \\
0
\end{bmatrix} = -
\begin{bmatrix}
\sigma^8_t \\
0\\
0
\end{bmatrix} \Xbetady_t.
\end{align*}

\textit{Step 2.} We now compute the futures hedging strategy. Using the linear transformation
from Proposition \ref{pro4.8} we obtain, for every $t\in [0,\Sd]$,
\begin{align*}
\phiv_t^{d,1} &= (\nud_t)^{-1}\big( \psi^{d,1}_t - \psi^{d,3}_t\, (\nu^{\cq,3}_t)^{-1}\nu^{\cq,1}_t \big), \\
\phiv_t^{d,2} &= (\nufq_t)^{-1}\big( \psi^{d,2}_t - \psi^{d,3}_t\, (\nu^{\cq,3}_t)^{-1}\nu^{\cq,2}_t \big), \\
\phiv_t^{d,3} &= (\nu^{\cq,3}_t)^{-1} \psi^{d,3}_t
\end{align*}
where $\nu^{\cq,1}_t=-\btT\Fq_t,\,\nu^{\cq,2}_t=\hbtT\Fq_t$ and $\nu^{\cq,3}_t=\barsig\Fq_t$ (see Remark \ref{rem4.3}). Furthermore, from Remark \ref{rem4.2}, we know that
\begin{align*}
\nud_t=- \delta^{-1}(1+\delta\Fd_t)\btUT
\end{align*}
and, from Remark \ref{rem4.2} and Remark \ref{rem5.1}, we obtain
\begin{align*}
\nufq_t =-\delta^{-1}(1+\delta \Ff_t)\hbtUT Q_t.
\end{align*}
Consequently, for every $t\in [\Sd, \Ud]$,
\begin{align*}
\phiv_t^{d,1} &= (\nud_t)^{-1}\psi^{d,1}_t, \quad \phiv_t^{d,2} = \phiv_t^{d,3} = 0.
\end{align*}
Furthermore, for every $t\in [\Ud, \Td]$,
\begin{align*}
\phiv_t^{d,1} &= (\nud_t)^{-1}\psi^{d,1}_t, \quad \phiv_t^{d,2} = \phiv_t^{d,3} = 0
\end{align*}
Since $\psi^{d,2}_t=\psi^{d,3}_t=0$ for every $t \in [\Sd,\Td]$ it is clear that $\phiv^{d,1}_t =\phiv^{d,1}_t=0$ for
every $t \in [\Sd,\Ud]$. It is known from Remark \ref{rem4.2} that the futures rate $\Fd$ satisfies
$\diff\Fd_t=\nud_t\diff \Zone_t$ where $\nud_t=-\delta^{-1}(1+\delta\Fd_t)\btUT$ for every $t\in[0,\Ud]$. Straightforward computations show that the asserted equalities are valid.

\subsection{Proof of replication of the foreign leg}   \label{AppD2}

Similar to replication of domestic leg, we will also proceed in two steps but we will omit some details to avoid repetitions. To find explicit expressions for the processes $\psi^1,\psi^2$ and $\psi^3$ for the foreign leg at time $\Td$, it suffices to apply the It\^o product rule to the pricing formulae established in Proposition \ref{pro5.1}.

\textit{Step 1.} For every $t\in [0,\Ud]$,
\begin{align*}
\diff \wt{X}^{f,\beta}_t = \diff \big((B^{\beta}_t)^{-1} \Xbetaf_{t} \big)&=(B^{\beta}_t)^{-1}\gm(t)\LqtT  \Big( Y^1_t Y^2_t \big(\zeta^1_t+\zeta^2_t\big)
-Y^1_t Y^3_t \big(\zeta^1_t+\zeta^3_t\big) \Big) \diff Z_t \\
&=(B^{\beta}_t)^{-1}\gm(t)\LqtT \Big( Y^1_t Y^2_t [0,\sigma^2_t,\sigma^1_t]
-Y^1_tY^3_t [0,\sigma^3_t,\sigma^1_t] \Big)\diff Z_t.
\end{align*}
Next, for every $t\in [\Ud,\Td]$,
\begin{align*}
d\wt{X}^{f,\beta}_t&=(B^{\beta}_t\big)^{-1}\gm(t) \Big(- \LqtT  Y^1_tY^3_t \big(\zeta^1_t+\zeta^3_t\big)\Big)\diff Z_t \\
&=(B^{\beta}_t)^{-1}\gm(t)\Big(\LqtT  Y^1_t(\Bf_{\Ud})^{-1} \Bf_t [0,0,\sigma^1_t]-\LqtT  Y^1_tY^3_t [0,\sigma^3_t,\sigma^1_t]\Big)\diff Z_t.
\end{align*}
We thus obtain, for every $t\in [0,\Ud]$,
\begin{align*}
\begin{bmatrix}
\psi^{f,1}_t\\
\psi^{f,2}_t\\
\psi^{f,3}_t
\end{bmatrix}
= \gm (t)
\begin{bmatrix}
0 \\
Y^1_t Y^2_t\sigma^2_t -Y^1_tY^3_t\sigma^3_t \\
Y^1_t Y^2_t\sigma^1_t -Y^1_tY^3_t\sigma^1_t
\end{bmatrix} =
\begin{bmatrix}
0 \\
\sigma^2_t\\
\sigma^1_t
\end{bmatrix} \Xbetafx_t -
\begin{bmatrix}
0 \\
\sigma^3_t\\
\sigma^1_t
\end{bmatrix} \Xbetafy_t,
\end{align*}
and, for every $t\in [\Ud, \Td]$,
\begin{align*}  
\begin{bmatrix}
\psi^{f,1}_t\\
\psi^{f,2}_t\\
\psi^{f,3}_t
\end{bmatrix}
= \gm (t)\LqtT
\begin{bmatrix}
0 \\
-Y^1_tY^3_t\sigma^3_t \\
Y^1_t(\Bf_{\Ud})^{-1}\Bf_t\sigma^1_t-Y^1_tY^3_t\sigma^1_t
\end{bmatrix} =
\begin{bmatrix}
0 \\
0\\
\sigma^1_t
\end{bmatrix} \Xbetafx_t -
\begin{bmatrix}
0 \\
\sigma^3_t\\
\sigma^1_t
\end{bmatrix} \Xbetafy_t .
\end{align*}

\textit{Step 2.} Using Proposition \ref{pro4.8}, we obtain, for $t\in [0, \Td]$:
\begin{align*}
\phiv_t^{f,1} &= - (\nud_t)^{-1}  \psi^{f,3}_t\, (\nu^{\cq,3}_t)^{-1}\nu^{\cq,1}_t, \\
\phiv_t^{f,2} &= (\nufq_t)^{-1}\big( \psi^{f,2}_t - \psi^{f,3}_t\, (\nu^{\cq,3}_t)^{-1}\nu^{\cq,2}_t \big), \\
\phiv_t^{f,3} &= (\nu^{\cq,3}_t)^{-1} \psi^{f,3}_t.
\end{align*}
Straightforward computations show that the stated equalities are satisfied.

\subsection{Proof of replication of the exchange of nominal principals}  \label{AppD3}
It remains to compute the hedging strategy for the exchange of nominal principals, which was analyzed in  Proposition \ref{pro5.2}.

\textit{Step 1.} To find explicit expressions for the processes $\psi^1,\psi^2$ and $\psi^3$ for the exchange of notional principals at time $\Td$, it suffices to apply the It\^o product rule to the pricing formulae established in Proposition \ref{pro5.2}. For brevity, we use the notation
\begin{align*}
[Y^1_t,Y^3_t,Y^4_t,Y^6_t,Y^8_t]=[Q_t,\whB_{\Td}(t,\rf_t),\whB_{\Sd}(t,\rf_t),\FB_{\Sd,\Td}(t,\rd_t),B_\Td(t,\rd_t)].
\end{align*}

Then we have
\begin{align*}
\begin{bmatrix}
\diff Y^1_t/Y^1_t \\
\diff Y^3_t/Y^3_t \\
\diff Y^4_t/Y^4_t \\
\diff Y^6_t/Y^6_t \\
\diff Y^8_t/Y^8_t
\end{bmatrix}
\mart
\begin{bmatrix}
0&0&\sigma^1_t\\
0&\sigma^3_t&0\\
0&\sigma^4_t&0\\
\sigma^6_t&0&0\\
\sigma^8_t&0&0
\end{bmatrix}
\begin{bmatrix}
\diff \Zone_t\\
\diff \Ztwo_t\\
\diff \Zthree_t
\end{bmatrix}
=
\begin{bmatrix}
0&0&\barsig \\
0&\hbtT &0\\
0&\hbtS &0\\
\btST&0&0\\
\btT &0&0\\
\end{bmatrix}
\begin{bmatrix}
\diff \Zone_t\\
\diff \Ztwo_t\\
\diff \Zthree_t
\end{bmatrix}.
\end{align*}
From Proposition \ref{pro5.2}, the arbitrage-free price $\Xbetap_t$ equals, for every $t\in [0,\Sd]$,
\begin{align*}
\Xbetap_t=\gm(t) \big(\LqtT  Y^1_t Y^3_t-\LqtS \GST  Y^1_t Y^4_t Y^6_t\big)
\end{align*}
and, for every $t\in[\Sd,\Td]$,
\begin{align*}
\Xbetap_t= \gm(t) \big(\LqtT  Y^1_t Y^3_t - Q_{\Sd} Y^8_t\big).
\end{align*}
Therefore, for every $t\in [0,\Sd]$,
\begin{align*}
\diff \wt{X}^{p,\beta}_t = \diff \big( \big(B^{\beta}_t\big)^{-1} \Xbetap_{t} \big)&=\big(B^{\beta}_t\big)^{-1}\gm(t) \LqtT  Y^1_t Y^3_t \big(\zeta^1_t+\zeta^3_t\big)\diff Z_t \\
&\quad - \big(B^{\beta}_t\big)^{-1}\gm(t) \LqtS \GST Y^1_t Y^4_t Y^6_t \big(\zeta^1_t+\zeta^4_t+\zeta^6_t\big)\diff Z_t
\end{align*}
or, equivalently,
\begin{align*}
\diff \wt{X}^{p,\beta}_t&=\big(B^{\beta}_t\big)^{-1}\gm(t)\LqtT Y^1_t Y^3_t\big([0,\sigma^3_t,\sigma^1_t]\big)\diff Z_t \\
&\quad -\big(B^{\beta}_t\big)^{-1}\gm(t)\LqtS\GST Y^1_t Y^4_t Y^6_t\big([\sigma^6_t,\sigma^4_t,\sigma^1_t]\big)\diff Z_t,
\end{align*}
and, for every $t\in[\Sd,\Td]$,
\begin{align*}
d\wt{X}^{p,\beta}_t=\big(B^{\beta}_t\big)^{-1}\gm(t)\Big(\LqtT Y^1_tY^3_t\big(\zeta^1_t+\zeta^3_t\big)
-Q_{\Sd} Y^8_t\zeta^8_t\Big)\diff Z_t
\end{align*}
or, equivalently,
\begin{align*}
d\wt{X}^{p,\beta}_t= \big(B^{\beta}_t\big)^{-1}\gm(t)\Big(\LqtT Y^1_tY^3_t\big[0, \sigma^3_t, \sigma^1_t]\big)
-Q_{\Sd} Y^8_t[\sigma^8_t, 0, 0]\Big)\diff Z_t.
\end{align*}
We deduce that, for every $t\in [0,\Sd]$,
\begin{align*}
\begin{bmatrix}
\psi_t^1\\
\psi_t^2\\
\psi_t^3
\end{bmatrix}
= \gm (t)
\begin{bmatrix}
-\LqtS \GST Y^1_t Y^4_t Y^6_t \sigma^6_t \\
\LqtT  Y^1_t Y^3_t \sigma^3_t - \LqtS \GST Y^1_t Y^4_t Y^6_t \sigma^4_t\\
\LqtT  Y^1_t Y^3_t \sigma^1_t - \LqtS \GST Y^1_t Y^4_t Y^6_t \sigma^1_t
\end{bmatrix} =
\begin{bmatrix}
0 \\
\sigma^3_t\\
\sigma^1_t
\end{bmatrix} \Xbetapx_t -
\begin{bmatrix}
\sigma^6_t \\
\sigma^4_t\\
\sigma^1_t
\end{bmatrix} \Xbetapy_t
\end{align*}
and, for every $t\in [\Sd, \Td]$,
\begin{align*}  
\begin{bmatrix}
\psi_t^1\\
\psi_t^2\\
\psi_t^3
\end{bmatrix} =
\begin{bmatrix}
0 \\
\sigma^3_t\\
\sigma^1_t
\end{bmatrix} \Xbetapx_t -
\begin{bmatrix}
\sigma^8_t \\
0\\
0
\end{bmatrix} \Xbetapy_t.
\end{align*}
\textit{Step 2.}  To identify the futures hedging strategy, we use once again Proposition \ref{pro4.8}. We find that,
for every $t\in [0,\Td]$,
\begin{align*}
\phiv_t^{1} &= - (\nud_t)^{-1}  \psi^{3}_t\, (\nu^{\cq,3}_t)^{-1}\nu^{\cq,1}_t, \\
\phiv_t^{2} &= (\nufq_t)^{-1}\big( \psi^{2}_t - \psi^{3}_t\, (\nu^{\cq,3}_t)^{-1}\nu^{\cq,2}_t \big), \\
\phiv_t^{3} &= (\nu^{\cq,3}_t)^{-1}\psi^3_t,
\end{align*}
and straightforward computations now suffice to confirm the stated equalities. For detailed computations using two alternative methods, the reader is referred to \cite{Ding2024wp}. It is crucial to observe that while the pricing formula is identical on the intervals $[0, \Sd]$ and $[\Sd, \Ud]$, the dynamics of futures differ so that the futures trading strategy also varies across these periods.

\bibliographystyle{plain}
\bibliography{CCBSbiblio}

@article{Backwell2021jumps,
author = {Backwell, Alex and Hayes, Joshua},
title = {Expected and unexpected jumps in the overnight rate: {C}onsistent management of the {Libor} transition},
journal = {Journal of Banking and Finance},
volume = {145},
pages = {106669},
year = {2022}
}

@article{Backwell2020affine,
author = {Backwell, Alex and Macrina, Andrea and Schl{\"o}gl, Erik and Skovmand, David},
title = {Term rates, multicurve term structures and overnight rate benchmarks: {A} roll-over risk approach},
journal = {Frontiers of Mathematical Finance},
volume = {2},
number = {3},
pages  = {340--384},
year={2023}
}

@article{Biagini2021unified,
title = {A unified approach to {xVA} with {CSA} discounting and initial margin},
author = {Biagini, Francesca and Gnoatto, Alessandro and Oliva, Immacolata},
journal = {SIAM Journal on Financial Mathematics},
volume = {12},
number = {3},
pages = {1013--1053},
year = {2021}
}

@misc{Bickersteth2025SOFR,
author = {Bickersteth, Matthew and Ding, Yining and Rutkowski, Marek},
title = {Pricing and hedging of {SOFR} derivatives},
year = {2025},
note = {Available at arXiv: \url{https://arxiv.org/abs/2112.14033}}
}

@article{BCR2018,
author = {Bielecki, Tomasz R. and Cialenco, Igor and Rutkowski, Marek},
title = {Arbitrage-free pricing of derivatives in nonlinear market models},
journal = {Probability, Uncertainty and Quantitative Risk},
volume  = {3},
number  = {1},
pages = {1229--1258},
year = {2018}
}

@article{BR2015,
author = {Bielecki, Tomasz R. and Rutkowski, Marek},
title = {Valuation and hedging of contracts with funding costs and collateralization},
journal = {SIAM Journal on Financial Mathematics},
volume = {6},
number = {1},
pages = {594--655},
year = {2015}
}

@article{Brace2024,
author = {Brace, Alan and Gellert, Karol and Schl\"{o}gl, Erik},
title = {{SOFR} term structure dynamics---{D}iscontinuous short
rates and stochastic volatility forward rates},
journal = {Journal of Futures Markets},
volume  = {44},
pages = {936--985},
year  = {2024}
}

@article{Brigo2014funding,
author = {Brigo, Damiano and Pallavicini, Andrea},
title  = {Non-linear consistent valuation of {CCP} cleared or {CSA} bilateral trades with initial
margins under credit, funding and wrong-way risks},
journal = {Journal of Financial Engineering},
volume  = {1},
pages = {1--60},
year  = {2014}
}

@article{Buehler2019deephedging,
author = {Buehler, Hans and Gonon, Lukas and Teichmann, Josef and Wood, Ben},
title = {Deep hedging},
journal = {Quantitative Finance},
volume = {19},
number = {8},
pages = {1271--1291},
year = {2019}
}

@misc{Castagna2013,
author = {Castagna, Antonio},
title = {Pricing of derivatives contracts under collateral agreements: {L}iquidity and funding value adjustments},
year  = {2013},
note  = {Available at SSRN: \url{https://ssrn.com/abstract=1974479}}
}

@article{Crepey2015bilateralV1,
author = {Cr{\'e}pey, St{\'e}phane},
title = {Bilateral counterparty risk under funding constraints -- {Part I}: Pricing},
journal = {Mathematical Finance},
volume = {25},
number = {1},
pages = {1--22},
year = {2015}
}

@article{Crepey2015bilateralV2,
author = {Cr{\'e}pey, St{\'e}phane},
title = {Bilateral counterparty risk under funding constraints -- {Part II}: {CVA}},
journal = {Mathematical Finance},
volume = {25},
number = {1},
pages = {23--50},
year = {2015}
}

@article{Dam2020inflation,
author = {Dam, Henrik and Macrina, Andrea and Skovmand, David and Sloth, David},
title = {Rational models for inflation-linked derivatives},
journal = {SIAM Journal on Financial Mathematics},
volume = {11},
number = {4},
pages = {974--1006},
year = {2020}
}

@misc{Ding2024wp,
author = {Ding, Yining and Liu, Ruyi and Rutkowski, Marek},
title = {Multi-curve approach to cross-currency basis swaps referencing backward-looking rates},
year = {2024},
note = {Available at arXiv: \url{https://arxiv.org/abs/2410.08477}}
}

@article{fontana2023affine,
author = {Fontana, Claudio},
title = {Caplet pricing in affine models for alternative risk-free rates},
journal = {SIAM Journal on Financial Mathematics},
volume = {14},
number = {1},
pages = {SC1--SC16},
year = {2023}
}

@article{Fontana2023spikes,
author = {Fontana, Claudio and Grbac, Zorana and Schmidt, Thorsten},
title = {Term structure modelling with overnight rates beyond stochastic continuity},
journal = {Mathematical Finance},
volume = {34},
number = {1},
pages = {151--189},
year = {2023}
}

@misc{fujii2010note,
author = {Fujii, Masaaki and Shimada, Yasufumi and Takahashi, Akihiko},
title = {A note on construction of multiple swap curves with and without collateral},
year = {2010},
note = {Available at SSRN: \url{https://ssrn.com/abstract=1440633}}
}

@misc{fujii2010multi,
author = {Fujii, Masaaki and Shimada, Yasufumi and Takahashi, Akihiko},
title = {On the term structure of interest rates with basis spreads, collateral
and multiple currencies},
year = {2010},
note = {Available at SSRN: \url{https://ssrn.com/abstract=1556487}}
}

@article{Fujii2011Choice,
author = {Fujii, Masaaki and Takahashi, Akihiko},
title = {Choice of collateral currency},
year  = {2011},
journal = {Risk},
volume = {January},
pages = {120--125}
}

@misc{Gellert2021,
author = {Gellert, Karol and Schl\"{o}gl, Erik},
title = {Short rate dynamics: A {F}ed {F}unds and {SOFR} perspective},
year = {2021},
note = {Available at arXiv: \url{https://arxiv.org/abs/2101.04308}}
}

@article{Gnoatto2021multi,
author = {Gnoatto, Alessandro and Seiffert, Nicole},
title = {Cross currency valuation and hedging in the multiple curve framework},
journal = {SIAM Journal on Financial Mathematics},
volume = {12},
number = {2},
pages = {967--1012},
year = {2021}
}

@misc{Gnoatto2023cchjm,
author = {Gnoatto, Alessandro and Lavagnini, Silvia},
title = {Cross currency {H}eath-{J}arrow-{M}orton framework in the multiple-curve setting},
year = {2023},
note = {Available at arXiv: \url{https://arxiv.org/abs/2312.13057}}
}

@misc{Henrard2018,
author = {Henrard, Marc P. A.},
title = {Overnight futures: {C}onvexity adjustment},
year = {2018},
note = {Available at SSRN: \url{https://ssrn.com/abstract=3134346}}
}

@misc{Hofmann2020,
author = {Hofmann, Karl F.},
title = {Implied volatilities for options on backward-looking term rates},
year = {2020},
note = {Available at SSRN: \url{https://ssrn.com/abstract=3593284}}
}

@misc{ISDA2003,
author = {{ISDA}},
title = {{C}ollateral {A}sset {D}efinitions},
note  = {{I}nternational {S}waps and {D}erivatives {A}ssociation, {I}nc., New York, 2003.
Available at: \url{https://www.isda.org/book/collateral-asset-definitions}}
}

@misc{ISDA2005,
author = {{ISDA}},
title = {{C}ollateral {G}uidelines},
note  = {{I}nternational {S}waps and {D}erivatives {A}ssociation, {I}nc., New York, 2005. Available at: \url{https://www.isda.org/book/2005-isda-collateral-guidelines}}
}

@article{Jamshidian1994,
author = {Jamshidian, Farshid},
title = {Hedging quantos, differential swaps and ratios},
journal = {Applied Mathematical Finance},
volume = {1},
number = {1},
pages = {1--20},
year = {1994}
}

@misc{Lyashenko2019SOFR,
author = {Lyashenko, Andrei and Mercurio, Fabio},
title = {Looking forward to backward-looking rates: {A} modeling framework for term rates replacing {LIBOR}},
year  = {2019},
note = {Available at SSRN: \url{https://ssrn.com/abstract=3330240}}
}

@article{Macrina2020rfr,
author = {Macrina, Andrea and Skovmand, David},
title = {Rational savings account models for backward-looking interest rate benchmarks},
journal = {Risks},
volume = {8, No. 23},
year = {2020},
number = {1}
}

@misc{Mercurio2018,
author = {Mercurio, Fabio},
title = {A simple multi-curve model for pricing {SOFR} futures and other derivatives},
year = {2018},
note = {Available at SSRN: \url{https://ssrn.com/abstract=3225872}}
}

@book{MR2005,
author = {Musiela, Marek and Rutkowski, Marek},
title  = {Martingale Methods in Financial Modelling},
edition = {2nd},
publisher = {Springer, Berlin Heidelberg},
year  = {2005}
}

@book{oksendal2003sde,
author = {{\O}ksendal, Bernt},
title = {Stochastic Differential Equations: An Introduction with Applications},
edition = {6th},
publisher = {Springer, Berlin Heidelberg},
year = {2003}
}

@misc{pallavicini2011funding,
author={Pallavicini, Andrea and Perini, Daniele and Brigo, Damiano},
title={{F}unding {V}aluation {A}djustment: {A} consistent framework including {CVA},
 {DVA}, collateral, netting rules and re-hypothecation},
year={2011},
note = {Available at arXiv: \url{https://arxiv.org/abs/1112.1521}}
}

@book{Pelsser2000,
author = {Pelsser, Antoon},
title  = {Efficient Methods for Valuing Interest Rate Derivatives},
publisher = {Springer, London Berlin Heidelberg},
year  = {2000}
}

@article{Pelsser2003,
author = {Pelsser, Antoon},
title = {Mathematical foundation of convexity correction},
journal = {Quantitative Finance},
volume = {3},
number = {1},
pages = {59--65},
year = {2003}
}

@article{Piterbarg2010funding,
author = {Piterbarg, Vladimir},
title = {Funding beyond discounting: {I}mpact of stochastic funding and collateral
agreements and derivatives pricing},
year = {2010},
journal = {Risk},
volume = {February},
pages = {98--102}
}

@article{piterbarg2012cooking,
author = {Piterbarg, Vladimir},
title = {Cooking with collateral},
year = {2012},
journal = {Risk},
volume = {August},
pages = {58--63}
}

@misc{Piterbarg2020,
author = {Piterbarg, Vladimir},
title = {Interest rates benchmark reform and options markets},
year = {2020},
note = {Available at SSRN: \url{https://ssrn.com/abstract=3537925}}
}

@article{Rogers2010hedge,
author = {Rogers, L. C. G. and Singh, Surbjeet},
title = {The cost of illiquidity and its effects on hedging},
journal = {Mathematical Finance},
volume = {20},
number = {4},
pages = {597--615},
year = {2010}
}

@article{Skov2021rer,
author = {Skov, Jacob Bjerre and Skovmand, David},
title = {Dynamic term structure models for {SOFR} futures},
journal = {Journal of Futures Markets},
volume = {41},
YEAR = {2021},
number = {10},
pages = {1520--1544}
}

@article{Turfus2020,
author = {Turfus, Colin},
title = {Caplet pricing with backward-looking rates},
journal = {Wilmott Magazine},
volume = {September},
pages = {106--109},
year = {2022},
note = {Available at SSRN: \url{https://ssrn.com/abstract=3527091}}
}

@article{Vasicek1977,
author = {Vasicek, Oldrich},
title = {An equilibrium characterization of the term structure},
journal = {Journal of Financial Economics},
volume = {5},
number = {2},
pages = {177--188},
year = {1977},
issn = {0304-405X},
}

@article{Wolf2022,
author = {Wolf, Felix L. and Grzelak, Lech A. and Deelstra, Griselda},
title = {Cheapest-to-deliver collateral: a common factor approach},
journal = {Quantitative Finance},
volume = {22},
number = {4},
pages = {707--723},
year = {2022}
}

\end{document}